\documentclass[11pt]{article}
\usepackage[dvips]{graphicx}
\usepackage{amsfonts}
\usepackage{amssymb}
\usepackage{amsmath}
\usepackage{epsfig}
\usepackage{multirow}
\usepackage{latexsym}
\usepackage{graphicx}
\usepackage{float}

\usepackage[nosort]{cite}

\newcommand{\nn}{\nonumber}
\newcommand{\be}{\begin{equation}}
\newcommand{\ee}{\end{equation}}
\newcommand{\bea}{\begin{eqnarray}}
\newcommand{\eea}{\end{eqnarray}}

\newcommand{\lsim}{\lesssim}
\newcommand{\vphi}{\varphi}

\newcommand{\beqa}{\begin{eqnarray}}
\newcommand{\eeqa}{\end{eqnarray}}

\renewcommand{\a}{{\alpha}}

\newcommand{\f}{{\bf 5}}
\newcommand{\fb}{{\bf \bar{5}}}
\newcommand{\te}{{\bf 10}}

\newcommand{\op}{\oplus}

\topmargin
-0.7cm
\textwidth
15.5cm
\textheight
22.8cm
\oddsidemargin
0.7cm
\evensidemargin
1.2cm

\begin{document}

%----------------------------------------------------------------------%
%  numbering equations with section number
%----------------------------------------------------------------------%
\makeatletter
\@addtoreset{equation}{section}
\makeatother
\renewcommand{\theequation}{\thesection.\arabic{equation}}
%----------------------------------------------------------------------%
%  title page
%----------------------------------------------------------------------%
\pagestyle{empty}
%\vspace*{0.1cm}
\rightline{CPHT-RR065.0911}
\rightline{CERN-PH-TH/2011-243}
\rightline{ICCUB-11-170}
\vspace{2.2cm}
\begin{center}
\Large{\bf Massive wavefunctions, proton decay and FCNCs in local F-theory GUTs \\[12mm]}
\large{Pablo G. C\'amara$^{1,2}$,\; Emilian Dudas$^{3,4}$,\; Eran Palti$^3$ \\[5mm]}
\small{$^1$ CERN, PH-TH Division, CH-1211 Geneve 23, Switzerland}\\
\vspace{0.15cm}
\small{$^2$ Departament de F{\'\i}sica Fonamental and
Institut de Ci\`encies del Cosmos, Universitat de Barcelona,
Mart\'{\i} i Franqu\`es 1, E-08028 Barcelona, Spain}\\
\vspace{0.15cm}
\small{$^3$ Centre de Physique Th´eorique, Ecole Polytechnique, CNRS, F-91128 Palaiseau, France.}\\
\vspace{0.15cm}
\small{$^4$ LPT, UMR du CNRS 8627, Bˆat 210, Universit´e de Paris-Sud, F-91405 Orsay Cedex, France.} \\[3mm]

\small{E-mail: pcamara@cern.ch, emilian.dudas@cpht.polytechnique.fr, palti@cpht.polytechnique.fr} \\[12mm]
\small{\bf Abstract} \\[3mm]
\end{center}
\begin{center}
\begin{minipage}[h]{16.0cm}
We study the coupling of MSSM fields to heavy modes through cubic superpotential interactions in F-theory SU(5) GUTs. The couplings are calculated by integrating the overlap of two massless and one massive wavefunctions. The overlap integral receives contributions from only a small patch around a point of symmetry enhancement thereby allowing the wavefunctions to be determined locally on flat space, drastically simplifying the calculation. The cubic coupling between two MSSM fields and one of the massive coloured Higgs triplets present in SU(5) GUTs is calculated using a local eight-dimensional SO(12) gauge theory. We find that for the most natural regions of local parameter space the coupling to the triplet is comparable to or stronger than in minimal four-dimensional GUTs thereby, for those regions, reaffirming or strengthening constraints from dimension-five proton decay. We also identify possible regions in local parameter space where the couplings to the lightest generations are substantially suppressed compared to minimal four-dimensional GUTs. We further apply our results and techniques to study other phenomenologically important operators arising from coupling to heavy modes. In particular we calculate within a toy model flavour non-universal soft masses induced by integrating out heavy modes which lead to FCNCs.
\end{minipage}
\end{center}
\newpage
%----------------------------------------------------------------------%
%  Resetting of counters
%----------------------------------------------------------------------%
\setcounter{page}{1}
\pagestyle{plain}
\renewcommand{\thefootnote}{\arabic{footnote}}
\setcounter{footnote}{0}
%----------------------------------------------------------------------%
%  Paper begins
%----------------------------------------------------------------------%

%\vspace*{1cm}

\tableofcontents

%\newpage

%%%%%%%%%%%%%%%%%%%%%%%%%%%%%%%%%%%%%%%%%%%%%%%%%%%%%%%%%%%%%%%%%%%%%%%%%%%%%
\section{Introduction}
\label{sec:intro}
%%%%%%%%%%%%%%%%%%%%%%%%%%%%%%%%%%%%%%%%%%%%%%%%%%%%%%%%%%%%%%%%%%%%%%%%%%%%%

Although string theory is primarily motivated as a fundamental unified theory because of its ultraviolet behaviour, phenomenological model building within string theory often concerns only the infrared spectrum. This is a natural first step given the expected hierarchy between the string and electroweak scales. However, heavy modes play a crucial role in our understanding of much of the physics which is relevant to the Standard Model and extensions of it, for example by inducing higher dimension operators in the infrared. The fact that studying such modes explicitly requires a good understanding of the ultraviolet physics means that this is one of the subjects where string phenomenology can play an important role.
Heavy modes are particularly important in the case of Grand Unified Theories (GUTs). For instance, the result of gauge coupling unification at the GUT scale is sensitive to threshold corrections from heavy modes, and one of the classic constraints on GUTs comes from dimension-five proton decay operators that are induced by integrating out heavy modes.

Studying detailed properties of these fields, such as their wavefunction profile, is typically a difficult prospect because of the complicated geometry associated to realistic models of particle physics in string theory. String modes can only be concretely studied in simple geometries where a world-sheet description is available. Kaluza-Klein (KK) modes, and the closely related Landau-levels\footnote{These are sometimes referred to as \emph{gonions} in the intersecting brane literature \cite{Aldazabal:2000cn}.}, are typically difficult to solve for within some complicated Calabi-Yau (CY) geometry. However, in some models, and in particular F-theory (or type IIB) GUTs, many important operators of the theory are associated to only a small patch within the full geometry. The extreme example of this are Yukawa couplings, which are associated to just a single point in the geometry. Analogous to the Yukawa couplings there are triple couplings between heavy modes and massless modes which can be locally studied within a small region around a point. Since locally the complicated global CY geometry is decoupled and essentially we can work on flat space, many properties of heavy modes become accessible. In this paper we use this local approach to study the coupling of heavy modes to massless modes through such a triple coupling operator. This is done by solving for the local form of  wavefunctions of massive and massless modes and calculating their triple overlap.

The particular operator that we study, coupling one heavy mode to two massless ones, plays a key role in GUTs. One of the general features of SU(5) GUTs is that associated to the MSSM Higgs doublets there are coloured triplets which complete a GUT representation. These modes have to obtain a mass, leading to the so called doublet-triplet splitting problem. Similarly, associated to the Yukawa couplings there are also triple couplings between one heavy triplet and two MSSM fields. Once the heavy triplets are integrated out these couplings induce dimension-five baryon and lepton number violating operators that lead to proton decay. Thus, understanding such couplings and their flavour structure is of crucial importance for placing constraints on GUT models. In minimal field-theory SU(5) GUTs these couplings are the same as Yukawa couplings and therefore are exactly known. However, in string theory GUTs this is not the case and the couplings can be completely different in nature. This means that without knowledge of how the triplets couple to the matter fields and in particular whether the coupling to the lightest generations is suppressed in a manner similar to the Yukawa couplings it is not possible to use dimension-five proton stability to constrain model building. The primary aim of this paper is to study the nature of the triplet couplings within a realistic string setup thereby performing this crucial step in imposing phenomenological constraints on string theory GUTs.

More generally the paper aims to show that much important physics can be extracted by similar calculations of couplings to heavy modes. Indeed, in section \ref{sec:pheno} we present a toy model where such a calculation allows to extract Flavour-Changing-Neutral-Current (FCNC) terms which, like proton stability, form one of the important observational constraints on ultraviolet physics. We also discuss how our calculations apply to string theoretic realisations of the Froggatt-Nielsen mechanism for generating flavour structure this being yet another mechanism which relies on higher dimension operators.

Our focus is on local F-theory GUTs \cite{Vafa:1996xn,Morrison:1996na,Donagi:2008ca,Beasley:2008dc,Beasley:2008kw,Donagi:2008kj,Denef:2008wq}. Within this framework a 7-brane carrying an SU(5) gauge group wraps a 4-dimensional surface $S$ inside a CY four-fold. Other 7-branes intersecting this brane are locally modeled by an enhancement of the gauge symmetry over loci in $S$: along complex curves on $S$ the group enhances by rank 1, to SU(6) or SO(10), while on the points where complex curves intersect it enhances by at least rank 2, to SU(7), SO(12) or E$_6$. We are particularly interested in a point of enhancement to SO(12) as it is there that the down-type Yukawa interaction is localised. To describe the physics near such point we consider an 8-dimensional gauge theory, which is just super Yang-Mills twisted to account for the embedding into the CY four-fold \cite{Donagi:2008ca,Beasley:2008dc,Conlon:2009qq}, with SO(12) gauge group broken down to SU(5)$\times$U(1)$\times$U(1) by a spatially varying Higgs field. Matter localises onto complex curves where the Higgs vev vanishes and at the SO(12) enhancement point three such matter curves intersect giving rise to a cubic coupling in the 4-dimensional effective theory. This coupling can be calculated directly from dimensional reduction of the 8-dimensional theory by integrating the overlap of the internal wavefunctions of localised fields. Yukawa couplings are calculated by overlaps of wavefunctions of three massless modes \cite{Cremades:2004wa} and have been extensively studied in \cite{Heckman:2008qa, Hayashi:2009ge, Font:2009gq, Cecotti:2009zf, Conlon:2009qq, Hayashi:2009bt, Aparicio:2011jx} (see also \cite{Abe:2008fi, Conlon:2008qi, DiVecchia:2008tm, Antoniadis:2009bg} in the context of magnetised D-branes). In this paper we calculate the wavefunctions for massive modes around an SO(12) point. Similar calculations of massive mode wavefunctions for other models were performed in \cite{Camara:2009xy, Marchesano:2010bs, Aparicio:2011jx}. Once we obtain the wavefunctions for massive modes we can study their overlap with massless wavefunctions, thereby probing the cubic coupling discussed above.

An important property of the dimension-five proton decay operator we are studying is that it is a superpotential operator. Since in type IIB string theory and F-theory the superpotential does not receive $\alpha'$ corrections, and since integrating out massive string oscillator modes induces $\alpha'$ corrections, we do not expect the operator to be induced by exchanging massive string oscillator modes. Thus, all the relevant heavy modes which participate in dimension-five proton decay are captured within the effective gauge theory described above.

\begin{figure}[ht!]
\vspace{0.6cm}
\begin{center}
\includegraphics[width=7cm]{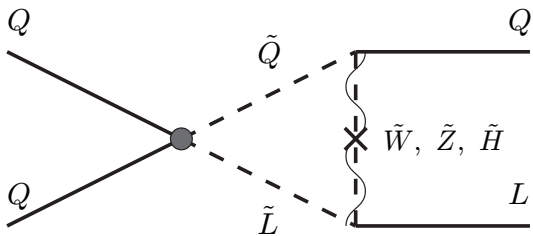}
\caption{Schematic structure of Feynman diagrams leading to proton decay through the dimension-five effective operators (\ref{supprotdec}). Similar diagrams exist involving right-handed states.\label{figdim5}}
\end{center}
\end{figure}

The calculation of the coupling to massive modes at an SO(12) point is only a part of the full calculation required to understand dimension-five proton stability. It is therefore worth discussing how the present calculation fits within the full picture of dimension-five proton stability in F-theory SU(5) GUTs. We begin by reviewing the constraints on the relevant operators. The effective superpotential couplings take the schematic form
\begin{multline}
W \supset Y^u_{ij} H^u  Q_i U_j + Y^d_{ij} H^d \left(Q_i D_j + L_i E_j\right)
\, +\,  \hat{Y}^u_{ij} T^u \left( Q_i Q_j + U_i E_j \right)\\
\, +\, \hat{Y}^d_{ij} T^d \left( Q_i L_j + U_i D_j \right)
\, +\,  M T^u T^d \;.
\end{multline}
Here $Y^u_{ij}$ and $Y^d_{ij}$ denote the up- and down-type Yukawa couplings with the MSSM superfields expressed in standard notation. The coloured triplets are denoted by $T^u$ and $T^d$ and their associated triple couplings are $\hat{Y}^u_{ij}$ and $\hat{Y}^d_{ij}$. In minimal 4-dimensional GUTs we have $Y^{u,d}_{ij}=\hat{Y}^{u,d}_{ij}$.\footnote{This relation and the above superpotential may be slightly modified by more complicated theories where the cubic couplings arise after fields in non-trivial GUT representations obtain a vev, as proposed for example in \cite{Georgi:1979df} to fix the GUT mass relations.} The scale $M$ is related to the mass of the triplets and is expected to be at or below the GUT scale, $M_{\rm GUT}$. Integrating out the heavy triplets leads to dimension-five operators
\be
\label{supprotdec}
W \supset \frac{\hat{Y}^u_{ij}\hat{Y}^d_{kl}}{M} \left( Q_i Q_j Q_k L_l + U_i E_j U_k D_l \right) \;.
\ee
There are a number of diagrams that lead to proton decay and involve
these operators.\footnote{There are a large number of papers which
  study nucleon decay in 4-dimensional supersymmetric GUTs. We refer to \cite{Weinberg:1981wj, Dimopoulos:1981dw, Sakai:1981pk, Ellis:1981tv, Ellis:1983qm, Ibanez:1991pr, Hisano:1992jj, Goto:1998qg, Murayama:2001ur, Bajc:2002bv, Bajc:2002pg, Raby:2002wc, EmmanuelCosta:2003pu} for a subset. Note that a number of these papers were using old experimental results on the proton lifetime which has since increased by 2 - 3 orders of magnitude.}
At TeV scale the diagrams involve a 4-point interaction coming from (\ref{supprotdec}) which has two fermions and two scalar superpartners, and a loop factor involving wino or Higgsino exchange to turn the scalars into fermions (c.f. figure \ref{figdim5}). This results in nucleon decay to kaons primarily (due to the need for a strange quark because of the anti-symmetric colour index). To discuss the constraints on the operators (\ref{supprotdec}) let us fix $M=M_{\rm GUT}$ and quote limits on $\hat{Y}^u_{ij}\hat{Y}^d_{kl}$ for different generation indices.\footnote{Note that using the results of \cite{Conlon:2007zza} that higher dimension superpotential operators are expected to be suppressed by the winding scale, and those of \cite{Conlon:2009xf,Conlon:2009kt,Conlon:2009qa} showing that the winding scale is also the unification scale, implies that $M_{\rm GUT}$ is a quite natural suppression scale. Of course the arguments given are simply scaling arguments and should not be taken to hold to significant accuracy. Nevertheless a suppression mass scale larger than $M_{\rm GUT}$ seems unlikely given that there are always some heavy modes at or below this scale.}
The precise constraints depend on a number of factors such as the soft masses, the size of the $\mu$-term and tan $\beta$ (with small tan $\beta$ and large soft masses giving generally weaker constraints). Instead of going into the details of these studies, we can concentrate on the most relevant aspect for the study in this paper: the difference between the Higgs and the coloured triplet couplings. In 4-dimensional field-theory analysis these couplings are taken as equal and this leads to an approximate bound $M_{\rm GUT}/M \lsim 10^{-2}-10^{-4}$. Thus, if coloured triplet couplings are suppressed with respect to Yukawa couplings by a factor larger than the 4-dimensional constraint on $M_{\rm GUT}/M$, dimension-five proton stability constraints can be satisfied for triplet masses of the order of the GUT scale or above. We can estimate the relevant parameters involving the down-type triplet couplings in the minimal 4-dimensional field-theory analysis,
\begin{equation}
\frac{\hat{Y}^d_{\tilde{t}b}}{Y^d_{bb}} = 1 \;, \qquad \frac{\hat{Y}^d_{\tilde{t}s}}{Y^d_{bb}} \sim 10^{-1} \;, \qquad \frac{\hat{Y}^d_{\tilde{t}d}}{Y^d_{bb}} \sim \frac{\hat{Y}^d_{\tilde{c}s}}{Y^d_{bb}} \sim 10^{-2} \;, \qquad \frac{\hat{Y}^d_{\tilde{c}d}}{Y^d_{bb}} \sim \frac{\hat{Y}^d_{\tilde{u}d}}{Y^d_{bb}} \sim 10^{-3} \;.\label{4dratios}
\end{equation}
This simply comes from using the 4-dimensional field-theory equalities, for example $\hat{Y}^d_{\tilde{c}s} = Y^d_{ss}$, and the measured quark masses and mixings. We will calculate precisely these ratios for local F-theory SU(5) GUTs and compare to the above values to see if there is enough additional suppression to avoid proton decay or, alternatively, if there is an enhancement thereby making proton decay constraints more severe.\footnote{It is important to note that in order to suppress proton decay all the ratios in (\ref{4dratios}) must be suppressed since, probing a superpotential coupling, we are working in a weak eigenstate basis. A single large coupling in the weak basis can lead to several large couplings in the mass eigenstate basis. Thus, to enhance the rate of proton decay it is sufficient that only one of the ratios is larger than in minimal 4-dimensional GUTs. Note also that some of the couplings may vanish through other selection rules such as additional symmetries or the fact that the colour index in the dimension-five operator must be anti-symmetric and so it cannot involve all the same generation. In such cases the ratios involving those operators would not be constrained and it would suffice to suppress only the other ratios.}

Calculating the parameters in (\ref{4dratios}) manifestly requires a theory of flavour. The study of flavour structures within F-theory GUTs has been an active research area in the recent years \cite{Beasley:2008kw,Heckman:2008qa,Hayashi:2009ge,Font:2009gq,Cecotti:2009zf,Conlon:2009qq,Hayashi:2009bt,Aparicio:2011jx,Font:2008id,Marchesano:2009rz,Dudas:2009hu,King:2010mq,Dudas:2010zb,
Leontaris:2010zd,Ludeling:2011en,Callaghan:2011jj}. We make use of the theory of flavour first proposed in \cite{Heckman:2008qa} and subsequently elucidated in \cite{Hayashi:2009ge,Font:2009gq,Cecotti:2009zf,Conlon:2009qq,Marchesano:2009rz,Aparicio:2011jx}.\footnote{There are two key motivations for studying this proposal as opposed to say that of \cite{Dudas:2009hu} which was based on a Froggatt-Nielsen mechanism with additional U(1) symmetries. The first is a practical one: it is not possible to study the flavour structure of \cite{Dudas:2009hu} locally near an SO(12) point. The second is that within the structure of \cite{Dudas:2009hu} the relations (\ref{4dratios}) are always at least as strong as in minimal field-theory GUTs since the suppression by the U(1) symmetries acts on the triplets in the same way as on the Higgs doublets.} The structure is such that all three generations are localised on a single matter curve and arise from the degeneracy of massless Landau-levels in the presence of flux.
This theory of flavour, however, requires ingredients which are not
present in our setup. More precisely, it was shown in
\cite{Cecotti:2009zf,Conlon:2009qq,Marchesano:2009rz} that to generate
non-vanishing Yukawa couplings for anything other than the heaviest
generation requires a non-commutative deformation of the theory
induced by closed string fluxes or non-perturbative effects. We
discuss this in more detail in section \ref{sec:pheno} but for now it
is sufficient to state that this does not affect the calculation we
are performing. Unlike Yukawa couplings, the coupling of one massive
mode to two massless ones is non-vanishing even in the absence of the
required non-commutative deformation. Turning on the additional,
necessarily small, such deformation will only perturb slightly our present calculation thereby maintaining its validity. Note that the fact that in the concrete setup we are using the Yukawa couplings are rank one in generation space while the triplet couplings are rank three highlights the fundamental difference between these couplings.

The computation that we perform is a necessary one to understand a
number of phenomenological issues. Regarding the particular problem of
dimension-five proton stability, it is worth discussing some
alternative solutions that have been proposed within the context of
F-theory. One way to avoid inducing proton decay is by having a
symmetry which forbids it. One such candidate symmetry is a U(1)
symmetry, which we label as U(1)$_{\rm PQ}$. Such a (massive) symmetry
has been studied in detail in F-theory GUT models, see for example
\cite{Heckman:2008qt,Bouchard:2009bu,Donagi:2009ra,Heckman:2009mn,Marsano:2009gv,Marsano:2009wr,Hayashi:2010zp,Grimm:2010ez,Pawelczyk:2010xh,Marsano:2010sq,Dolan:2011iu,Dolan:2011aq,Grimm:2011tb,Dudas:2009hu,Dudas:2010zb}.
Although at the GUT level it was found that many models can exhibit
such a symmetry, it was shown in a series of papers
\cite{Marsano:2009gv,Marsano:2009wr,Dudas:2009hu,Dudas:2010zb,Marsano:2010sq,Dolan:2011iu}
that the use of hypercharge flux to break the GUT group and induce
doublet-triplet splitting is incompatible with such a
symmetry.\footnote{As pointed out in \cite{Anderson:2011ns}, the same
  problem arises for Wilson-line GUT breaking.} The precise statement
is that the presence of a U(1)$_{\rm PQ}$ symmetry necessarily implies
the presence of exotic non-MSSM states in the massless spectrum. The
mass of the exotic states is set by the scale at which the U(1)$_{\rm
  PQ}$ symmetry is broken and therefore the constraints coming from
dimension-five proton decay translate to constraints on the mass of
the exotic states. The phenomenology associated to different masses
for the exotic states was studied in detail for a number of models in
\cite{Dudas:2010zb,Dolan:2011iu,Dolan:2011aq}. Since the exotic states do not form
complete GUT multiplets, the most immediate constraints on their
masses come from gauge coupling unification. The tension between a
large exotics mass to maintain gauge coupling unification and a small
mass to preserve an approximate U(1)$_{\rm PQ}$ symmetry implies that
it is difficult to practically realise the full suppression necessary
for dimension-five proton stability using such a symmetry alone. The
suppression due to a U(1)$_{\rm PQ}$ symmetry is additive to that
studied in this work and therefore whether we find additional
suppression or alternatively an enhancement of coupling to massive
modes can allow for or rule out a
number of proposed models.

An alternative possibility for suppressing dimension-five proton decay even without a U(1)$_{\rm PQ}$ is keeping the matter curves associated to the up and down Higgs fields in the same homology class but still geometrically separated. The interaction between the up and down triplets may then be suppressed by their small wavefunction overlap, although explicitly studying this would require a calculation of massive wavefunctions similar to that presented in this paper. Apart from the fact that this rather complicated setup has yet to be realised explicitly, there are a number of phenomenological problems with such a setup. The first is that the use of hypercharge flux for doublet-triplet splitting is difficult since it acts in the same way on both the Higgs curves. Another problem is that the theory of flavour of \cite{Heckman:2008qa} is based on local geometric symmetries which means that in order to correlate the up- and down-type Yukawas, as is required by a realistic CKM matrix, the geometric separation between them should be small. Indeed this is one of the primary motivations presented in \cite{Heckman:2009mn} for a proposed point of E$_8$ unification.

The outline of this paper is as follows. In section \ref{sec:effthe} we introduce the effective theory that we will be using. Following this, in sections \ref{sec:wave} and \ref{sec:overlaps} we present the actual calculations of the relevant wavefunctions and their overlaps. In section \ref{sec:pheno} the results and their phenomenological applications are discussed in detail. We summarise our findings in section \ref{sec:summary}. In appendix \ref{app:oblique} we present the wavefunctions and overlaps for a more general set of background fluxes, consisting also of oblique fluxes. In appendix \ref{app:norm} we study in more detail the normalisation of the wavefunctions.

%%%%%%%%%%%%%%%%%%%%%%%%%%%%%%%%%%%%%%%%%%%%%%%%%%%%%%%%%%%%%%%%%%%%%%%%%%%%%
\section{The effective theory}
\label{sec:effthe}
%%%%%%%%%%%%%%%%%%%%%%%%%%%%%%%%%%%%%%%%%%%%%%%%%%%%%%%%%%%%%%%%%%%%%%%%%%%%%

We consider F-theory on an elliptically fibered Calabi-Yau 4-fold $X$, with the degrees of freedom of an SU(5) GUT localised on a codimension-2 singularity. In the infrared those are described by a twisted 8-dimensional $\mathcal{N}=1$ gauge theory, with gauge group $G$ and support on $\mathbb{R}^{1,3}\times S$, where $S$ is a 4-dimensional K\"ahler sub-manifold of $X$ \cite{Donagi:2008ca, Beasley:2008dc}.\footnote{If  $S$ is shrinkable (more formally it has an ample normal bundle), the resulting 4-dimensional gauge coupling $\alpha_{\rm GUT}$ can be tuned independently of $M_{\rm Planck}$. However we do not necessarily assume this property in our analysis.} In this section we describe the 8-dimensional effective theory and how its massless and massive localised spectrum is calculated. Related computations to the ones that we describe here have also recently appeared in \cite{Aparicio:2011jx}.

%%%%%%%%%%%%%%%%%%%%%%%%%%%%%%%%%%%%%%%%%%%%%%%%%%%%%%%%%%%%%%%%%%%%%%%%%%%%%
\subsection{The 8-dimensional effective theory}
\label{sec:8d}
%%%%%%%%%%%%%%%%%%%%%%%%%%%%%%%%%%%%%%%%%%%%%%%%%%%%%%%%%%%%%%%%%%%%%%%%%%%%%

For convenience, we arrange the 8-dimensional fields in adjoint valued, $S$-valued, 4-dimensional $\mathcal{N}=1$ multiplets
\bea
{\bf A}_{\bar{m}} &=& \left( A_{\bar{m}}, \psi_{\bar{m}}, \mathcal{G}_{\bar{m}}\right) \;, \label{hol}\\
{\bf \Phi}_{mn} &=& \left( \varphi_{mn}, \chi_{mn}, \mathcal{H}_{mn} \right) \;, \nn\\
{\bf V} &=& \left( \eta, A_{\mu}, \mathcal{D} \right) \;.\nn
\eea
The subindices on the fields denote their local differential structure on $S$. Thus, for instance $A_{\bar{m}} \in \bar{\Omega}^1_{S}\otimes\mathrm{ad}(P)$ where $\Omega^p_S$ denotes the space of holomorphic $p$-forms on $S$ and $P$ is the principal bundle (in the adjoint representation) associated to the gauge group $G$.
Here ${\bf A}$ and ${\bf \Phi}$ are chiral multiplets with respective F-terms $\mathcal{G}$ and $\mathcal{H}$. ${\bf V}$ is a vector multiplet with D-term $\mathcal{D}$. $A_{\bar{m}}$ and $\varphi_{mn}$ are complex scalars while $\psi_{\bar{m}}$, $\chi_{mn}$, and $\eta$ are fermions.

The action for the effective theory was given in \cite{Beasley:2008dc}. For the bosonic components of the multiplets it reads,\footnote{This action can be shown to be equivalent to 8-dimensional super Yang-Mills theory with a non-trivial Higgs bundle \cite{Conlon:2009qq}.}
\begin{multline}
S_{\rm 8d}=M_*^4\int_{\mathbb{R}^{1,2}\times S}d^4x \ \textrm{Tr}\left[\omega\wedge\omega\left(\frac12 \mathcal{D}^2-\frac14F_{\mu\nu}F^{\mu\nu}\right)-D_\mu\varphi\wedge D^\mu\bar\varphi+2i\omega\wedge \mathcal{G}\wedge \overline{\mathcal{G}}\right.\\
\left.+\mathcal{H}\wedge\overline{\mathcal{H}}-F^{(2,0)}\wedge \overline{\mathcal{H}}-F^{(0,2)}\wedge \mathcal{H}-\overline{\mathcal{G}}\wedge \partial_A\bar\varphi-\mathcal{G}\wedge\bar\partial_A\varphi\right.\\
\left.+2\left(\omega\wedge F^{(1,1)}+\frac{i}{2}[\varphi,\bar\varphi]\right)\mathcal{D}-2i\omega \wedge F_{S\mu}^{(1,0)}\wedge F_{S}^{(0,1)\mu}+\ldots\right]\label{8dact}
\end{multline}
where $\omega$ is the K\"ahler form of $S$. Our conventions are such that $\omega$ is dimensionless, $A$ and $\varphi$ have dimensions of mass and the auxiliary fields $\mathcal{D}$, $\mathcal{G}$ and $\mathcal{H}$ have dimensions of $\textrm{(mass)}^2$. $M_*$ denotes the UV cutoff of the theory. In the weakly coupled type IIB limit this is related to the string scale as
\begin{equation}
M_*^4=(2\pi)^{-5}\alpha'^{-2} \;.
\end{equation}
Above this scale, corrections to eq.~(\ref{8dact}) in the form of higher derivative couplings become important and keeping only the leading term in the $M_*$ expansion, eq.~(\ref{8dact}), is not a valid approximation. Thus, in what follows we shall stick to the regime where, at every point of $\mathbb{R}^{1,3}\times S$,
\begin{equation}
\langle \partial A\rangle, \ \langle \partial\varphi\rangle \ll M_*^2. \label{alpha}
\end{equation}

Ideally we would like to dimensionally reduce the 8-dimensional effective action (\ref{8dact}) on $S$ in order to obtain the spectrum of 4-dimensional fields with masses smaller than $M_*$. However, such a program would require the precise knowledge of the geometry of $S$, which in general is only available for few highly symmetric spaces such as $T^4$, $\mathbb{P}^1\times \mathbb{P}^1$ or $\mathbb{P}^2$ (see for instance \cite{Conlon:2008qi}). Alternatively, we can solve the equations of motion in a local patch around a particular point of $S$ where the energy density of a set of charged modes localises. This approach has been extensively used in recent phenomenological studies of Yukawa couplings in F-theory GUTs (see e.g. \cite{Heckman:2008qa,  Hayashi:2009ge, Font:2009gq, Cecotti:2009zf, Conlon:2009qq, Hayashi:2009bt, Aparicio:2011jx}). In what follows we describe it in detail.

%%%%%%%%%%%%%%%%%%%%%%%%%%%%%%%%%%%%%%%%%%%%%%%%%%%%%%%%%%%%%%%%%%%%%%%%%%%%%%%%%%%%%%%%%
\subsection{Equations of motion for localised fields}
%%%%%%%%%%%%%%%%%%%%%%%%%%%%%%%%%%%%%%%%%%%%%%%%%%%%%%%%%%%%%%%%%%%%%%%%%%%%%%%%%%%%%%%%%

Let us first consider 4-dimensional massless fields. Setting the 4-dimensional variations of the fields to zero, the equations of motion that follow for their internal wavefunctions are \cite{Beasley:2008dc}
\bea
& &\mathcal{H} - F^{(2,0)} = 0 \;, \label{eom1} \\
& &i\left[ \varphi, \bar{\varphi} \right] + 2\omega \wedge F^{(1,1)} + \star_{S} \mathcal{D} = 0 \;, \label{eom2} \\
& &2i\omega \wedge \bar{\mathcal{G}} - \bar{\partial}_{A} \varphi = 0 \;, \label{eom3} \\
& &- \partial \bar{\mathcal{H}} + 2 \omega \wedge \bar{\partial} \mathcal{D} + \bar{\mathcal{G}} \wedge \bar{\varphi} - \bar{\chi} \wedge \bar{\psi}- i 2\sqrt{2} \omega \wedge \eta \wedge \psi = 0 \;,  \label{eom4} \\
& &\omega \wedge \partial_{A} \psi + \frac{i}{2} \left[ \bar{\varphi}, \chi \right] = 0 \;,  \label{eom5} \\
& &\bar{\partial}_A \chi - 2i\sqrt{2} \omega \wedge \partial_A \eta - \left[ \varphi, \psi \right] = 0 \;,  \label{eom6} \\
& &\bar{\partial}_{A} \psi - \sqrt{2} \left[ \bar{\varphi}, \eta\right] = 0 \;, \label{eom7} \\
& &-\sqrt{2}\left[\bar{\eta},\bar{\chi}\right] - \bar{\partial}_{A} \mathcal{G} - \frac12 \left[\psi,\psi\right]= 0 \;. \label{eom8}
\eea
where we have also included the equations of motion for the fermionic fluctuations. Eqs.~(\ref{eom1}) and (\ref{eom2}) are usually dubbed as the F-term and the D-term conditions for the flux, respectively.

Generically, the gauge group $G$ is broken by $\langle\varphi\rangle$ and $\langle A\rangle$ to a smaller subgroup $G'\subset G$. We are interested in vacua where $\langle \varphi\rangle$ and $\langle A\rangle$ take values in a subgroup $G_H=U(1)\times\ldots\times U(1)$ belonging to the Cartan of $G$,
\begin{equation}
G\to G'\times G_H
\end{equation}
with $G'$ the commutant of $G$ and $G_H$. In that case eq.~(\ref{eom3}) can be simply satisfied by requiring  $\langle\varphi\rangle$ to be holomorphic
\be
\bar{\partial}_{A} \langle\varphi\rangle = \bar{\partial} \langle\varphi\rangle + \left[ A , \langle\varphi\rangle \right] = \bar{\partial} \langle\varphi\rangle  = 0 \;.
\ee
Note also that to preserve 4-dimensional Poincar\'e invariance we must impose
\be
\langle \chi \rangle = \langle \psi \rangle = \langle \eta \rangle = 0 \;, \label{vanishvev}
\ee
which, making use of eqs.~(\ref{eom1})-(\ref{eom3}), imply that eqs.~(\ref{eom4}) and (\ref{eom8}) are automatically satisfied.

Modes charged under $G_H$ arrange into Landau levels and are localised around points in $S$ where $\langle \varphi\rangle=\langle A\rangle=0$. In order to have a description of these modes it therefore suffices to consider a local patch around a localisation point.
We can take the K\"ahler form to be given in the local patch by the expansion
\be
\omega = \frac{i}{2} \left(dz_1 \wedge d\bar{z}_{\bar{1}} + dz_2 \wedge d\bar{z}_{\bar{2}} \right)+\ldots
\ee
where the dots denote higher order terms in the two local complex coordinates $z_1$ and $z_2$.  Our conventions are such that coordinates are dimensionless, $dz_i$ denote local vielbein 1-forms and the origin of coordinates is at the localisation point.

Similarly, we expand\footnote{Note that most generally we could have also considered constant terms in these local expansions.}
\begin{align}
\langle A\rangle&=-\frac{M_*}{R_{\parallel}}\,\textrm{Im}( M^a_{ij} z_id\bar z_j)Q_a+\ldots\ , \label{a}\\ \langle\varphi\rangle&=M_*R_{\perp}\, m^a_{i} z_i \, Q_a dz_1\wedge dz_2+\ldots\ ,\label{phi}
\end{align}
where $M^a_{ij}$ and $m^a_{i}$ are arbitrary numbers related to the quanta of gauge and Higgs fluxes and $Q_a$ denote the different Abelian generators of $G_H$. We have also introduced the standard lengths $R_{\parallel}$ and $R_{\perp}$ in the local patch and its transverse space respectively, measured in $M_*^{-1}$ units, which for simplicity we have taken to be the same in all directions. Note that we have chosen to parameterise the dimensionful part of $\vphi$ with the transverse (``winding'') scale $R_{\perp}M_*$. This is not an arbitrary choice but follows from embedding the 8-dimensional theory into a 10-dimensional theory where the Higgs would correspond to deformations of the 7-brane into the normal directions. In the 8-dimensional theory the Higgs kinetic term arises from the pull-back of the 10-dimensional metric normal to the brane,
\begin{equation}
S_{\rm 7-brane}\supset M_*^4\int d^8x\ g_{3\bar 3}\partial_\mu\phi^3 \partial^\mu\bar \phi^{\bar 3} = M_*^4\int d^8x\ M_*^2R_{\perp}^2\partial_\mu\phi^3 \partial^\mu\bar \phi^{\bar 3} =M_*^4\int d^4x\ \partial_\mu\varphi \wedge \partial^\mu\bar\varphi
\end{equation}
where $g_{3\bar 3}$ is the metric transverse to the 7-brane and $\phi^3$ is the complex scalar parameterizing geometric deformations of the 7-brane along the holomorphic normal vector. In order to have a canonically normalised quasi-topological term (c.f. eq.~(\ref{8dact})), in the last equality we have redefined,
\begin{equation}
\varphi\equiv M_*R_\perp \iota_\phi \Omega= M_*R_\perp \phi^3 dz^1\wedge dz^2 \;,
\end{equation}
with $\Omega$ the local holomorphic 3-form of the 3-fold base $B_3$ of $X$. Hence, the factor $R_{\perp}M_*$ is the appropriate one for the canonically normalised Higgs field in the 8-dimensional theory.\footnote{The above scalings with $R_\parallel$ and $R_\perp$ can also be understood from the T-dual setup with magnetised D9-branes. Indeed, T-dualising along the transverse space to the 7-brane, the Higgs and gauge fluxes are mapped respectively to gauge fluxes $F_{i\alpha}$ and $F_{ij}$ on a stack of D9-branes \cite{Aparicio:2011jx}. In a vielbein basis the components of the flux are respectively,
\begin{equation*}
F_{i\alpha}\simeq \frac{m M_*^2}{R_{\parallel}R^{D9}_{\perp}}\ , \qquad F_{ij}\simeq \frac{M M_*^2}{R_{\parallel}^2} \;. \end{equation*}
T-dualising along the transverse directions, these become the Higgs and gauge fluxes on the 7-brane
\begin{equation*}
\partial\langle\varphi\rangle \simeq \frac{m M_*^2R^{D7}_{\perp}}{R_{\parallel}}\ , \qquad \partial\langle A\rangle \simeq \frac{M M_*^2}{R_{\parallel}^2} \;,
\end{equation*}
where $R_{\perp}^{D7}=\left(R_{\perp}^{D9}\right)^{-1}$, in agreement with (\ref{a}) and (\ref{phi}).}

Plugging the above local expansions into eqs.~(\ref{eom5})-(\ref{eom7}) we obtain that the relevant F-term equations for a given massless 4-dimensional fermionic field read,
to leading order in the coordinates,
\begin{equation}
\mathbb{D}\Psi=0\label{z1} \;,
\end{equation}
with,
\begin{equation}
\mathbb{D}=\begin{pmatrix}0& D_1& D_2 & D_3\\
-D_1& 0& D_3^\dagger& -D_2^\dagger\\
-D_2& -D_3^\dagger&0&D_1^\dagger\\
-D_3& D_2^\dagger&-D_1^\dagger&0 \end{pmatrix}\ , \qquad \Psi=\begin{pmatrix}\sqrt{2}\eta\\ \psi_{\bar 1}\\ \psi_{\bar 2}\\ \chi\end{pmatrix} \;,
\label{z2}
\end{equation}
and
\begin{align}
D_i&\equiv\frac{M_*}{R_{\parallel}}\left(\partial_i-\frac{1}{2}q_a(M^a_{ji})^*\bar{z}_j\right)& D_i^{\dagger}&\equiv\frac{M_*}{R_{\parallel}}\left(\bar \partial_i+\frac{1}{2}q_aM^a_{ji}z_j\right)\, \qquad i=1,2\label{gaugecov}\\
D_3&\equiv-M_*R_{\perp}\, q_a m^a_{i} \bar{z}_i & D_3^\dagger&\equiv M_*R_{\perp}\, q_a(m^a_{i})^* z_i \;.
\end{align}
In these expressions $\vec q$ is the vector of $G_H$-charges for the localised mode and we have relabeled $\varphi_{12} \rightarrow \varphi$ and $\chi_{12} \rightarrow \chi$ to simplify the notation. To obtain a finite set of solutions, we have to supplement these equations with a set of boundary conditions encoding the global obstruction from the topology of $S$ and which, in particular, determine the degeneracy of the zero modes.

Massive 4-dimensional fields can similarly be accounted for by the 8-dimensional effective theory. In that case, one obtains the more general set of equations
\begin{equation}
\mathbb{D}^\dagger\mathbb{D}\Psi=|m_\lambda|^2\Psi\label{massiveeq} \;,
\end{equation}
where $m_\lambda$ is the mass of the 4-dimensional field and the same definitions above hold. As we explicitly show in next subsection, these are the equations of motion for a set of three complex quantum harmonic oscillators which can be solved by means of standard techniques in quantum mechanics.

%%%%%%%%%%%%%%%%%%%%%%%%%%%%%%%%%%%%%%%%%%%%%%%%%%%%%%%%%%%%%%%%%%%%%%%%%%%%%%%%%%%%%%
\subsection{Localised fields and supersymmetric quantum mechanics}
\label{sublocal}
%%%%%%%%%%%%%%%%%%%%%%%%%%%%%%%%%%%%%%%%%%%%%%%%%%%%%%%%%%%%%%%%%%%%%%%%%%%%%%%%%%%%%%

Let us first solve the equations of motion for 4-dimensional massless fields, eq.~(\ref{z1}) or equivalently eq.~(\ref{massiveeq}) with $m_\lambda=0$.  For that we closely follow the techniques developed in \cite{Camara:2009xy, Marchesano:2010bs, Aparicio:2011jx}.\footnote{See also \cite{Oikonomou:2011kd} for related work.}

Different matter representations have different charges under the gauge group generators and therefore different equations governing their wavefunctions. We take $\Psi$ to transform in a representation $\mathcal{R}$ of the gauge group, with $D_i$ the corresponding gauge covariant derivatives defined in eq.~(\ref{gaugecov}). From eq.~(\ref{z2}) we observe that the operator which appears in the left-hand-side of eq.~(\ref{massiveeq}) can be written as
\begin{equation}
\mathbb{D}^\dagger\mathbb{D}=-\triangle\mathbb{I}+\mathbb{B}\label{dd} \;,
\end{equation}
where
\begin{equation}
\triangle\equiv \sum_{i=1,2,3}D_i^\dagger D_i \;,
\end{equation}
and
\begin{equation}
\mathbb{B}=\begin{pmatrix}0&0&0&0\\
0&[D_2^\dagger,D_2]&[D_2,D_1^\dagger]&[D_3,D_1^\dagger]\\
0&[D_1,D_2^\dagger]&[D_1^\dagger,D_1]&[D_3,D_2^\dagger]\\
0&[D_1,D_3^\dagger]&[D_2,D_3^\dagger]&[D_2^\dagger,D_2]+[D_1^\dagger,D_1]\end{pmatrix}\label{bmatrix} \;.
\end{equation}
We have made use of the F-term equations for the background, that we take to preserve $\mathcal{N}\geq 1$ supersymmetry in 4-dimensions, in order to simplify the hermitian matrix $\mathbb{B}$. Imposing also the D-term condition on the background it is easy to check that $\mathbb{B}$ is traceless.

A suitable approach to obtain the zero mode wavefunctions is therefore to make a change of basis which diagonalises $\mathbb{B}$ and to solve the equations of motion in that basis. Let $\mathbb{J}$ be the matrix which diagonalises $\mathbb{B}$ and has canonically normalised column vectors,
\begin{equation}
\mathbb{J}^{-1}\cdot\mathbb{B}\cdot\mathbb{J}=\left(\frac{M_{*}}{R_{\parallel}}\right)^2 \textrm{diag}(0,\lambda_1,\lambda_2,\lambda_3)\label{jj} \;,
\end{equation}
where $\lambda_1+\lambda_2+\lambda_3=0$. The dimensionless eigenvalues $\lambda_p$ are given by the three roots of the characteristic polynomial of the non-trivial part of $(R_{\parallel}/M_{*})^2\mathbb{B}$, which is a depressed cubic equation. We can rotate the operator $\mathbb{D}$ to the diagonal basis by taking,
\begin{equation}
\tilde{\mathbb{D}}\equiv(\mathbb{J}^{-1})^*\cdot\mathbb{D}\cdot\mathbb{J} \;.
\end{equation}
Notice that $\tilde{\mathbb{D}}$ has again the same structure as in (\ref{z2}) but  in the new basis covariant derivatives are given by
\begin{equation}
\tilde D_p=\sum_{k=1}^3\mathbb{J}_{kp}D_k=\frac{1}{||\xi_p||}\sum_{k=1}^3\xi_{p,\;k}D_k\label{tilded} \;,
\end{equation}
with $\xi_p$ the $p$-th eigenvector of $\mathbb{B}$ and $||\xi_p||$ its norm. In particular, it is simple to check that the only non-vanishing commutators of the rotated covariant derivatives are the diagonal ones,
\begin{equation}
[\tilde D_p^\dagger,\tilde D_p]=-\left(\frac{M_{*}}{R_{\parallel}}\right)^2\lambda_p\ , \quad p=1,2,3 \label{diagalg} \;.
\end{equation}
As we have advanced, this is the algebra for the ladder operators of a set of three quantum harmonic oscillators.

Generically we can distinguish four towers of solutions to eq.~(\ref{massiveeq}), one per eigenvector of $\mathbb{B}$. These four towers can be identified with the four complex fermions of a (broken) $\mathcal{N}=4$ supermultiplet.
In particular, if the three non-trivial eigenvalues $\lambda_p$ are different from zero, there is a massless chiral $\mathcal{N}=1$ supermultiplet in the 4-dimensional spectrum, which corresponds to the ground state of one of the above four towers. To see this more explicitly, let us assume for a while that $\lambda_1\lambda_2\lambda_3<0$ with two positive and one negative real eigenvalues. We take $\lambda_1$ to be the negative eigenvalue. The localised normalisable solution satisfies the equations
\begin{equation}
\tilde{\mathbb{D}}\cdot\begin{pmatrix}0\\ \varphi\\ 0\\ 0\end{pmatrix}= 0\quad \Leftrightarrow \quad \begin{cases}\tilde D_1\varphi&=0\\
\tilde D_2^\dagger\varphi&=0\\
\tilde D_3^\dagger\varphi&=0\end{cases} \;.\label{zero00}
\end{equation}
We can identify the raising $\hat{a}^{\dagger}$ and lowering $\hat{a}$ operators as,
\begin{align}
\hat{a}_1 &\equiv i\tilde D_1\;, &
\hat{a}_2 &\equiv i\tilde D_2^{\dagger} \;, &
\hat{a}_3 &\equiv i\tilde D_3^{\dagger} \;, \\
\hat{a}^{\dagger}_1 &\equiv i\tilde D^{\dagger}_1\;,  &
\hat{a}^{\dagger}_2 &\equiv i\tilde D_2\;, &
\hat{a}^{\dagger}_3 &\equiv i\tilde D_3\;, \nn
\end{align}
so that the function $\varphi$ in eq.~(\ref{zero00}) is annihilated by the three $\hat a_i$ operators. More generically, for fields transforming in the representation $\mathcal{R}$, we have in the diagonal basis
\begin{equation}
\mathcal{R}\ : \qquad  \tilde{\mathbb{D}}^\dagger\tilde{\mathbb{D}}=\sum_{i=1,2,3}{\hat{a}_i^\dagger \hat{a}_i} \;\mathbb{I}+\left(\frac{M_*}{R_\parallel}\right)^2\textrm{diag}(-\lambda_1,0,\lambda_2-\lambda_1,\lambda_3-\lambda_1)\label{ddtilde} \;.
\end{equation}
Since the ground state in each of the four towers of fermions is by definition annihilated by all lowering operators $\hat a_i$, the four entries in the last term correspond to the masses of these ground states. Their wavefunctions are given in terms of the function $\varphi$ as,
\begin{equation}
\Psi_p = \frac{\xi_p}{N}\varphi(z_1,z_2,\bar z_1,\bar z_2)\ , \qquad p=0,1,2,3 \label{ground}
\end{equation}
where $N$ is a normalisation constant. Similarly, wavefunctions for the heavier modes in each tower are obtained by acting  on the corresponding ground state wavefunction with the raising operators. We can label these fields by three quantum numbers, $n$, $m$ and $l$, according to
\begin{equation}
\Psi_{p,(n,m,l)}=\frac{(R_\parallel/M_*)^{n+l+m}}{\sqrt{m!n!l!}\left(-\lambda_1\right)^{n/2}\lambda_2^{m/2}\lambda_3^{l/2}}(\tilde{D}_1^\dagger)^{n}(\tilde D_2)^m(\tilde D_3)^l\Psi_p \;,\label{massiverep}
\end{equation}
where the particular pre-factor ensures the correct normalisation. Thus, calculating massive wavefunctions is a simple task of applying differential operators to functions. The mass of the resulting 4-dimensional fields is given by
\begin{align}
M^2_{\Psi_{0,(n,m,l)}} &= \left(\frac{M_{*}}{R_{\parallel}}\right)^2 \left[ -(n+1)\lambda_1+m\lambda_2+l\lambda_3 \right] \;, \label{mas}\\
M^2_{\Psi_{1,(n,m,l)}} &= \left(\frac{M_{*}}{R_{\parallel}}\right)^2 \left( -n\lambda_1+m\lambda_2+l\lambda_3 \right) \;, \nn\\
M^2_{\Psi_{2,(n,m,l)}} &= \left(\frac{M_{*}}{R_{\parallel}}\right)^2 \left[ -(n+1)\lambda_1+(m+1)\lambda_2+l\lambda_3 \right] \;, \nn\\
M^2_{\Psi_{3,(n,m,l)}} &= \left(\frac{M_{*}}{R_{\parallel}}\right)^2 \left[ -(n+1)\lambda_1+m\lambda_2+(l+1)\lambda_3 \right] \;. \nn
\end{align}
In particular, $\Psi_{1,(0,0,0)}$ denotes the wavefunction for the massless chiral fermion transforming in the $\mathcal{R}$ representation.

Whereas this description is complete for massless chiral fields, massive fields contain both chiralities and the above wavefunctions only represent half of their degrees of freedom, namely those transforming in the $\mathcal{R}$ representation of the gauge group. Wavefunctions for the $\bar{\mathcal{R}}$ components of the massive fields can be worked out following the same procedure, taking care of the change of sign in the charges. One may easily check that the analogous operator to (\ref{ddtilde}) for fields transforming in the $\bar{\mathcal{R}}$ representation is
\begin{equation}
\bar{\mathcal{R}}\ : \qquad \tilde{\mathbb{D}}^\dagger\tilde{\mathbb{D}}=\sum_{i=1,2,3}{\hat{a}_i^\dagger \hat{a}_i} \;\mathbb{I}+\left(\frac{M_*}{R_\parallel}\right)^2\textrm{diag}(-\lambda_1,-2\lambda_1,\lambda_3,\lambda_2) \;.
\end{equation}
Wavefunctions are therefore given again by the same functions $\Psi^i_{p,(n,m,l)}$, with $p=0,1,2,3$, but the corresponding masses are shifted with respect to eq.~(\ref{mas}). Massive components transforming in the $\mathcal{R}$ and the $\bar{\mathcal{R}}$ representations pair up non-trivially. For instance, the first excited states of the massless mode in the $\mathcal{R}$ representation pair up with same-mass ground states in the $\bar{\mathcal{R}}$ representation.\footnote{This non-trivial pairing has its origin in the fact that the 4-dimensional mass term comes from the 8-dimensional kinetic term, and the latter contains a $\Gamma^iD_i$ operator acting non-trivially on $\Psi$.} Note also that there is no massless fermion transforming in the $\bar{\mathcal{R}}$ representation, as expected.

Wavefunctions for the scalar fields can be worked out in a similar way and, in particular, they can be shown to be identical to the ones of their corresponding fermionic superpartners, as a consequence of supersymmetry and flatness of the local patch.

Summarizing, we have shown that at each localisation point in $S$ there are four towers of fields with equal gauge charges, corresponding to the degrees of freedom of a broken $\mathcal{N}=4$ supermultiplet. In these conventions, for $\lambda_1\lambda_2\lambda_3 < 0$ there is a localised massless $\mathcal{N}=1$ chiral supermultiplet transforming in the $\mathcal{R}$ representation of the gauge group. The degeneracy of this field is only globally determined. It is also easy to check  that for $\lambda_1\lambda_2\lambda_3 > 0$ the roles of $\mathcal{R}$ and $\bar{\mathcal{R}}$ are exchanged and there is instead a massless $\mathcal{N}=1$ chiral supermultiplet transforming in the $\bar{\mathcal{R}}$ representation. These two possibilities are separated by a wall of marginal stability at $\lambda_1\lambda_2\lambda_3=0$. At this wall at least one of the three eigenvalues $\lambda_p$ vanishes and 4-dimensional fields arrange into $\mathcal{N}=2$ supermultiplets (or $\mathcal{N}=4$ supermultiplets if all eigenvalues are zero) with conserved Kaluza-Klein momentum. In that case, the wavefunction of the fields is no-longer localised along the matter curve, and their mass is determined by the particular topology of the curve. We present in section \ref{sec:wave} an example of this type.

%%%%%%%%%%%%%%%%%%%%%%%%%%%%%%%%%%%%%%%%%%%%%%%%%%%%%%%%%%%%%%%%%%%%%%%%%%%%%%%%%%%%%%%%%
\subsection{Validity of the local approach}
\label{sec:validity}
%%%%%%%%%%%%%%%%%%%%%%%%%%%%%%%%%%%%%%%%%%%%%%%%%%%%%%%%%%%%%%%%%%%%%%%%%%%%%%%%%%%%%%%%%

The wavefunctions and consequently, through their overlaps, the cubic couplings in the 4-dimensional theory depend on the parameters of the 8-dimensional theory such as the fluxes and the local scales $R_{\parallel}$ and $R_{\perp}$. It is therefore important to quantify the possible range of these parameters which is consistent with the local effective theory being used.

The first constraint we must impose is $(\ref{alpha})$ which ensures that higher derivative corrections to the 8-dimensional effective action are negligible. Using the expressions (\ref{a}) and (\ref{phi}) this gives
\begin{equation}
\frac{M^a_{ij}}{R_{\parallel}^2}\ll 1\ , \qquad \frac{R_{\perp}}{R_{\parallel}} m^a_{i}\ll 1\label{localalpha} \;.
\end{equation}
These amount to small intersection angles and small flux densities.
The flux parameters $M^a_{ij}$ and $m^a_{i}$ would be integer quantised in a homogenous setup but in the local setup need not be so. However, generically they are expected to be of order one and we shall therefore take them as so while keeping in mind that the local freedom to adjust the fluxes allows for some flexibility in satisfying the consistency constraints. Taking the fluxes as such we can rephrase (\ref{localalpha}) in terms of geometric constraints. We define
\begin{equation}
R \equiv R_{\parallel} R_{\perp} \;, \qquad
\varepsilon \equiv \frac{R_{\perp}}{R_{\parallel}} \;, \label{redef}
\end{equation}
using which we can write
\begin{equation}
\varepsilon \ll 1 \;,\qquad \frac{\varepsilon}{R} \ll 1 \;. \label{erconst}
\end{equation}
We can also consider these as constraints on mass scales:
it is simple to check that for large $R$ the eigenvalues $\lambda_p$ scale as,
\begin{equation}
\lambda_1,\ \lambda_2 \sim R\ , \qquad \lambda_3\sim 1\label{eigenscaling} \;,
\end{equation}
and are independent of $\varepsilon$. Therefore, from eqs.~(\ref{mas}) we observe two types of massive modes with masses scaling as $M \sim M_*/R_{\parallel}$ or $M \sim M_*\sqrt{R}/R_{\parallel}$. The mass of these modes should be kept below the cutoff scale of the theory which is consistent with the constraints (\ref{erconst}). For generic order one fluxes the two above constraints can be simultaneously satisfied by taking
\begin{equation}
R_{\parallel} \gg R_{\perp}\ , \qquad R_{\parallel} \gg 1 \;. \label{largerprp}
\end{equation}
In this limit there is a large number of 4-dimensional massive fields below the cutoff scale $M_*$. Note that, although not necessarily required, these constraints also allow for length scales $R_\perp < 1$. The stability of the 8-dimensional effective theory  against $1/M_{\rm Planck}$ corrections for such small values of $R_\perp$ depends on the particular connection between the local and global scales, which we now discuss.

The relation between the local scales $R_\parallel$ and $R_\perp$ and the global ones is model dependent and generically too complicated to be computed explicitly in given models. Whereas the limit (\ref{largerprp}) can always be taken in the local setup, once we begin to relate  local scales to global ones new phenomenological constraints are expected to arise from the observed values of $\alpha_{\rm GUT}$ and $M_{\rm Planck}$. For instance, if $S$ is completely homogenous then we have approximately
\begin{equation}
R_{\parallel} \sim \alpha_{GUT}^{-\frac14} \;. \label{agutrel}
\end{equation}
Since we observe $\alpha_{GUT}^{-1} \sim 24$ this implies that $R_{\parallel}$ cannot be too large. The relation is approximate and the space $S$ is in general not homogenous, but nonetheless it is difficult to conceive a departure of the local $R_{\parallel}$ scale too far from (\ref{agutrel}). We must therefore keep in mind that although formally our calculations can be made very precise by taking the limit (\ref{largerprp}), in a phenomenologically viable setup there will be corrections (essentially $\alpha'$ corrections) that are not hugely suppressed. Taking this into account, and that the constraints are only approximate up to order one factors, in what follows we allow ourselves to take $R$ in the range $1<R<25$, with $\varepsilon$ also in the range $1<\varepsilon^{-1}<25$ but chosen appropriately such that for each value of $R$ eqs.~(\ref{erconst}) are satisfied. Note that the most natural values are towards the lower end of the range, however, due to the strong model dependence of the relation between local and global scales this range is only an approximate one and some flexibility should be allowed.

Similarly, we expect the local scale $R_\perp$ to be related to the global ones and in particular to $M_{\rm Planck}$. The particular relation strongly depends on the geometry of the CY base $B_3$. In the case of a torus $B_3=T^6$ we have
\begin{equation}
 R_\perp \sim \frac{g_sM_{\rm Planck}\alpha_{\rm GUT}^{1/2}}{M_*}\;.
\end{equation}
It is therefore important to note that in a torus,  and more generally in a near homogeneous setup, the observed values of $M_{\rm Planck}$ and $\alpha_{\rm GUT}$ are not compatible with the constraints (\ref{erconst}) and we expect higher derivative corrections to the effective 8-dimensional theory coming from large brane intersection angles. At a deeper level this can be taken as motivation for local models based on contractible cycles as then the scaling with respect to the Planck scale is expected to be modified to the schematic form $R_{\perp}\sim \left(M_{\rm Planck}/M_*\right)^{1/3}$. More generally the geometry can lead to differences between $R_{\perp}$ and the global scales either coming from inhomogeneities of the divisors or from the geometry allowing a decoupling of the intersecting brane setup from the overall volume. Given this in general we do not attempt to relate the local scales  with the global ones.

%%%%%%%%%%%%%%%%%%%%%%%%%%%%%%%%%%%%%%%%%%%%%%%%%%%%%%%%%%%%%%%%%%%%%%%%%%%%%
\section{The SO(12) enhancement point}
\label{sec:wave}
%%%%%%%%%%%%%%%%%%%%%%%%%%%%%%%%%%%%%%%%%%%%%%%%%%%%%%%%%%%%%%%%%%%%%%%%%%%%%

%%%%%%%%%%%%%%%%%%%%%%%%%%%%%%%%%%%%%%%%%%%%%%%%%%%%%%%%%%%%%%%%%%%%%%%%%%%%%
\subsection{The SO(12) point and background fluxes}
\label{sec:wave12}
%%%%%%%%%%%%%%%%%%%%%%%%%%%%%%%%%%%%%%%%%%%%%%%%%%%%%%%%%%%%%%%%%%%%%%%%%%%%%

We now apply the procedure described in the previous section to the point in $S$ where the down-type Yukawa coupling localises. At that locus there is an SO(12) enhancement of the gauge symmetry which can be seen by decomposing the adjoint
\bea
\textrm{SO(12)} &\supset& \textrm{SU(5)} \times \textrm{U(1)}_1 \times \textrm{U(1)}_2 \;, \\
\bf{66} &\rightarrow& \bf{24}^{(0,0)} \op \bf{1}^{(0,0)} \op \bf{1}^{(0,0)} \op \left(\f^{(-1,0)} \op \f ^{(1,1)} \op
\te^{(0,1)} \ \op \ c.c.\right) \nn \label{so12dec} \;.
\eea
The spontaneous breaking of the SO(12) symmetry away from the enhancement point can be obtained by turning on a background for the Higgs scalar,
\be
\left< \varphi \right> = M_{*}R_\perp\left(\frac{z_1}{v_1} Q_1 + \frac{z_2}{v_2} Q_2 \right)\;, \label{vevphi}
\ee
where $v_1$ and $v_2$ are dimensionless parameters. The generators $Q_1$ and $Q_2$ are those corresponding to the U(1) factors in the decomposition of SO(12). This Higgs background describes three sets of intersecting 7-branes with localised matter on their intersection curves
\bea
\te_M &:& z_2=0 \;, \nn \\
\fb_M &:& z_1=0 \;, \nn \\
\fb_H &:& v_2z_1+v_1z_2=0 \;.
\eea
We will be solving for the wavefunctions of modes localised along these curves.

In order to get chiral matter, U(1) flux must be turned on along the generators $Q_1$ and $Q_2$. Moreover, to break the GUT group and induce doublet-triplet splitting also flux must be turned on along the hypercharge direction of SU(5) so that,
\begin{align*}
\textrm{SU(5)}&\ \to \ \textrm{SU(3)}\times \textrm{SU(2)}_L\times \textrm{U(1)}_Y \;,\\
\fb &\ \to \ (\bar{\mathbf{3}},\mathbf{1})_{1/3}\ \oplus\ (\mathbf{1},\mathbf{2})_{-1/2} \;,\\
\te &\ \to \ (\bar{\mathbf{3}},\mathbf{1})_{-2/3}\ \oplus\ (\mathbf{3},\mathbf{2})_{1/6}\ \oplus\ (\mathbf{1},\mathbf{1})_{1} \;.
\end{align*}
The geometric properties of these fluxes are dependent not only on $S$ but also on the full CY four-fold $X$. Let us recall some of the defining global properties of the fluxes. First, the hypercharge flux must be turned on along a cycle which is homologically non-trivial when pulled back to the GUT divisor $S$, but trivial in the full CY so that U(1)$_Y$ is massless \cite{Buican:2006sn,Donagi:2008ca,Beasley:2008kw,Donagi:2008kj}. Secondly, the fluxes must be turned on such that they induce the correct chiral matter spectrum, which is determined as
\bea
n_{(\mathbf{3},\mathbf{1})_{-1/3}} - n_{(\bar{\mathbf{3}},\mathbf{1})_{1/3} } &=& M_{5} \;, \nn \\
n_{(\mathbf{1},\mathbf{2})_{1/2}} - n_{(\mathbf{1},\mathbf{2})_{-1/2} } &=& M_{5} + N \;, \label{chiralflux5}
\eea
for the $\fb$ curves and
\bea
n_{(\mathbf{3},\mathbf{2})_{1/6}} - n_{(\bar{\mathbf{3}},\mathbf{2})_{-1/6} } &=& M_{10} \;, \nn \\
n_{(\bar{\mathbf{3}},\mathbf{1})_{-2/3}} - n_{(\mathbf{3},\mathbf{1})_{2/3} } &=& M_{10} - N \;, \nn \\
n_{(\mathbf{1},\mathbf{1})_{1}} - n_{(\mathbf{1},\mathbf{1})_{-1} } &=& M_{10} + N\;, \label{chiralflux10}
\eea
for the $\te$ curves, where $n_{\mathcal{R}}$ denotes the number of massless 4-dimensional fields transforming in the $\mathcal{R}$ representation of SU(3)$\times$SU(2)$_L\times$U(1)$_Y$. Here the fluxes $M_{5}$, $M_{10}$ and $N$ are specified by fractional line-bundles $L_{Y}$, $V_{10}$ and $V_{5}$ such that $M_{10}=\mathrm{deg}\left(L_{Y}^{1/6}\otimes V_{10}\right)$, $M_{5}=\mathrm{deg}\left(L_{Y}^{-1/3}\otimes V_{5}\right)$ and $N=\mathrm{deg}\left(L_{Y}^{5/6}\right)$. To maintain complete representations on the matter curves we want $N=0$ for those curves, while to induce doublet-triplet splitting we want $M_{5}=0$ and $N=1$ for the Higgs curve.

Working locally near a point on $S$ we are not sensitive to the full global structure of the fluxes. Indeed, all the geometry of the cycles locally reduces to the four possible components of the flux along $dz_1\wedge d\bar{z}_1$, $dz_2\wedge d\bar{z}_2$, $dz_1\wedge d\bar{z}_2$ and $dz_2\wedge d\bar{z}_1$. These components are constrained by the local D-term condition
\begin{equation}
\omega\wedge F^{(1,1)} = 0 \;,
\end{equation}
whereas the F-term condition simply requires $F^{(2,0)}=F^{(0,2)}=0$.

We assume that locally the flux takes a constant profile in $S$, neglecting a possible spatial dependence of the flux. This can be expected to be a decent approximation if the curvature around the enhancement point is small, as in that case it can be thought as the leading term of a Taylor expansion in the local coordinates, as we have argued in the previous section. Taking varying flux into account would lead to technical difficulties, in particular D-terms would generally not be solved by a flat-space profile, which would imply having to work in curved space \cite{Conlon:2009qq}. Moreover, for simplicity, in the main part of the paper we do not turn on flux along the oblique components $dz_1\wedge d\bar{z}_2$ and $dz_2\wedge d\bar{z}_1$. Oblique fluxes turn out to not affect the physics of the wavefunction in a qualitatively important way. We relegate a treatment of the more general flux including such components to appendix \ref{app:oblique}.
A general suitable choice for the U(1) flux is therefore
\begin{equation}
F^{(1,1)}=\frac{2iM_{*}^2}{R_\parallel^2}(dz_1\wedge d\bar z_1-dz_2\wedge d\bar z_2)(-M_1Q_1+M_2Q_2 + \gamma Q_Y) \;, \label{fluxansatz}
\end{equation}
where $M_1$, $M_2$ and $\gamma$ are dimensionless real constants and the generator $Q_Y$ is along the hypercharge direction in SU(5)$_{\rm GUT}$.
The gauge potential associated with this flux reads
\begin{equation}
A=\frac{iM_*}{R_\parallel}(z_1 d\bar z_1-\bar z_1 dz_1-z_2 d\bar z_2+\bar z_2 d z_2)(-M_1Q_1+M_2Q_2+ \gamma Q_Y) \;.
\end{equation}

The relation between the local values of the flux $M_1$, $M_2$ and $\gamma$ and the global integrated values in (\ref{chiralflux5}) and (\ref{chiralflux10}) is subtle. The connection is that the local fluxes $M_1$, $M_2$ and $\gamma$ determine the chirality of the localised fields, which generically arrange in $\mathcal{N}=1$ supermultiplets as we have described in previous section. At a given localisation point, eq.~(\ref{z1}) has an infinite number of solutions. Consistency with the topological data of the flux and $S$ however selects a finite subset of size given by the global integrated values in (\ref{chiralflux5}) and (\ref{chiralflux10}), in accordance with standard index theorems.

Using the flux (\ref{fluxansatz}) the chirality for a given localised mode is determined by the analogous local expressions to eqs.~(\ref{chiralflux5}) and (\ref{chiralflux10}). That is, for the $\fb_{M}$ curve,
\bea
\delta_{(\mathbf{3},\mathbf{1})_{-1/3}}  &=& \mathrm{sign}\left[-M_1 + \frac13 \gamma\right]  \;, \nn \\
\delta_{(\mathbf{1},\mathbf{2})_{1/2}} &=& \mathrm{sign}\left[\left(-M_1 + \frac13\gamma \right) - \frac56 \gamma\right] \;, \label{localchiralflux5}
\eea
for the $\te_{M}$ curve,
\bea
\delta_{(\mathbf{3},\mathbf{2})_{1/6}} &=& \mathrm{sign}\left[M_2+\frac16\gamma \right]\;, \nn \\
\delta_{(\bar{\mathbf{3}},\mathbf{1})_{-2/3}} &=& \mathrm{sign}\left[\left(M_2+\frac16\gamma\right) - \frac56 \gamma \right]\;, \nn \\
\delta_{(\mathbf{1},\mathbf{1})_{1}} &=& \mathrm{sign}\left[\left(M_2+\frac16\gamma\right) + \frac56 \gamma \right]\;, \label{localchiralflux10}
\eea
and for the $\fb_{H}$ curve,
\bea
\delta_{(\mathbf{3},\mathbf{1})_{-1/3}}  &=& \mathrm{sign}\left[M_1-M_2 + \frac13 \gamma\right]  \;, \nn \\
\delta_{(\mathbf{1},\mathbf{2})_{1/2}}  &=& \mathrm{sign}\left[\left(M_1-M_2 + \frac13\gamma \right) - \frac56 \gamma \right]\;. \label{localchiralfluxhiggs}
\eea
where $\delta_{\mathcal{R}}=+1$ ($\delta_{\mathcal{R}}=-1$) means that the corresponding set of localised massless 4-dimensional fields transforms in the $\mathcal{R}$ ($\bar{\mathcal{R}}$) representation of SU(3)$\times$SU(2)$_L\times$U(1)$_Y$ and we have taken $v_2>v_1$ without loss of generality.\footnote{The relative signs between the components in expressions (\ref{localchiralflux5}), (\ref{localchiralflux10}) and (\ref{localchiralfluxhiggs}) are determined from the group theory charges of the states while the overall sign for each curve is determined by studying the form of the wavefunctions in section \ref{sec:massless}, such that given the sign of the fluxes the correct state localises.}
In this regard, it is also worth stressing that if the local flux vanishes along the matter curve for a given representation the wavefunction does not localise, as we have already commented in the previous section.

With these relations we see that there are some constraints on the local fluxes that one has to satisfy in order to properly model the massless spectrum. In particular, we shall require that the expressions in (\ref{localchiralflux5}) are all negative and that the expressions in (\ref{localchiralflux10}) are all positive. This can be implemented by taking $M_1$ and $M_2$ positive and much larger than $\gamma$. On the Higgs curve we require that the second expression of (\ref{localchiralfluxhiggs}) is negative. The sign of the first expression of (\ref{localchiralfluxhiggs}) determines whether locally there is a massless triplet, a massless anti-triplet or a vector-like pair, with the mass of the latter depending on the particular topology of the matter curve,
\bea
M_1-M_2 + \frac13 \gamma &>& 0 \;,\;\;\mathrm{Massless\;}(\mathbf{3},\mathbf{1})_{-1/3}  \label{notripletchiral} \\
M_1-M_2 + \frac13 \gamma &<& 0 \;,\;\;\mathrm{Massless\;}(\bar{\mathbf{3}},\mathbf{1})_{1/3} \nn \\
M_1-M_2 + \frac13 \gamma &=& 0 \;,\;\;\mathrm{Massless\ or\ massive\;}(\bar{\mathbf{3}},\mathbf{1})_{1/3}\oplus(\mathbf{3},\mathbf{1})_{-1/3} \;. \nn
\eea
We will study all three possibilities in the next sections. The most appealing case is the third one, as the mass of such a vector-like mode is only determined globally. If the flux vanishes not only locally but everywhere along the matter curve the vector-like pair is massless whenever $h_0\left(\Sigma,K_{\Sigma}^{1/2}\right) \neq 0$, where $\Sigma$ denotes the matter curve and $K$ its canonical bundle \cite{Donagi:2008ca,Beasley:2008dc}. In the interesting case where the $\fb_H$ curve is a $\mathbb{P}^1$ on a del-Pezzo such a vector-like pair is massive because of the twisting. The other two cases in (\ref{notripletchiral}) are  different from the third in that the only way to give them a mass would be to somehow deform the local geometry. Such a deformation could be due, for instance, to the presence of another nearby enhancement point with a localised massless triplet of opposite chirality, such that the massless mode (and their Landau replicas) develop a supersymmetric $\mu$-term. If the shift in the masses is much smaller than $M_{*}/R_\parallel$ we may think of the first two cases in eq.~(\ref{notripletchiral}) as a good approximation to that setup.

%%%%%%%%%%%%%%%%%%%%%%%%%%%%%%%%%%%%%%%%%%%%%%%%%%%%%%%%%%%%%%%%%%%%%%%%%%%%%
\subsection{Wavefunctions for the SO(12) point}
\label{sec:massless}
%%%%%%%%%%%%%%%%%%%%%%%%%%%%%%%%%%%%%%%%%%%%%%%%%%%%%%%%%%%%%%%%%%%%%%%%%%%%%

We now determine the wavefunctions for the fields localised near an SO(12) enhancement point. For that we solve eq.~(\ref{massiveeq}) with the above background, closely following the method described in section \ref{sublocal}. The solution to eqs.~(\ref{zero00}) can be shown to have the general expression
\be
\vphi = f\left(-k_{2}z_1 + k_{1}z_2\right)e^{-p_1 |z_1|^2 - p_2 |z_2|^2 + p_3 \bar{z}_1z_2 + p_4 \bar{z}_2 z_1} \;,\label{vphi}
\ee
where $k_m$, $m=1,2$, and $p_i$, $i=1,2,3,4$, are constants depending on the background.
In this expression $f$ is a general holomorphic function of the particular combination of variables that is denoted. In particular, depending on the form of $f$, the wavefunction can become delocalised along one real direction within the matter curve. We return to this point soon but first we note that locally we can always take $f$ to be of the form
\be
f(z) = z^{l} \;,\label{fz}
\ee
where $l$ counts the massless Landau level degeneracy, which is the number of generations and so ranges over $l=0,1,2$. This choice is well-motivated from the existence of local geometric selection rules in the evaluation of cubic superpotential couplings, as we discuss in section \ref{sec:overlaps}.

Thus, from eq.~(\ref{ground}) we have that the ground state wavefunctions can be generically expressed as,
\be
\Psi^l_p = \frac{\xi_p}{N_{p}^l} \left(-k_{2}z_1 + k_{1}z_2\right)^{l}
e^{-p_1 |z_1|^2 - p_2 |z_2|^2 + p_3 \bar{z}_1z_2 + p_4 \bar{z}_2 z_1}\label{massless1}
\ee
where $N_{p}^l$ is such that
\be
\int_S \Psi^i_p \left(\Psi_q^j\right)^{\dagger} = \delta^{ij}\delta_{pq} \;.
\ee
This normalisation factor can be evaluated using standard gaussian formulae,
a particularly useful such integral takes the form
\be \label{int1}
I\left(n_1,n_2,n_3,n_4;p_1,p_2,p_3,p_4\right) \equiv \int_S \left(w\bar{w}\right)^{n_1}\left(u\bar{u}\right)^{n_2}\left(u\bar{w}\right)^{n_3}\left(w\bar{u}\right)^{n_4}e^{-p_1|w|^2-p_2|u|^2+p_3u\bar{w}+p_4w\bar{u}}
\ee
where the $n_i$ are positive integers. This integral is simply evaluated to be
\be
I\left(n_1,n_2,n_3,n_4;p_1,p_2,p_3,p_4\right) = \pi^2R_{\parallel}^4 \left(\prod_{i=1}^4 (\partial_{p_i})^{n_i}\right) \frac{(-1)^{n_1+n_2}}{p_1 p_2 - p_3 p_4} \;. \label{Ianswer}
\ee
In particular,
\begin{equation}
N_{p}^l=\frac{\pi R_\parallel^2\sqrt{l!}}{2^{\frac{2+l}{2}}}\frac{\left(|k_1|^2\textrm{Re}(p_1)+|k_2|^2\textrm{Re}(p_2)-\textrm{Re}\left[\bar k_1k_2( \bar p_3+p_4)\right]\right)^{\frac{l}{2}}}{\left(\textrm{Re}(p_1)\textrm{Re}(p_2)-\frac14|p_3+\bar p_4|^2\right)^{\frac{1+l}{2}}}||\xi_p||\label{norma1} \;.
\end{equation}

An important observation regarding the form of the wavefunction (\ref{vphi}), with $f$ given in eq.~(\ref{fz}), is that the combined effect of the Higgs profile and the flux generically induces localisation along all directions in $S$. However, complete localisation does not always occur and this can be seen by the fact that the arbitrary holomorphic function when extended over the full matter curve can cancel the exponential localisation in one real direction thereby delocalising the wavefunction along that direction. The expression (\ref{fz}) is obtained as the leading term of a local expansion of the holomorphic prefactor of the wavefunctions. It is still valid for use within the triple overlap since the combination of the wavefunctions of different matter curves completely localises that integral. However, for the wavefunction normalisation this potential partial delocalistion introduces an ambiguity. As we discuss in appendix \ref{app:norm}, for fluxes of order one, as we take in our examples, this ambiguity in the amount of localisation along the curve is not very large. Therefore in the main text we use the normalisation form (\ref{norma1}) and in appendix \ref{app:norm} we present a more quantitative analysis of the effect on our results that a possible wavefunction delocalisation along one real direction can have.

In what follows we work out the precise expression for $k_m$ and $p_i$ in eq.~(\ref{massless1}) in terms of the background, for each of the curves in the SO(12) enhancement point.  Since we have already described the procedure with great detail in section \ref{sublocal}, here we are rather schematic. Readers which are not interested in the precise form of the parameters may want to skip the remaining of this section and jump directly to section \ref{sec:overlaps}.

%%%%%%%%%%%%%%%%%%%%%%%%%%%%%%%%%%%%%%%%%%%%%%%%%%%%%%%%%%%%%%%%%%%%%%%%%%%%%
\subsubsection{Wavefunctions for the $\fb_M$ matter curve}
\label{sec:masslessz1}
%%%%%%%%%%%%%%%%%%%%%%%%%%%%%%%%%%%%%%%%%%%%%%%%%%%%%%%%%%%%%%%%%%%%%%%%%%%%%

For matter localised on the curve $z_1=0$ the gauge covariant derivatives appearing in eq.~(\ref{massiveeq}) are
\begin{equation}
D_1=\frac{M_*}{R_\parallel}\left(\partial_1 + \tilde{M}_1\bar z_1\right)\ , \label{zz1} \qquad
D_2=\frac{M_*}{R_\parallel}\left(\partial_2 - \tilde{M}_1\bar z_2\right)\ , \qquad
D_3=-\frac{M_*R_\perp}{v_1} \bar z_1 \;,
\end{equation}
where $\tilde{M}_1 \equiv M_1 + q_Y \gamma$, and $q_Y$ denotes the hypercharge of the localised state. Hermitian conjugation is obtained in these conventions by complex conjugation in addition to flipping the charge of the mode, $q \rightarrow -q$.

The matrix $\mathbb{B}$, defined in eq.~(\ref{bmatrix}), reads
\begin{equation}
\mathbb{B}=\frac{M_*^2}{R_\parallel^2}\begin{pmatrix}0&0&0&0\\
0& -2\tilde{M}_1 & 0 & \frac{R}{v_1} \\
0& 0 & 2\tilde{M}_1 & 0 \\
0& \frac{R}{v_1} & 0 & 0 \end{pmatrix}
\end{equation}
with dimensionless eigenvalues
\begin{equation}
\lambda_0^{\fb_M}=0\ , \qquad \lambda_1^{\fb_M}=-\rho_1-\tilde{M}_1\ ,\qquad \lambda_2^{\fb_M}=\rho_1-\tilde{M}_1\ , \qquad \lambda_3^{\fb_M}=2\tilde{M}_1\ ,
\end{equation}
and where $\rho_1\equiv\sqrt{\tilde{M}_1^2+\left(\frac{R}{v_1}\right)^2}$.
The corresponding eigenvectors are,
\begin{equation}
\xi_0^{\fb_M}=\begin{pmatrix}1\\ 0\\ 0 \\ 0\end{pmatrix}\ , \quad \xi_1^{\fb_M}=\begin{pmatrix}0\\ -\rho_1-\tilde{M}_1\\ 0\\ \frac{R}{v_1}\end{pmatrix}\ , \quad \xi_2^{\fb_M}=\begin{pmatrix}0\\ \rho_1-\tilde{M}_1\\ 0 \\ \frac{R}{v_1}\end{pmatrix}\ , \quad \xi_3^{\fb_M}=\begin{pmatrix}0\\ 0\\ 1\\ 0\end{pmatrix}\ .
\end{equation}
Solving for eqs.~(\ref{zero00}) then leads to the wavefunctions for the ground state of each of the four towers of fields localised in the $\fb_M$ matter curve, which are given by eq.~(\ref{massless1}) with
\begin{equation}
p_1^{\fb_M}=\rho_1\ , \qquad p_2^{\fb_M}=\tilde{M}_1\ , \qquad k_1^{\fb_M}=1\ , \qquad p_3^{\fb_M}=p_4^{\fb_M}=k_2^{\fb_M}=0\ ,
\end{equation}
where, following the discussion in the previous subsection, we have assumed $\tilde{M}_1$ to be positive.

%%%%%%%%%%%%%%%%%%%%%%%%%%%%%%%%%%%%%%%%%%%%%%%%%%%%%%%%%%%%%%%%%%%%%%%%%%%%%
\subsubsection{Wavefunctions for the $\te_M$ matter curve}
\label{sec:masslessz2}
%%%%%%%%%%%%%%%%%%%%%%%%%%%%%%%%%%%%%%%%%%%%%%%%%%%%%%%%%%%%%%%%%%%%%%%%%%%%%

The simplest way to deduce the wavefunctions for the $\te_M$ curve along $z_2=0$, is to use a particular symmetry of the equations of motion (\ref{z2}) for our choice of background. This symmetry acts with the following transformations
\be
z_1 \leftrightarrow z_2 \;,\qquad
\tilde{M}_1 \leftrightarrow \tilde{M}_2 \;,\qquad
v_1 \leftrightarrow v_2 \;,\qquad
D_1 \leftrightarrow D_2 \;,\label{trans}
\ee
where $\tilde{M}_2\equiv M_2+q_Y\gamma$. Acting with this symmetry on the $\fb_M$ wavefunctions directly gives the wavefunctions for the $\te_M$ curve. These are given by eq.~(\ref{massless1}) with
\begin{equation}
p_1^{\te_M}=\tilde{M}_2\ , \qquad p_2^{\te_M}=\rho_2\ , \qquad k_2^{\te_M}=1\ , \qquad p_3^{\te_M}=p_4^{\te_M}=k_1^{\te_M}=0
\end{equation}
and $\rho_2\equiv\sqrt{\tilde{M}_2^2+\left(\frac{R}{v_2}\right)^2}$. We have assumed $\tilde{M}_2$ to be positive, following again the discussion in the previous subsection.

Dimensionless eigenvalues and their corresponding eigenvectors are given respectively by
\begin{equation}
\lambda_0^{\te_M}=0\ , \qquad \lambda_1^{\te_M}=-\rho_2-\tilde{M}_2\ ,\qquad \lambda_2^{\te_M}=\rho_2-\tilde{M}_2\ , \qquad \lambda_3^{\te_M}=2\tilde{M}_2\ ,
\end{equation}
and,
\begin{equation}
\xi_0^{\te_M}=\begin{pmatrix}1\\ 0\\ 0 \\ 0\end{pmatrix}\ , \quad \xi_1^{\te_M}=\begin{pmatrix}0\\ 0\\ -\rho_2-\tilde{M}_2\\ \frac{R}{v_2}\end{pmatrix}\ , \quad \xi_2^{\te_M}=\begin{pmatrix}0\\ 0\\ \rho_2-\tilde{M}_2\\ \frac{R}{v_2}\end{pmatrix}\ , \quad \xi_3^{\te_M}=\begin{pmatrix}0\\ 1\\ 0\\ 0\end{pmatrix}\ .
\end{equation}

%%%%%%%%%%%%%%%%%%%%%%%%%%%%%%%%%%%%%%%%%%%%%%%%%%%%%%%%%%%%%%%%%%%%%%%%%%%%%
\subsubsection{Wavefunctions for the $\fb_H$ Higgs curve: non-vanishing flux density}
\label{sec:masslesshiggs}
%%%%%%%%%%%%%%%%%%%%%%%%%%%%%%%%%%%%%%%%%%%%%%%%%%%%%%%%%%%%%%%%%%%%%%%%%%%%%

As we have already commented, we should consider three different possibilities for the wavefunctions localised in the $\fb_H$ Higgs curve, depending on whether $\tilde{M}_{12}$ is smaller, bigger or equal to zero, with
\be
\label{mt12def}
\tilde{M}_{12}\equiv M_1-M_2-q_Y\gamma \;.
\ee
In this subsection we compute the wavefunctions for the first two cases, whereas the third case is addressed in the next subsection. For concreteness we take $\tilde{M}_{12}<0$ in eqs.~(\ref{localchiralfluxhiggs}). We follow exactly the same procedure as in previous subsections. However, we find convenient to express the result in a different coordinate system
\begin{align}
w&=\frac{1}{\sqrt{2}}(z_1+z_2) \;,& \psi_{\bar w}&=\frac{1}{\sqrt{2}}(\psi_{\bar 1}+\psi_{\bar 2})\label{uwbasis} \;,\\
u&=\frac{1}{\sqrt{2}}(z_1-z_2) \;,& \psi_{\bar u}&=\frac{1}{\sqrt{2}}(\psi_{\bar 1}-\psi_{\bar 2})\nn \;.
\end{align}
In these coordinates the Higgs curve is given by the equation $v_+u+v_-w=0$. The equations of motion in the $(u,w)$-basis are again given by eq.~(\ref{massiveeq}) but
\begin{equation}
\mathbb{D}=\begin{pmatrix}0& D_u& D_w & D_3\\
-D_u& 0& D_3^\dagger& -D_w^\dagger\\
-D_w& -D_3^\dagger&0&D_u^\dagger\\
-D_3& D_w^\dagger&-D_u^\dagger&0 \end{pmatrix}\ , \qquad \Psi=\begin{pmatrix}\sqrt{2}\eta\\ \psi_{\bar u}\\ \psi_{\bar w}\\ \chi\end{pmatrix} \;.
\end{equation}
Gauge covariant derivatives are,
\begin{equation}
D_u = \frac{M_*}{R_\parallel}\left(\partial_u - \tilde{M}_{12} \bar{w} \right) \;, \quad
D_w = \frac{M_*}{R_\parallel}\left(\partial_w - \tilde{M}_{12} \bar{u} \right) \;, \quad
D_3 = M_*R_\perp\left( \frac{\bar{w}}{v_+} + \frac{\bar{u}}{v_-}\right) \;,\label{gauged}
\end{equation}
where we have introduced the following definition
\begin{equation}
\frac{1}{v_\pm} \equiv\frac{1}{\sqrt{2}}\left(\frac{1}{v_1}\pm\frac{1}{v_2} \right) \;.
\end{equation}
Recall that the orientation is fixed as $v_2>v_1$ so that $v_->0$.
In this basis the matrix $\mathbb{B}$ reads
\begin{equation}
\mathbb{B}=\frac{M_*^2}{R_\parallel^2}\begin{pmatrix}0&0&0&0\\
0&0&2\tilde{M}_{12}& -\frac{R}{v_-} \\
0&2\tilde{M}_{12}&0&-\frac{R}{v_+}\\
0&-\frac{R}{v_-}&-\frac{R}{v_+}&0\end{pmatrix}\label{bh} \;,
\end{equation}
with eigenvalues given by the three roots of the cubic equation,
\begin{equation}
\left(\lambda^{\fb_H}_p\right)^3 - \lambda^{\fb_H}_p\left(\frac{R^2}{v_+^2} + \frac{R^2}{v_-^2} + 4\tilde{M}_{12}^2 \right) - \frac{4R^2}{v_+v_-} \tilde{M}_{12} =0 \;, \qquad p=1,2,3 \label{lcubich}
\end{equation}
and $\lambda^{\fb_H}_0=0$. The corresponding eigenvectors are,
\begin{equation}
\xi^{\fb_H}_0=\begin{pmatrix}1\\ 0\\ 0\\ 0\end{pmatrix}\ , \qquad \xi^{\fb_H}_p=\begin{pmatrix} 0\\ k^{\fb_H}_{(p,u)} \\ k^{\fb_H}_{(p,w)} \\ -4\tilde{M}_{12}^2+\left(\lambda_p^{\fb_H}\right)^2  \end{pmatrix} \;,\quad p=1,2,3
\end{equation}
where
\begin{equation}
k^{\fb_H}_{(p,u)} = -R\left(\frac{2\tilde{M}_{12}}{v_+}+\frac{\lambda_p^{\fb_H}}{v_-}\right) \;, \qquad
k^{\fb_H}_{(p,w)} =  -R\left(\frac{2\tilde{M}_{12}}{v_-}+\frac{\lambda_p^{\fb_H}}{v_+}\right)\;.
\end{equation}
With this information at hand we can solve eqs.~(\ref{zero00}). For $\tilde{M}_{12}<0$ we find that the ground state wavefunctions for each of the four towers of fields localised in the $\fb_H$ curve are given by
\be
\Psi^{\fb_H,\, l}_{p} = \frac{\xi_p^{\fb_H}}{N_{p}^l} \left(-k_{u}^{\fb_H}w + k_{w}^{\fb_H}u\right)^{l}
e^{-p_1^{\fb_H} |w|^2 - p_2^{\fb_H} |u|^2 + p_3^{\fb_H} \bar{w}u + p_4^{\fb_H} \bar{u}w}\label{higgsgen}
\ee
with $k^{\fb_H}_{u}=k^{\fb_H}_{(1,u)}$, $k^{\fb_H}_{w}=k^{\fb_H}_{(1,w)}$ and
\begin{align}
p^{\fb_H}_1 &= -\frac{(2\tilde{M}_{12}v_+v_- +  \lambda^{\fb_H}_1v_-^2)\lambda^{\fb_H}_1}{4\tilde{M}_{12}v_+v_-  + (v_+^2 + v_-^2)\lambda^{\fb_H}_1} \;, &
p^{\fb_H}_2 &= -\frac{(2\tilde{M}_{12}v_+v_- +  \lambda^{\fb_H}_1v_+^2)\lambda^{\fb_H}_1}{4\tilde{M}_{12}v_+v_-  + (v_+^2 + v_-^2)\lambda^{\fb_H}_1} \;,  \\
p^{\fb_H}_3 &= \frac{(2\tilde{M}_{12}v_+^2 +  \lambda^{\fb_H}_1v_+v_-)\lambda^{\fb_H}_1}{4\tilde{M}_{12}v_+v_-  + (v_+^2 + v_-^2)\lambda^{\fb_H}_1} - \tilde{M}_{12}\;, &
p^{\fb_H}_4 &= \frac{(2\tilde{M}_{12}v_-^2 +  \lambda^{\fb_H}_1v_+v_-)\lambda^{\fb_H}_1}{4\tilde{M}_{12}v_+v_-  + (v_+^2 + v_-^2)\lambda^{\fb_H}_1} - \tilde{M}_{12} \;. \label{higgswavepi}\nn
\end{align}

Wavefunctions for heavier modes can be obtained by acting with the gauge covariant derivatives (\ref{gauged}) on the ground state wavefunctions, as in eq.~(\ref{massiverep}). The corresponding masses are given in eqs.~(\ref{mas}). Thus, for $\tilde{M}_{12}<0$ there is a massless chiral fermion coming from the $\fb_H$. Wavefunctions for fields in the $\mathbf{5}_H$ representation are also given by the same expressions, but their masses are shifted in such a way that there is no zero mode, in agreement with the general discussion of subsection \ref{sublocal}.

Similarly we can work out the wavefunctions for the case on which $\tilde{M}_{12}>0$. In that case,
\begin{equation}
\lambda_1^{\fb_H}\lambda_2^{\fb_H}\lambda_3^{\fb_H}=\frac{4\tilde{M}_{12}R^2}{v_+v_-}>0 \;,
\end{equation}
as can be deduced from the determinant of the non-trivial part of eq.~(\ref{bh}), and there are one positive and two negative eigenvalues. Thus, wavefunctions are again given by the same expressions as in the case $\tilde{M}_{12}<0$, but with the change $\tilde{M}_{12}\to-\tilde{M}_{12}$ and where $\lambda_1^{\fb_H}$ now denotes the positive eigenvalue. The roles of the $\fb_H$ and the $\mathbf{5}_H$ representations is also exchanged in such a way that the massless chiral multiplet now comes from the $\mathbf{5}_H$ representation.

%%%%%%%%%%%%%%%%%%%%%%%%%%%%%%%%%%%%%%%%%%%%%%%%%%%%%%%%%%%%%%%%%%%%%%%%%%%%%
\subsubsection{Wavefunctions for the $\fb_H$ Higgs curve: vanishing flux density}
\label{sec:wavehiggsnoflux}
%%%%%%%%%%%%%%%%%%%%%%%%%%%%%%%%%%%%%%%%%%%%%%%%%%%%%%%%%%%%%%%%%%%%%%%%%%%%%

Let us now consider the case on which $\tilde{M}_{12}=0$ for the triplet mode on the $\fb_H$ curve. In that case the eigenvalues of the matrix $(R_\parallel/M_*)^2\mathbb{B}$ in eq.~(\ref{bh}) become
\begin{equation}
\lambda^{\fb_H}_0=\lambda^{\fb_H}_3=0\ , \qquad \lambda^{\fb_H}_1=-R\sqrt{\frac{1}{v_+^2}+\frac{1}{v_-^2}}\ , \qquad \lambda^{\fb_H}_2=R\sqrt{\frac{1}{v_+^2}+\frac{1}{v_-^2}} \;,
\end{equation}
and the corresponding eigenvectors
\begin{equation}
\xi^{\fb_H}_0=\begin{pmatrix}1\\ 0\\ 0\\ 0\end{pmatrix}\ , \qquad \xi^{\fb_H}_1=\begin{pmatrix} 0\\ v_+ \\ v_- \\ \sqrt{v_+^2+v_-^2}  \end{pmatrix} \ ,\qquad \xi^{\fb_H}_2=\begin{pmatrix} 0\\ v_+ \\ v_- \\ -\sqrt{v_+^2+v_-^2}  \end{pmatrix}\ , \qquad \xi^{\fb_H}_3=\begin{pmatrix} 0\\ -v_- \\ v_+ \\ 0  \end{pmatrix} \;.
\end{equation}
In particular, $[\tilde{D}_3^\dagger,\tilde D_3]=0$ and there is a conserved complex Kaluza-Klein momentum associated to that commutator instead of a quantum harmonic oscillator. We denote by $k_{\rm KK}$ the conserved quantum number.

Wavefunctions are obtained by following the same procedure as in previous sections. Thus, we obtain
\begin{equation}
\Psi_{p,(n,m,k_{\rm KK})}=\frac{(R_\parallel/M_*)^{m+n}(\tilde{D}_1^\dagger)^{n}(\tilde D_2)^m}{\sqrt{m!n!}(-\lambda_1^{\fb_H})^{n/2}\left(\lambda_2^{\fb_H}\right)^{m/2}}\Psi_{p,(0,0,k_{\rm KK})} \;,
\end{equation}
with ground state wavefunction
\begin{equation}
\Psi_{p,(0,0,k_{\rm KK})}=\frac{\xi_p}{N_{p}}\textrm{exp}\left[\frac{1}{\sqrt{v_+^2+v_-^2}}\left(-\frac{R}{v_+v_-}|v_+u+v_-w|^2+2i\textrm{Im}\left[k_{\rm KK}(-v_-u+v_+w)\right]\right)\right]
\end{equation}
where $N_p$ is the normalisation constant. The masses of these modes are
\begin{align}
M^2_{\Psi_{0,(n,m,k_{\rm KK})}} &= \left(\frac{M_*}{R_\parallel}\right)^2 \left[ -(m+n+1)\lambda_1^{\fb_H}+|k_{\rm KK}|^2 \right] \;, \\
M^2_{\Psi_{1,(n,m,k_{\rm KK})}} &= \left(\frac{M_*}{R_\parallel}\right)^2 \left[ -(m+n)\lambda_1^{\fb_H}+|k_{\rm KK}|^2 \right] \;, \nn\\
M^2_{\Psi_{2,(n,m,k_{\rm KK})}} &= \left(\frac{M_*}{R_\parallel}\right)^2 \left[ -(m+n+2)\lambda_1^{\fb_H}+|k_{\rm KK}|^2 \right] \;, \nn\\
M^2_{\Psi_{3,(n,m,k_{\rm KK})}} &= \left(\frac{M_*}{R_\parallel}\right)^2 \left[ -(m+n+1)\lambda_1^{\fb_H}+|k_{\rm KK}|^2 \right] \;. \nn
\end{align}
Wavefunctions are only localised along the directions which are transverse to the Higgs curve and arrange in vector-like pairs. Thus, the normalisation constant $N_p$ and the correct quantisation of $k_{\rm KK}$ are only globally determined. A rough estimate of $N_p$ can be given by assuming the Higgs curve to be a completely homogeneous space, resulting in
\begin{equation}
\label{wavenofluxnorm}
N_p\simeq R_\parallel\left(\frac{\pi v_+v_-\textrm{Vol}_{\fb_H}}{2R}\right)^{1/2}||\xi_p||\simeq \frac{1}{\varepsilon}\left(\frac{\pi v_+v_- R}{2}\right)^{1/2}||\xi_p||
\end{equation}
where $\textrm{Vol}_{\fb_H}$ denotes the volume of the Higgs curve and $\varepsilon$ was defined in eqs.~(\ref{erconst}).

%%%%%%%%%%%%%%%%%%%%%%%%%%%%%%%%%%%%%%%%%%%%%%%%%%%%%%%%%%%%%%%%%%%%%%%%%%%%%
\section{Wavefunction overlaps}
\label{sec:overlaps}
%%%%%%%%%%%%%%%%%%%%%%%%%%%%%%%%%%%%%%%%%%%%%%%%%%%%%%%%%%%%%%%%%%%%%%%%%%%%%

Our aim is now to compute the coefficient of the cubic couplings in the 4-dimensional effective action which results from dimensionally reducing the 8-dimensional theory of section \ref{sec:8d}.
The relevant operator in the 8-dimensional theory is given by \cite{Beasley:2008dc}
\be
W \supset \int_S {\bf A} \wedge {\bf A} \wedge {\bf \Phi} \;. \label{supyuk}
\ee
Thus, having calculated the wavefunctions for the massless and massive charged fields, cubic couplings in the 4-dimensional effective theory are given by integrating the triple overlap of the associated wavefunctions over the internal 4-cycle $S$. Since wavefunctions are localised within the local patch so is their overlap, which means that effectively we can perform the integral over $S$ as an integral over ${\mathbb C}^2$.

Physical cubic couplings are determined from overlaps of normalised wavefunctions, the normalisation condition assuring 4-dimensional canonically normalised kinetic terms. Wavefunctions for the charged fields are given by
\bea
{\bf A}_{\bar{1}} &=& \psi^{\fb_M}_{\bar{1}} t^{\fb_M} + \psi^{\te_M}_{\bar{1}} t^{\te_M} + \psi^{\fb_H}_{\bar{1}} t^{\fb_H} \;, \nn \\
{\bf A}_{\bar{2}} &=& \psi^{\fb_M}_{\bar{2}} t^{\fb_M} + \psi^{\te_M}_{\bar{2}} t^{\te_M} + \psi^{\fb_H}_{\bar{2}} t^{\fb_H} \;, \nn \\
{\bf \Phi}_{12} &=& \chi^{\fb_M} t^{\fb_M} + \chi^{\te_M} t^{\te_M} + \chi^{\fb_H} t^{\fb_H} \;,
\eea
where the matrices $t^{\fb_M}$, $t^{\te_M}$ and $t^{\fb_H}$ are along the corresponding bi-fundamental generators in $\mathfrak{su}(5)\times \mathfrak{u}(1)\times \mathfrak{u}(1)\subset \mathfrak{so}(12)$. Substituting into (\ref{supyuk}) we get
\bea
Y^{(i,j)}_{p,(n,m,l)} &=& \frac16 \int_S \left[ \psi^{\fb_H}_{\bar{1}\, p,(n,m,l)} \psi^{\te_M,j}_{\bar{2}} \chi^{\fb_M,i} + \psi^{\fb_H}_{\bar{2}\, p,(n,m,l)} \psi^{\fb_M,i}_{\bar{1}} \chi^{\te_M,j} \right. \nn \\
& & \quad - \chi^{\fb_H}_{p,(n,m,l)} \psi^{\fb_M,i}_{\bar{1}} \psi^{\te_M,j}_{\bar{2}} - \psi^{\fb_H}_{\bar{1}\, p,(n,m,l)} \psi^{\fb_M,i}_{\bar{2}} \chi^{\te_M,j}  \nn \\
& & \quad\left. - \psi^{\fb_H}_{\bar{2}\, p,(n,m,l)} \chi^{\fb_M,i} \psi^{\te_M,j}_{\bar{1}} + \chi^{\fb_H}_{p,(n,m,l)} \psi^{\fb_M,i}_{\bar{2}} \psi^{\te_M,j}_{\bar{1}} \right] \;.
\label{yukexpterms}
\eea
In this expression we have dropped the overall group theory factor $\mathrm{Tr}\left\{\left[t^{\fb_M},t^{\te_M}\right]t^{\fb_H}\right\}$ and reinstated the generation indices $i$ and $j$ counting the massless Landau level degeneracy and the quantum numbers $(n,m,l)$ labeling massive modes on each of the three towers $p=1,2,3$ of fields localised in the $\fb_H$ curve.\footnote{Cubic couplings between two massless matter fields and the vector multiplets $\Psi_{0,(n,m,l)}$ vanish because of the $\mathcal{N}\geq 1$ supersymmetry preserved by the background. They can only be generated after supersymmetry breaking.} For the case discussed in subsection \ref{sec:wavehiggsnoflux} where the effective local flux density vanishes, relabeling $(n,m,l)\to (n,m,k_{\rm KK})$ is understood in the above expression.

Note that $Y^{(i,j)}_{p,(n,m,l)}$ implicitly depends on the hypercharge of the fields involved in the coupling. Yukawa couplings correspond to the case of three massless fields and are thus given by $Y^{(i,j)}_{1,(0,0,0)}$, where the wavefunctions $\psi_{\bar{1}\, 1,(0,0,0)}^{\fb_H}$, $\psi_{\bar{2}\, 1,(0,0,0)}^{\fb_H}$ and $\chi_{1,(0,0,0)}^{\fb_H}$ have $q_Y=-1/2$. On the other hand, triple couplings between two massless matter fields and one heavy coloured triplet are given by $Y^{(i,j)}_{p,(m,n,l)}$, where now  $\psi_{\bar{1}\, 1,(0,0,0)}^{\fb_H}$, $\psi_{\bar{2}\, 1,(0,0,0)}^{\fb_H}$ and $\chi_{1,(0,0,0)}^{\fb_H}$ have $q_Y=1/3$. Integrating out the three towers of massive (anti-)triplets leads to non-renormalisable operators in the infrared, which often put strong experimental bounds on the particular F-theory GUT model. Performing this integration is however beyond the scope of this work, especially since for the case of proton decay through dimension-five operators it would also require knowledge of the up-type cubic couplings localised at E$_6$ enhancement points of $S$. Instead, in this section we analyze the down-type cubic couplings, eq.~(\ref{yukexpterms}), between two massless matter fields and the leading contributors in the three towers of massive triplets. We split the analysis into two subsections, concerning the cases where the Higgs curve has non-vanishing flux density and where it has vanishing flux density and therefore massive fields carry some conserved KK momentum.

%%%%%%%%%%%%%%%%%%%%%%%%%%%%%%%%%%%%%%%%%%%%%%%%%%%%%%%%%%%%%%%%%%%%%%%%%%%%%%%%%%%%%%%%%%
\subsection{$\fb_H$ Higgs curve with non-vanishing flux density}
\label{sec:waveoverhiggsflux}
%%%%%%%%%%%%%%%%%%%%%%%%%%%%%%%%%%%%%%%%%%%%%%%%%%%%%%%%%%%%%%%%%%%%%%%%%%%%%%%%%%%%%%%%%%

We first consider the case where triplets feel a non-vanishing flux density and therefore all massive modes are localised Landau levels with no conserved KK momentum. The relevant wavefunctions were presented in subsection \ref{sec:masslesshiggs} (and the more general versions in appendix \ref{app:oblique}).

In order to evaluate eq.~(\ref{yukexpterms}), it is useful to recall some geometric selection rules that identify the particular Landau levels coupling to each matter generation. For that, we can assign a global U(1) charge to each 4-dimensional field corresponding to the following local holomorphic rotation of the internal space
\be
z_1 \rightarrow e^{i\theta} z_1 \ , \qquad z_2 \rightarrow e^{i\theta}z_2 \;. \label{u1}
\ee
For constant fluxes $M_1$, $M_2$ and $\gamma$, the exponential factor of the wavefunctions is invariant under this transformation and so the wavefunctions transform with a phase. The charge of the wavefunction is such that in the polynomial prefactor each power of a holomorphic coordinate $z_1$ or $z_2$, contributes a $+1$ charge while each power of an anti-holomorphic coordinate contributes a $-1$ charge.

Massless wavefunctions only involve holomorphic prefactors and so can only have positive charges. The charge is essentially labeled by the generation number since the holomorphic polynomial prefactors are given by holomorphic coordinates raised to the power of the generation index. As studied in \cite{Heckman:2008qa, Font:2009gq, Cecotti:2009zf, Conlon:2009qq, Aparicio:2011jx} the heaviest generation is usually associated to the constant prefactor and so to vanishing charge, whereas second and first generations are associated to charges $+1$ and $+2$ respectively.\footnote{Another theory of flavour was proposed in \cite{Dudas:2009hu} for which the generation structure cannot be studied locally at an SO(12) point. We discuss the implications for this model in subsections \ref{sec:waveoverhiggsnoflux} and \ref{sec:phenoyuk}.}

Massive Landau replicas are obtained by acting with the raising operators on the massless fields, as described in subsection \ref{sublocal}. Ladder operators carry a definite charge under the above U(1) symmetry. More precisely, for creation operators we have
\begin{equation}
Q_{U(1)}\left(\tilde D^{\dagger}_1\right) = +1 \;, \qquad
Q_{U(1)}\left(\tilde D_2\right) = -1 \;, \qquad
Q_{U(1)}\left(\tilde D_3\right) = -1 \;,
\end{equation}
with opposite charges for the annihilation operators. Thus, the global U(1) charge of a massive field is given by\footnote{Note that, since we only require one generation for the $\fb_H$ curve, we assume that the global structure is such that the generation index vanishes for this curve.}
\be
Q_{U(1)}\left(\Psi^i_{p,(n,m,l)}\right) = i+n-m-l \;.
\ee

The usefulness of assigning such charges becomes apparent when considering the triple wavefunction overlap integral. Since the integration measure is invariant under this symmetry, the product of the three wavefunctions must also be invariant in order for the integral to be non-vanishing. Thus, the only non-vanishing couplings $Y^{(i,j)}_{p,(n,m,l)}$ are such that $i+j+n-m-l=0$. This in particular implies that all Yukawa couplings vanish except for the one of the $b$ quark, that is $Y^{(0,0)}_{1,(0,0,0)}$ \cite{Heckman:2008qa,Cecotti:2009zf,Conlon:2009qq}. It is however important to emphasise that this U(1) symmetry is broken by non-trivial local metric or flux profiles, closed string fluxes \cite{Cecotti:2009zf} or non-perturbative effects \cite{Marchesano:2009rz}. We should therefore think of this symmetry as applying at leading order in an expansion in the spatial variation of the metric and fluxes.

The above selection rule also constrains the possible couplings between two massless matter fields and one heavy (anti-)triplet. We have summarised in table \ref{table1} the allowed cubic couplings between right-handed quarks and squarks and the lightest massive anti-triplets $T_{p,(n,m,l)}$. The remaining couplings to heavy anti-triplets allowed by the selection rule are formed by acting on these lightest states with `vector-like' combinations of raising operators, in the sense of adding no net U(1) charge. Thus, for instance $\tilde t_{\rm R} b_{\rm R}$ couples to the infinite set of heavy anti-triplets
\be
T_{p,(0,0,0)}\ , \ \ T_{p,(1,1,0)}\ , \ \ T_{p,(1,0,1)}\ , \ \ T_{p,(2,2,0)}\ , \ \ T_{p,(2,1,1)}\ , \ \ T_{p,(2,0,2)} \ ,\ \ \ldots  \label{vecsum}
\ee
and similarly for other pairs of right-handed quarks and squarks.

\begin{table}[!ht]
\begin{center}
\begin{tabular}{|c||c|c|c|}
\hline
& $\tilde u_{\rm R}$ & $\tilde c_{\rm R}$ & $\tilde t_{\rm R}$\\
\hline \hline
$d_{\rm R}$ & {\footnotesize $T_{(0,4,0)}$, $T_{(0,3,1)}$, $T_{(0,2,2)}$, $T_{(0,1,3)}$, $T_{(0,0,4)}$} & {\footnotesize $T_{(0,3,0)}$, $T_{(0,2,1)}$, $T_{(0,1,2)}$, $T_{(0,0,3)}$}& {\footnotesize $T_{(0,2,0)}$, $T_{(0,1,1)}$, $T_{(0,0,2)}$}\\
\hline
$s_{\rm R}$& {\footnotesize $T_{(0,3,0)}$, $T_{(0,2,1)}$, $T_{(0,1,2)}$, $T_{(0,0,3)}$}& {\footnotesize $T_{(0,2,0)}$, $T_{(0,1,1)}$, $T_{(0,0,2)}$}& {\footnotesize $T_{(0,1,0)}$, $T_{(0,0,1)}$} \\
\hline
$b_{\rm R}$ & {\footnotesize $T_{(0,2,0)}$, $T_{(0,1,1)}$, $T_{(0,0,2)}$}& {\footnotesize $T_{(0,1,0)}$, $T_{(0,0,1)}$}& {\footnotesize $T_{(0,0,0)}$}\\
\hline
\end{tabular}
\caption{Non-vanishing cubic couplings $Y^{(i,j)}_{p,(n,m,l)}$ between right-handed quarks and squarks and the lowest laying massive anti-triplets $T_{p,(n,m,l)}$ according to the geometric U(1) selection rule. We have omitted the subindex $p$ in order to simplify the notation. Similar couplings are possible between up-quarks, sleptons and the lowest laying massive anti-triplets.\label{table1}}
\end{center}
\end{table}

In what follows we perform the explicit calculation of the couplings which are shown in table \ref{table1}. For large values of $R$, these are the most relevant couplings, not only because they involve the lightest triplets but also because cubic couplings to fields involving a larger number of raising operators are suppressed by higher powers of $R$, as we show below.

It is worth noting that in the limit where the local U(1) symmetry is exact, and where the up-type triplet coupling associated to a point of E$_6$ and the down-type coupling of the SO(12) point are completely coincident, so that they obey the same local geometric selection rules, we expect that dimension-five proton decay operators do not arise. This is because all the massless modes have a positive or vanishing charge, which means that only the dimension-five operator involving the heaviest generations can be U(1) neutral (or put another way, the coupling between massive up-type and down-type triplets is forbidden by the U(1) selection rules). But this operator vanishes by the anti-symmetrisation of the colour indices. The relation between the local geometric U(1) symmetries of the SO(12) and E$_6$ points is beyond the scope of this work and so we cannot quantify this possibility further. However, we note some general things. Firstly it is natural that the local geometric U(1) symmetries of the SO(12) and E$_6$ points be strongly correlated as this is required for an appropriate CKM matrix. However we also know that the U(1) symmetry must be broken at scales of order $\sim 0.2$ in order to induce significant generation mixing in the Yukawa couplings. Therefore we do not expect a significant suppression from such a symmetry. Further note that the symmetry may be broken even more strongly in the massive sector: although in the case studied in this section where the Higgs curve has non-vanishing flux the massive modes are Landau-levels and so have a definite U(1) charge, in the next section we study the case where the Higgs curve has vanishing flux and then massive modes carry a non-trivial conserved KK momentum which completely break the U(1) selection rules. In that case we expect no suppression of dimension-five operators due to the local geometric U(1) symmetry.

Starting from (\ref{yukexpterms}) we can pull out an overall factor by expressing the wavefunctions $\psi_{\bar{1}}$, $\psi_{\bar{2}}$ and $\chi$ in terms of their respective functions $\vphi$, as in eq.~(\ref{ground}). This gives
\begin{multline}
Y^{(i,j)}_{p,(n,m,l)} = \\
\frac{N_{p}^{\rm cubic}(R_\parallel/M_*)^{n+m+l}}{N_{1}^{\fb_{M},i}N_{1}^{\te_M,j}N^{\fb_H,0}_{p}\sqrt{n!m!l!}\left(-\lambda^{\fb_H}_1\right)^{n/2}\left(\lambda^{\fb_H}_2\right)^{m/2}\left(\lambda^{\fb_H}_3\right)^{l/2}}
\int_S \vphi^{\fb_M,i}\vphi^{\te_M,j}\vphi^{\fb_H}_{(n,m,l)}\;, \label{yukvphi}
\end{multline}
where we have defined
\be
\vphi^{\fb_H}_{(n,m,l)} = (\tilde{D}_1^{\fb_H\dagger})^{n}(\tilde D^{\fb_H}_2)^m(\tilde D^{\fb_H}_3)^l\vphi^{\fb_H} \;,
\ee
and the overall factor $N_{p}^{\rm cubic}$ is given by
\begin{multline}
N_{p}^{\rm cubic} = \frac16 \left[ -\frac{R}{\sqrt{2}v_1}\left(k^{\fb_H}_{(p,u)} + k^{\fb_H}_{(p,w)}\right) \left(\rho_2 + \tilde{M}_2 \right)
				-\frac{R}{\sqrt{2}v_2}\left(-k^{\fb_H}_{(p,u)} + k^{\fb_H}_{(p,w)}\right) \left(\rho_1 + \tilde{M}_1 \right) \right.  \\
		 \left.+ \left(4\tilde{M}_{12}^2 - \left(\lambda_p^{\fb_H}\right)^2  \right) \left(\rho_1 + \tilde{M}_1 \right) \left(\rho_2 + \tilde{M}_2 \right) \right]\;.
		 \label{Ncubicover}
\end{multline}
To calculate the couplings of table \ref{table1} we rewrite the polynomial prefactor of the wavefunctions $\vphi^{\fb_M,i}$, $\vphi^{\te_M,j}$, $\vphi^{\fb_H}_{(n,m,l)}$ in the $(u,w)$ coordinate basis (c.f. eq.~(\ref{uwbasis})) and collect powers of $w\bar{w}$, $u\bar{u}$, $u\bar{w}$ and $w\bar{u}$. Making use of the integral (\ref{int1}) we can then express the relevant coupling constant as
\begin{multline}
Y^{(p,q)}_{r,(0,m,l)} = \frac{N^{\rm cubic}_r(-1)^{p+q}}{N_{1}^{\fb_M,p}N_{1}^{\te_M,q}N^{\fb_H,0}_{r}\sqrt{2^{p+q}m!l!}\left(\lambda^{\fb_H}_2\right)^{m/2}\left(\lambda^{\fb_H}_3\right)^{l/2}}\times\\
 \sum^p_{k_a=0} \sum^q_{k_b=0} \sum^m_{k_c=0} \sum^l_{k_d=0}
 \left[\left(-1\right)^{k_a} \begin{pmatrix}p\\ k_a\end{pmatrix}\begin{pmatrix}q\\ k_b\end{pmatrix}\begin{pmatrix}m\\ k_c\end{pmatrix}\begin{pmatrix}l\\ k_d\end{pmatrix}
 \left(\tilde{d}^2_w\right)^{m-k_c} \left(\tilde{d}^2_u\right)^{k_c}
    \left(\tilde{d}^3_w\right)^{n-k_d} \left(\tilde{d}^3_u\right)^{k_d}\right. \\
 I\left(p+q-\mathrm{max}\left[k_a+k_b,k_c+k_d\right],\mathrm{min}\left[k_a+k_b,k_c+k_d\right], \right.
\delta_k\Theta\left(\delta_k\right), \\ \left.
 -\delta_k\Theta\left(-\delta_k\right) \ ; \ \mathfrak{p}_1,\mathfrak{p}_2,\mathfrak{p}_3,\mathfrak{p}_4 \right) \bigg]
\label{cubichiggsflux}
\end{multline}
where $\Theta$ is the Heaviside theta function, $\delta_k\equiv k_a+k_b-k_c-k_d$ and we have defined the quantities
\begin{align}
\label{piy}
\mathfrak{p}_1 &\equiv p_1^{\fb_H} + \frac12 \left( \rho_1 + \rho_2 + \tilde{M}_1 + \tilde{M}_2 \right) \ , &
\mathfrak{p}_2 &\equiv p_2^{\fb_H} + \frac12 \left( \rho_1 + \rho_2 + \tilde{M}_1 + \tilde{M}_2 \right) \;,  \\
\mathfrak{p}_3 &\equiv p_3^{\fb_H} + \frac12 \left( -\rho_1 + \rho_2 + \tilde{M}_1 - \tilde{M}_2 \right) \ , &
\mathfrak{p}_4 &\equiv p_4^{\fb_H} + \frac12 \left( -\rho_1 + \rho_2 + \tilde{M}_1 - \tilde{M}_2 \right) \;,\nn
\end{align}
as well as
\bea
\tilde{d}^i_w &\equiv & \frac{1}{||\xi^{\fb_H}_i||}\left[k^{\fb_H}_{(i,u)} \left(p_3^{\fb_H}- \tilde{M}_{12} \right) - p_1^{\fb_H} k^{\fb_H}_{(i,w)} - \frac{R}{v_+}\left(4\tilde{M}_{12}^2 - \left(\lambda^{\fb_H}_i \right)^2 \right)\right] \;,  \\
\tilde{d}^i_u &\equiv & \frac{1}{||\xi^{\fb_H}_i||}\left[k^{\fb_H}_{(i,w)} \left(p_4^{\fb_H}- \tilde{M}_{12} \right) - p_2^{\fb_H} k^{\fb_H}_{(i,u)} - \frac{R}{v_-}\left(4\tilde{M}_{12}^2 - \left(\lambda^{\fb_H}_i \right)^2 \right)\right]  \;.\nn
\eea
This expression specifies the cubic couplings to the lightest fields localised in the $\fb_H$ Higgs curve. Given some matter generation choice specified by $(p,q)$ the relevant couplings are all the combinations of $(m,l)$ such that $m+l=p+q$, as summarised in table \ref{table1}. Although the expressions are rather cumbersome they are simple to evaluate with the aid of a computer.

It is interesting to recall the dependence of eq.~(\ref{cubichiggsflux}) on the local scales $R_\parallel$ and $R_\perp$ or, equivalently, on the parameters $\varepsilon$ and $R$ introduced in eqs.~(\ref{erconst}). Such dependence can be obtained from the expression of the normalisation constants, raising operators and the integral (\ref{int1}). In particular, in the limit $R\gg 1 \gg \varepsilon$ where the local approach that we are using becomes reliable, $\lambda_r$ eigenvalues scale as in eq.~(\ref{eigenscaling}) and it is possible to show that the above cubic couplings scale as
\begin{equation}
Y^{(p,q)}_{r,(0,m,l)}\sim \frac{\varepsilon}{R^{\frac32+l+\frac{m}{2}}}\label{suppres} \;.
\end{equation}
The fact that cubic couplings to heavy triplets are suppressed by higher powers of $R_\parallel R_\perp$ than the Yukawa couplings has important phenomenological consequences for proton decay, as we discuss in detail in subsection \ref{sec:proton}.\footnote{Since the suppression of lighter generations coupling is related to the higher Landau-level number of the massive mode it is natural to question how this would be modified by fluxes which do not have a constant local profile and therefore break the local geometric U(1) selection rules allowing the lighter generations to couple to lower Landau-levels. We expect that the coupling would still be suppressed in a similar, and slightly stronger due to the small parameter associated to the flux spatial variation, fashion. Evidence for this can be found in section \ref{sec:waveoverhiggsnoflux} where there is no local U(1) selection rule and we indeed find such a suppression for the lighter generations.}

%%%%%%%%%%%%%%%%%%%%%%%%%%%%%%%%%%%%%%%%%%%%%%%%%%%%%%%%%%%%%%%%%%%%%%%%%%%%%%%%%%%%%%%%%%
\subsection{$\fb_H$ Higgs curve with vanishing flux density}
\label{sec:waveoverhiggsnoflux}
%%%%%%%%%%%%%%%%%%%%%%%%%%%%%%%%%%%%%%%%%%%%%%%%%%%%%%%%%%%%%%%%%%%%%%%%%%%%%%%%%%%%%%%%%%

In the case where the effective local density of flux felt by the triplets in the $\fb_H$ curve vanishes, the triplets carry some conserved KK momentum, as discussed in subsection \ref{sec:wavehiggsnoflux}. Their wavefunction overlap with massless matter fields may therefore differ significantly from the case with only Landau levels that we have discussed in the previous subsection. There are two key differences: the first is that geometric U(1) selection rules no longer apply since KK wavefunctions explicitly break this symmetry. Moreover, the overlap integral now includes a new exponential factor depending on the KK momentum.

The relevant triple overlaps are given by
\be
Y^{(p,q)}_{r,(n,m,k_{\rm KK})} =  \frac{N_r^{\rm cubic}(R_\parallel/M_*)^{n+m}}{N_{1}^{\fb_M,p}N_{1}^{\te_M,q}N^{\fb_H}_{r}\sqrt{n!m!}\left(-\lambda^{\fb_H}_1\right)^{n/2}\left(\lambda^{\fb_H}_2\right)^{m/2}}
\int_S \vphi^{\fb_M,i}\vphi^{\te_M,j}\vphi^{\fb_H}_{(n,m,k_{\rm KK})}\;, \label{yvanish}
\ee
where we define
\be
\vphi^{\fb_H}_{r,(n,m,k_{\rm KK})} = (\tilde{D}_1^{\fb_H\dagger})^{n}(\tilde D^{\fb_H}_2)^m\vphi^{\fb_H}_{k_{\rm KK}} \;,
\ee
and the overall factor $N_r^{\rm cubic}$ is now given by
\bea
N^{\rm cubic}_r &\equiv& \frac16 \left[ -\frac{R}{\sqrt{2}v_1} \left(v_+ + v_-\right)\left(\rho_2 + \tilde{M}_2\right) - \frac{R}{\sqrt{2}v_2} \left(v_- - v_+\right)\left(\rho_1 + \tilde{M}_1\right) \right. \nn \\
		& & \left.+(-1)^r \left(v_+^2 + v_-^2 \right)^{\frac12}\left(\rho_1 + \tilde{M}_1\right)\left(\rho_2 + \tilde{M}_2\right) \right]\;, \nn \qquad \qquad r=1,2\\
N^{\rm cubic}_3 &\equiv& \frac16 \left[ -\frac{R}{\sqrt{2}v_1} \left(v_+ - v_-\right)\left(\rho_2 + \tilde{M}_2\right) - \frac{R}{\sqrt{2}v_2} \left(v_- + v_+\right)\left(\rho_1 + \tilde{M}_1\right) \right]\;.		
\eea
To evaluate the above expression it is convenient to introduce the integral
\bea
& &I_{\rm KK}\left(n_1,n_2,n_3,n_4\, ;\, p_1,p_2,p_3,p_4\, ;\, a_1,a_2\right)
\nn \\
& &
\equiv \int_{S} w^{n_1}\bar{w}^{n_2}u^{n_3}\bar{u}^{n_4}e^{-p_1|w|^2-p_2|u|^2+p_3u\bar{w}+p_4\bar{u}w + 2i\mathrm{Im}\left(a_1w+a_2u\right)} \nn \\
& &= R^4_{\parallel}\left(-1\right)^{n_2+n_4} \partial^{n_1}_{a_1} \bar{\partial}^{n_2}_{\bar{a}_1} \partial^{n_3}_{a_2} \bar{\partial}^{n_4}_{\bar{a}_2}
\left[\frac{\pi^2}{p_1p_2-p_3p_4}
e^{-\frac{\left|a_1\right|^2 p_2 + \left|a_2\right|^2 p_1 + a_1 \bar{a}_2 p_3 +  a_2 \bar{a}_1 p_4}{p_1p_2-p_3p_4}} \right] \;.
\label{ikkdef}
\eea
In order not to overload this section, here we just present the explicit analytic expression for the particular case on which one of the two quantum numbers of the heavy triplet vanishes,\footnote{The expression for $Y^{(p,q)}_{r,(n,0,k_{\rm KK})}$ follows from this one by making the replacements $\tilde{d}^1_{\{u,w\}} \to \tilde{d}^2_{\{u,w\}}$ and using the function
\begin{equation*}
I_{\rm KK}\left(p+q-k_a-k_b,k_c,k_a+k_b,n-k_c\, ; \, \mathfrak{p}_1,\mathfrak{p}_2,\mathfrak{p}_3,\mathfrak{p}_4\, ; \, \frac{k_{\rm KK} v_+}{ \sqrt{v_+^2 + v_-^2}}, -\frac{k_{\rm KK} v_-}{\sqrt{v_+^2 + v_-^2}}\right) \;.
\end{equation*}}
\begin{multline}
Y^{(p,q)}_{r,(n,0,k_{\rm KK})} =  \frac{N_r^{\rm cubic}(-1)^{p+q}}{N_1^{\fb_M,p}N_1^{\te_{M},q}N^{\fb_H}_{r}\sqrt{2^{p+q}n!}\left(-\lambda^{\fb_H}_1\right)^{n/2}}
\times \\
\sum_{k_a}^p \sum_{k_b}^q \sum_{k_c}^n  \Bigg[\left(-1\right)^{k_a}\begin{pmatrix}p\\ k_a\end{pmatrix}\begin{pmatrix}q\\ k_b\end{pmatrix}\begin{pmatrix}n\\ k_c\end{pmatrix} \left(\tilde{d}^1_u\right)^{n-k_c} \left(\tilde{d}^1_w\right)^{k_c}  \nn \\
\left.I_{\rm KK}\left(p+q-k_a-k_b+k_c,0,k_a+k_b+n-k_c,0 \, ; \, \mathfrak{p}_1,\mathfrak{p}_2,\mathfrak{p}_3,\mathfrak{p}_4\, ; \, \frac{k_{\rm KK} v_+}{ \sqrt{v_+^2 + v_-^2}}, \frac{-k_{\rm KK} v_-}{\sqrt{v_+^2 + v_-^2}}\right)\right]
\end{multline}
where have introduced the quantities
\be
\tilde{d}^i_u \equiv - \frac{2R\sqrt{v_+^2 + v_-^2}}{v_-||\xi_i^{\fb_H}||} \ , \qquad
\tilde{d}^i_w \equiv - \frac{2R\sqrt{v_+^2 + v_-^2}}{v_+||\xi_i^{\fb_H}||} \;, \qquad i=1,2
\ee
and $\mathfrak{p}_k$ are given as in eqs.~(\ref{piy}) with the replacements
\be
p_1^{\fb_H} \rightarrow \frac{R v_-}{v_+\sqrt{v_-^2+v_+^2}} \;, \qquad
p_2^{\fb_H} \rightarrow \frac{R v_+}{v_-\sqrt{v_-^2+v_+^2}} \;, \qquad
p_3^{\fb_H},\, p_4^{\fb_H} \rightarrow - \frac{R}{\sqrt{v_-^2+v_+^2}}  \;.
\ee
Explicit analytic expressions for the couplings to other heavy triplets can be worked out in a similar fashion starting from eq.~(\ref{yvanish}).

An important qualitative feature of the coupling constants (\ref{ikkdef}) is that  there is an additional exponential suppression with respect to the case with no KK momentum discussed in the previous subsection. This suppression, however, cannot be significant for the lightest KK modes if we are to keep within the local approximation, as studied in section \ref{sec:validity}. It is simple to check this explicitly by varying the parameters.  Nevertheless, a more intuitive understanding is as follows. The exponential suppression comes from the oscillations of the $\fb_H$ wavefunction with KK momentum. If the gaussian decay envelope of the matter wavefunctions is much larger than the oscillation frequency of the KK state then the oscillations cancel, leading to an exponentially suppressed overlap integral. However, within the local approximation the gaussian width must be less than the length of the $\fb_H$ curve and so the lightest KK state should have at most one oscillation within the gaussian envelope. Thus, for the range of validity of the local approach, this extra exponential suppression on top of the polynomial one of eq.~(\ref{suppres}) is rather mild.

%%%%%%%%%%%%%%%%%%%%%%%%%%%%%%%%%%%%%%%%%%%%%%%%%%%%%%%%%%%%%%%%%%%%%%%%%%%%%
\section{Phenomenological implications}
\label{sec:pheno}
%%%%%%%%%%%%%%%%%%%%%%%%%%%%%%%%%%%%%%%%%%%%%%%%%%%%%%%%%%%%%%%%%%%%%%%%%%%%%

We now address a more quantitative analysis of the cubic couplings that we have presented in the previous section and discuss their possible phenomenological implications. The primary application of our results is to proton decay induced by dimension-five operators. More precisely we have computed the coupling of the right-handed quarks and squarks (or equivalently of the up-quarks and sleptons) to the massive down-type coloured triplets which mediate proton decay. Although this coupling does not contain the full information needed to calculate the coefficient of the dimension-five operator, it plays a key role in such a calculation. Suppressing this coupling is a sufficient but not necessary condition to avoid the present strong  experimental bounds on such dimension-five operators. One of the most important attributes of the coupling is that it is a superpotential term, which means that due to its holomorphic nature it cannot involve string oscillators. The latter therefore can only enter in the physical coupling through the normalisation of the fields.

Besides the most direct application to proton stability, the coupling of matter fields to heavy modes is also important in the context of other non-renormalisable operators of phenomenological interest. For instance, in F-theory implementations of the Froggatt-Nielsen mechanism \cite{Dudas:2009hu} Yukawa couplings arise from higher dimension superpotential couplings. Similarly, FCNC operators can often be induced by integrating out heavy fields and may play an important role in studies of supersymmetry breaking. We further discuss the implications of our results for these non-renormalisable operators in subsections \ref{sec:phenoyuk} and \ref{sec:phenoFCNC} respectively.

%%%%%%%%%%%%%%%%%%%%%%%%%%%%%%%%%%%%%%%%%%%%%%%%%%%%%%%%%%%%%%%%%%%%%%%%%%%%%
\subsection{Coupling to heavy modes: proton decay}
\label{sec:proton}
%%%%%%%%%%%%%%%%%%%%%%%%%%%%%%%%%%%%%%%%%%%%%%%%%%%%%%%%%%%%%%%%%%%%%%%%%%%%%

As discussed in section \ref{sec:wave}, the nature of heavy triplets differs according to whether or not they feel a non-vanishing flux density on the curve. We consider these two possibilities in subsections \ref{sec:protonnoKK} and \ref{sec:protonKK}. In both cases the strategy is similar: we evaluate the leading couplings of interest and study their behaviour as we vary the input parameters. Although there are many input parameters that go into the calculation, we find that the qualitative behaviour depends primarily on the ratio of the fluxes to the parameter $R$ defined in eqs.~(\ref{erconst}). Hence, a simplified strategy that we adopt here is to fix the flux parameters to some $O(1)$ values and to study the behaviour of the coupling as we vary $R$ for some fixed $\varepsilon<1$. The scanning range for $R$ is set by the limits for which the effective theory is under control, as studied in section \ref{sec:validity}, keeping in mind their approximate nature. For $R<1$ the local approach that we have adopted in our computations breaks down and higher derivative corrections to the 8-dimensional effective action (in the type IIB language, corrections from string oscillators to the normalisation of the fields) become important. On the other hand, too large values of $R$ are in tension with the observed value of $\alpha_{\rm GUT}$, requiring values of $\varepsilon$ close to the unity, for which higher derivative corrections become again relevant. Thus, given some region of values for $R$, the value of $\varepsilon$ is fixed within a range by the validity of the 8-dimensional effective theory and the relation between the local scales and $\alpha_{\rm GUT}$ in that particular model.

%%%%%%%%%%%%%%%%%%%%%%%%%%%%%%%%%%%%%%%%%%%%%%%%%%%%%%%%%%%%%%%%%%%%%%%%%%%%%
\subsubsection{Coupling for Higgs curve with non-vanishing flux}
\label{sec:protonnoKK}
%%%%%%%%%%%%%%%%%%%%%%%%%%%%%%%%%%%%%%%%%%%%%%%%%%%%%%%%%%%%%%%%%%%%%%%%%%%%%

The first case that we model is where the flux density on the $\fb_H$ Higgs curve is such that there is a massless anti-triplet. Without any other extra ingredient, this scenario therefore does not fully account for doublet-triplet splitting. The massless anti-triplet would have to be lifted by some deformation of the geometry, which in the effective field theory would correspond to a supersymmetric $\mu$-term or to a vev for a GUT singlet. If the resulting mass is well below the local KK scale we expect this to only induce a small correction to our calculations. The lifted massless anti-triplet, however, leads to a new extra contribution to proton decay which can in particular dominate over the one of the massive modes that we are computing. In this sense the contributions that we consider in this subsection should be regarded as only part of the total one in a complete model.\footnote{Nevertheless, the wavefunctions and overlaps for this setup are identical to an analogous setup where this problem is avoided. This is where the hypercharge flux is such that there is a massless triplet rather than anti-triplet. Still some deformation of the geometry is required in order to lift the massless mode, but now since it is a triplet rather than anti-triplet it does not couple to the massless matter fields through down-type Yukawa couplings and does not enhance the rate of proton decay. Therefore we regard this calculation as still useful for realistic examples.}

The relevant coupling was calculated in section \ref{sec:waveoverhiggsflux}. As discussed in the introduction our primary interest is in the ratios of the heaviest generation Yukawa coupling to the cubic couplings to heavy anti-triplets. This measures the departure from the naive 4-dimensional equality of Yukawa couplings and cubic couplings to heavy triplets. In order to present the results most concisely we define an average coupling to the dominant massive modes for each pair of generations
\begin{equation}
\left\langle\hat Y^{(r)}_{pq}\right\rangle \equiv \frac{1}{1+p+q}\sum_{n,m,l} Y^{(p,q)}_{r,(n,m,l)} \;,\label{average}
\end{equation}
where the sum runs over the dominant massive anti-triplets in the $r$-th tower of localised heavy fields in the $\fb_H$ Higgs curve. See table \ref{table1} for an explicit list of the terms.

\begin{figure}[ht!]
\begin{center}
\includegraphics[width=12cm]{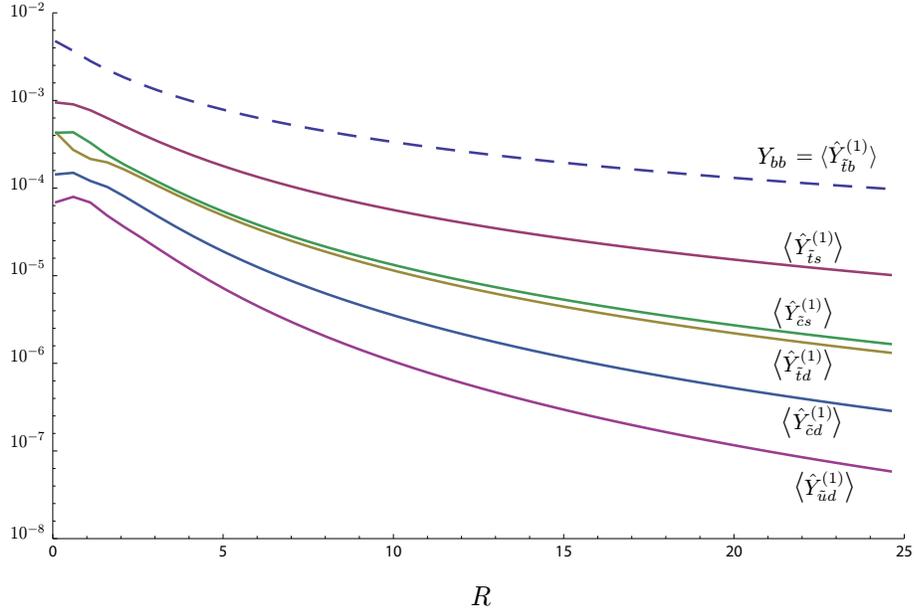}
\caption{Average couplings of different generations to massive anti-triplets in the first tower of localised fields, for flux values $v_1=5/6$, $v_2=5/4$, $M_1=1.6$, $M_2=2$, $\gamma=0$ and $\varepsilon=1/10$. For reference, we also show the Yukawa coupling for the $b$ quark (dashed line).\label{figtriplewithflux}
}
\end{center}
\end{figure}

\begin{figure}[ht!]
{\begin{center}
            \includegraphics[width=.45\textwidth]{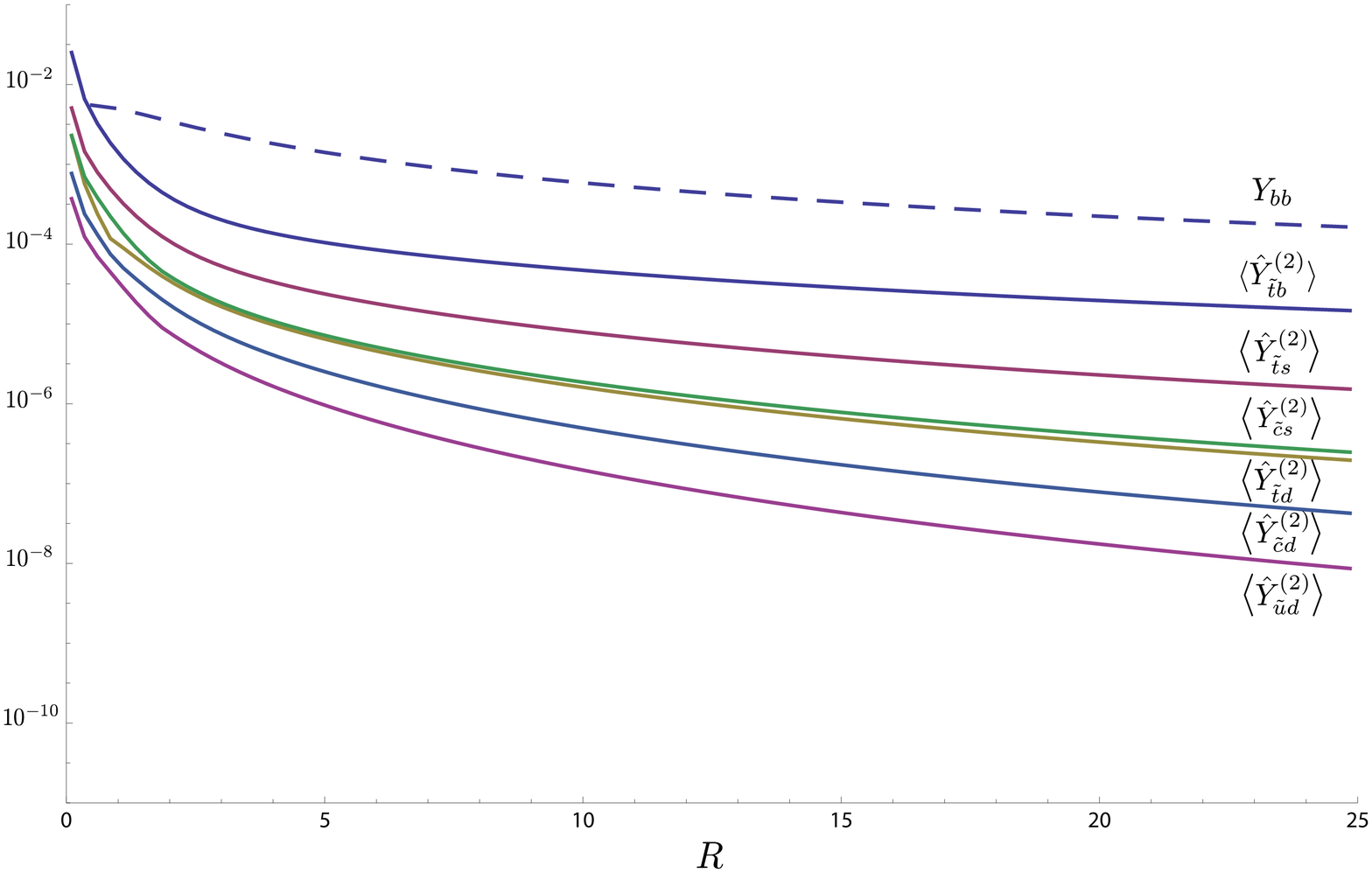}
\hspace{.3in}
 \includegraphics[width=.45\textwidth]{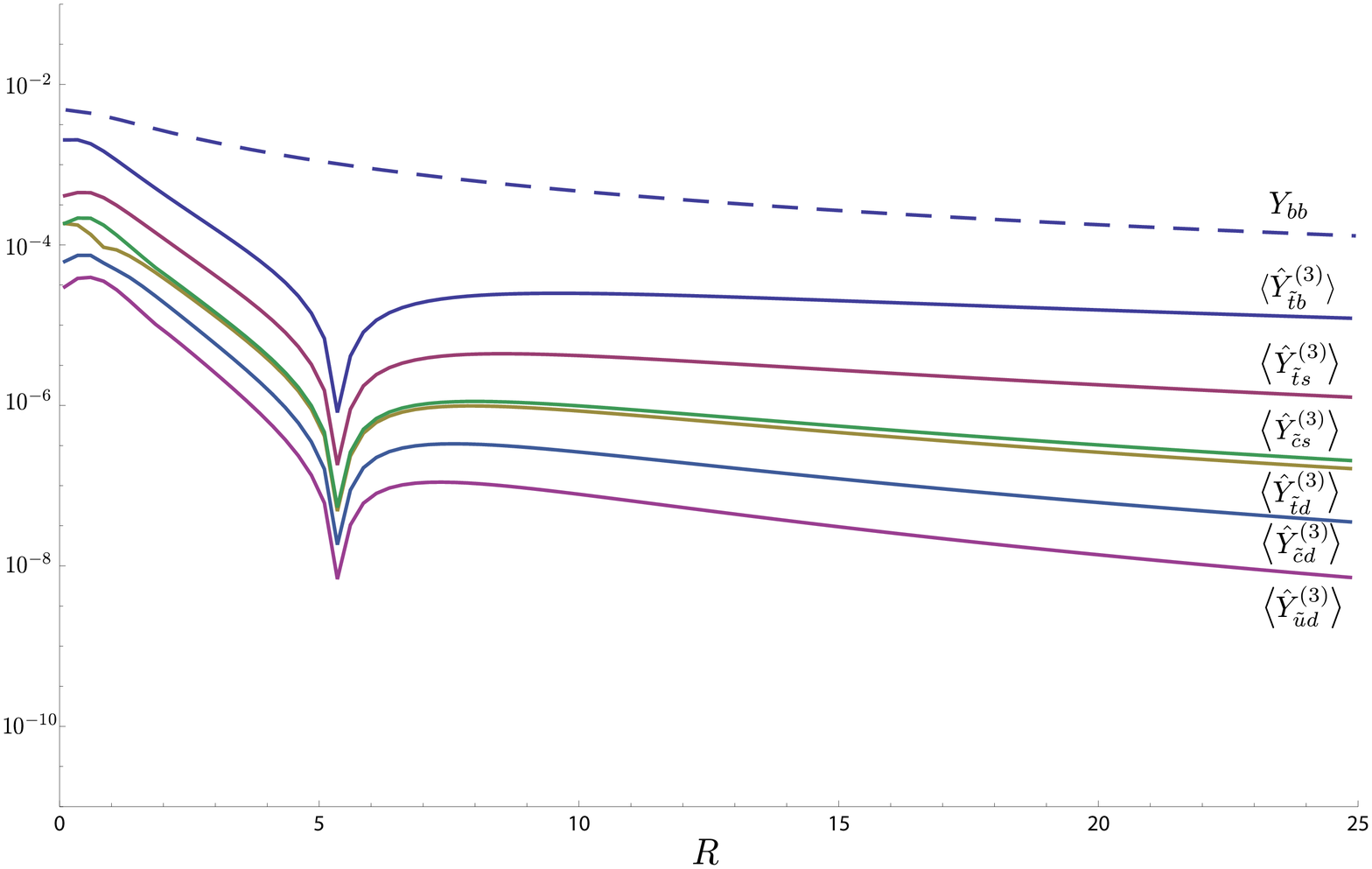}
 \caption{Average couplings of different generations to massive anti-triplets in the second (left) and third (right) towers of localised modes, for flux values $v_1=5/6$, $v_2=5/4$, $M_1=1.6$, $M_2=2$, $\gamma=0$ and $\varepsilon=1/10$.}
 \label{figtriplewithflux2}
 \end{center}}
\end{figure}

\begin{figure}[ht!]
\label{masses}
\begin{center}
\includegraphics[width=10cm]{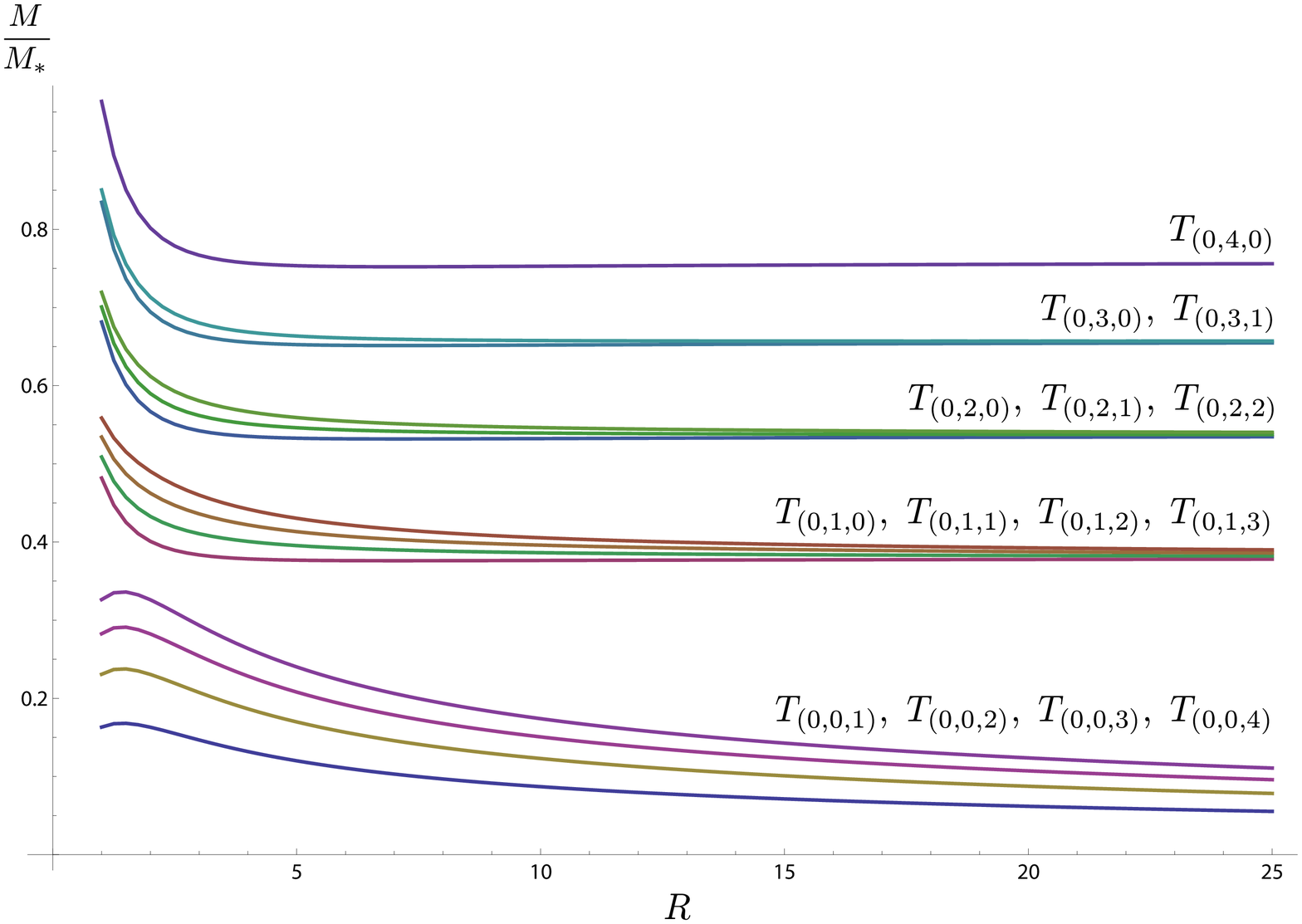}
\caption{Masses of the anti-triplets participating in the average couplings (\ref{average}), as summarised in table \ref{table1}, for the first tower of modes with flux values $v_1=5/6$, $v_2=5/4$, $M_1=1.6$, $M_2=2$, $\gamma=0$ and $\varepsilon=1/10$.\label{figmasses}}
\end{center}
\end{figure}

In figure \ref{figtriplewithflux} we plot the average coupling to the first tower of massive anti-triplets, the one generated by acting with raising operators on the massless modes, for values of the flux of order one. Since doublet-triplet splitting is not fully accounted for in this case we turn off the hypercharge flux for simplicity $\gamma=0$. The plot shows the coupling to the leading massive modes for different generations. The dashed line shows the triple coupling of massless modes, i.e. the Yukawa coupling of the heaviest generation.
The couplings to the other two towers, which do not have a zero mode, are similar in nature and are shown in figure \ref{figtriplewithflux2}.\footnote{The dip seen in the couplings to the third tower shown in figure \ref{figtriplewithflux2} goes all the way to zero, although due to the resolution of the plot it is not completely captured. The origin of this dip is on the overall factor (\ref{Ncubicover}) which develops a zero as a function of $R$ for that particular tower. A similar feature occurs in figure \ref{figtriplewithoutflux2}.}
For completeness we also plot in figure \ref{figmasses} the masses of the different coloured anti-triplets that participate in the average couplings (\ref{average}).

The behaviour of the cubic couplings to massive modes is relatively universal and robust with respect to small variations of the flux and $R$. We find that for small values of $R$, such that it is of order the flux parameters, the coupling to massive anti-triplets for the lighter generations is only slightly suppressed compared to the Yukawa coupling. This means that for this region of the parameter space the constraints on proton decay are stronger than the naive 4-dimensional field theory result using (\ref{4dratios}). As we increase $R$ relative to the flux the lighter generation couplings decrease more steeply than the heavier generations, such that for large enough values of $R$ the couplings of the lighter generations are significantly suppressed compared to the minimal 4-dimensional field theory guess. However, note that the bottom Yukawa coupling also decreases for large $R$ leading to very small tan~$\beta$ which eventually becomes incompatible with gauge coupling unification, which favours tan~$\beta>1$. Therefore in such a setup we expect that from the Yukawa coupling not being too suppressed that smaller values of $R$ are favoured.

%%%%%%%%%%%%%%%%%%%%%%%%%%%%%%%%%%%%%%%%%%%%%%%%%%%%%%%%%%%%%%%%%%%%%%%%%%%%%
\subsubsection{Coupling for Higgs curve with vanishing flux}
\label{sec:protonKK}
%%%%%%%%%%%%%%%%%%%%%%%%%%%%%%%%%%%%%%%%%%%%%%%%%%%%%%%%%%%%%%%%%%%%%%%%%%%%%

We study now the case where triplets in the $\fb_H$ Higgs curve feel a vanishing total effective flux and therefore, given the appropriate topology, there are no massless triplets and doublet-triplet splitting is accounted for. Massive vector-like pairs of triplets carry a conserved KK charge and their wavefunctions are delocalised along the Higgs matter curve. This means that we cannot calculate the wavefunction normalisation explicitly and have to estimate it, as we did in eq.~(\ref{wavenofluxnorm}). However, this is not a particularly serious shortcoming given the accuracy at which we are working, and we refer to appendix B for a more detailed study of this issue.

\begin{figure}[ht!]
\begin{center}
\includegraphics[width=12cm]{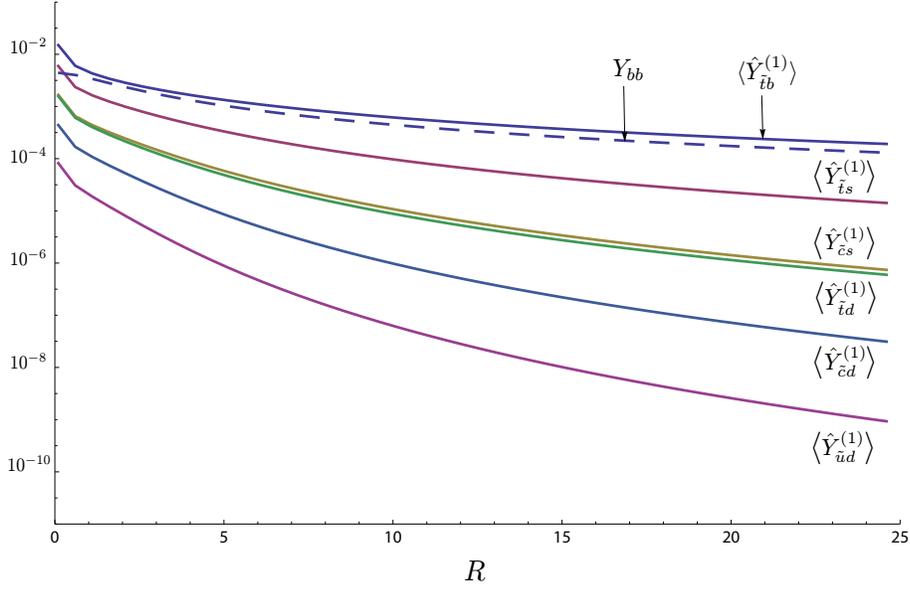}
\caption{Couplings to the lightest massive vector-like pair of triplets in the first tower of fields localised in the Higgs curve, for flux values $v_1=5/6$, $v_2=5/4$, $M_1=1.6$, $M_2=2$, $\gamma=1.2$, $\varepsilon=1/10$, $k_{\rm KK}=1$ and trivial Landau-level quantum numbers.
For reference, we also show the Yukawa coupling for the $b$ quark (dashed line). Note that, since hypercharge flux is non-vanishing the coupling is not GUT group universal. The coupling to the triplets that we plot is $(\bar{3},1)_{1/3}\otimes(1,2)_{-1/2}\otimes(3,2)_{1/6}$ while the Yukawa coupling plotted for comparison is $(1,2)_{-1/2}\otimes(\bar{3},1)_{1/3}\otimes(3,2)_{1/6}$.\label{figtriplewithoutflux}
}
\end{center}
\end{figure}

\begin{figure}[ht!]
{\begin{center}
                      \includegraphics[width=.45\textwidth]{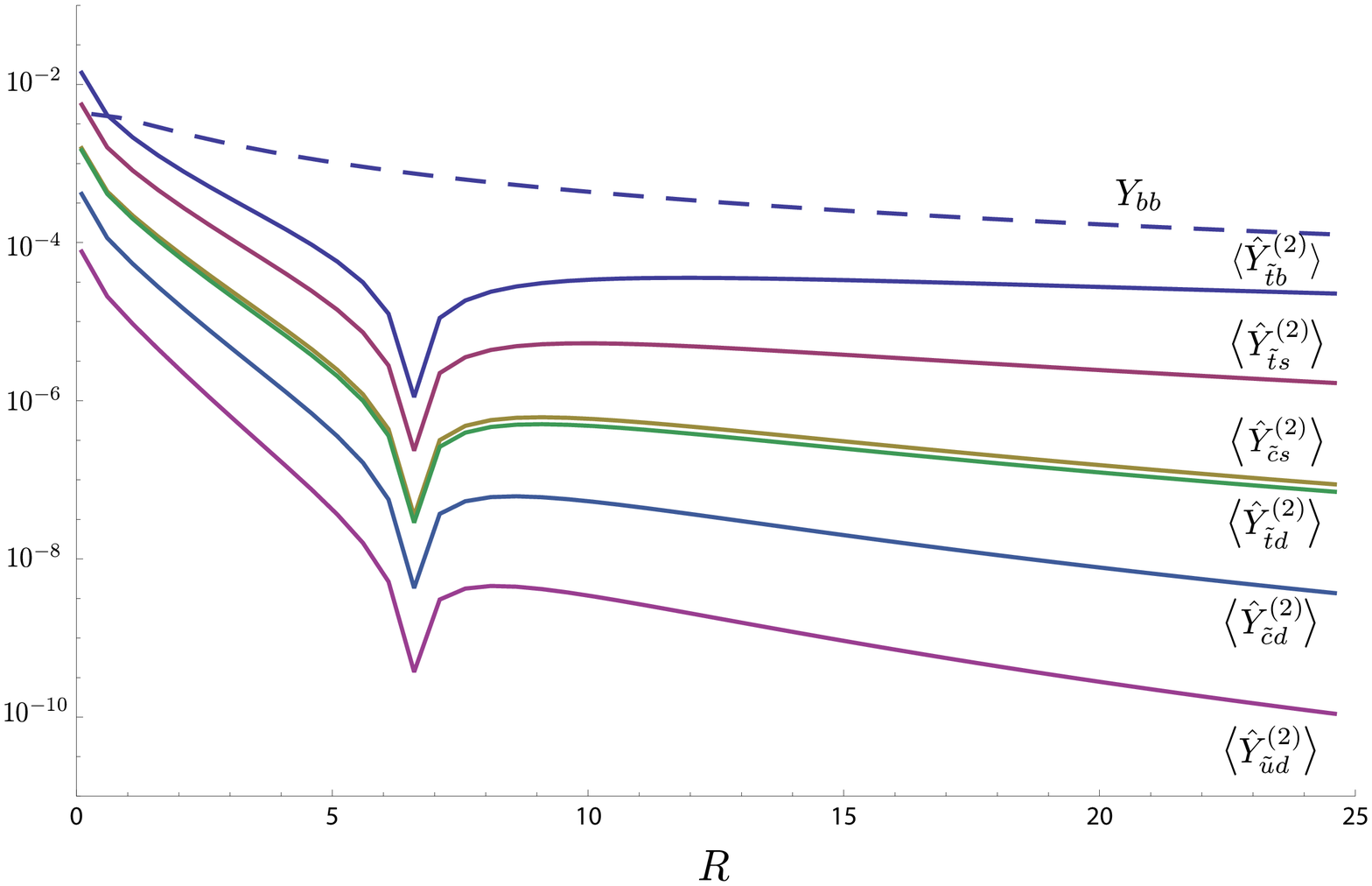}
\hspace{.3in}
 \includegraphics[width=.45\textwidth]{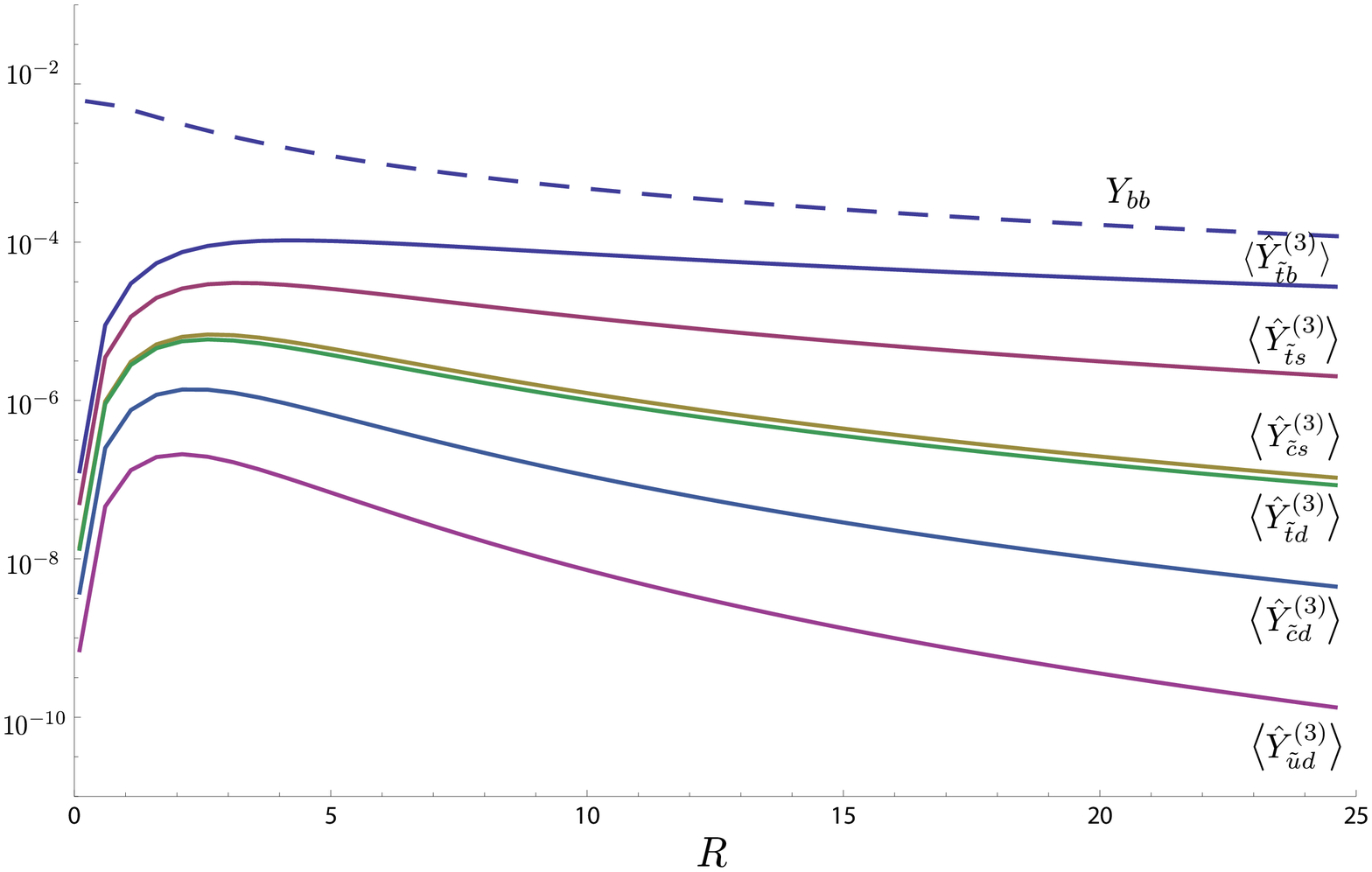}
 \caption{Couplings to the lightest massive vector-like pair of triplets in the second (left) and third (right) towers of fields localised in the Higgs curve, for flux values $v_1=5/6$, $v_2=5/4$, $M_1=1.6$, $M_2=2$, $\gamma=1.2$, $\varepsilon=1/10$, $k_{\rm KK}=1$ and trivial Landau-level quantum numbers.}
 \label{figtriplewithoutflux2}
 \end{center}}
\end{figure}

We plot in figure \ref{figtriplewithoutflux} the coupling of the different matter generations to the lightest massive vector-like pair of triplets in the first tower of localised fields, for values of the flux of order one.
The flux is chosen such that there is no total effective flux for the triplets, so that $\tilde{M}_{12}$ as defined in (\ref{mt12def}) vanishes. The massless chiral spectrum is determined by the expressions (\ref{localchiralflux5})-(\ref{localchiralfluxhiggs}) and is such that that there is a massless chiral Higgs doublet as well as the appropriate chiral matter fields. The massive vector-like pair of triplets that we consider is the lowest KK mode with vanishing Landau-level quantum numbers.
As discussed in section \ref{sec:wave}, the presence or absence of a vector-like pair of massless triplets is only determined by the global geometry. From the local approach we assume that such a massless pair is absent and therefore all triplets carry non-zero KK momentum. The coupling to the lightest vector-like pair of triplets in the other two towers of localised fields is plotted in figure \ref{figtriplewithoutflux2}.

The general behaviour of the couplings is similar to the setup discussed in the previous subsection. The more detailed differences being that the Yukawa coupling is slightly less sensitive to the values of $R$, allowing for a larger viable range. Also the suppression of the lighter generations is stronger than that in figure \ref{figtriplewithflux}. Both of these changes improve the prospects for suppressing proton decay operators.

\subsubsection{General overview}

Having studied both the possible cases for the Higgs triplets in the last two subsections we can attempt to draw some general conclusions from our analysis of the coupling of the heavy triplets to the different generations. Of course all conclusions are qualified with the fact that our analysis can only be taken to hold up to order 1 factors or so. Also the possible parameter space is restricted by our use of a local 8-dimensional field theory as discussed in section \ref{sec:validity} and it may be that leaving this framework could change the results in a quantitative and perhaps even qualitative way. With this in mind however it is also important to note that the behaviour we observe, up to order 1 factors, is quite robust against variations of the flux parameters and geometric scales within their respective allowed regimes that allow for the perturbative field theory analysis we have performed.

Perhaps the most general and important property of the couplings we have calculated is that there is a suppression of the coupling of lighter generations to the heavy triplets in analogy with the minimal 4-dimensional field theory GUTs behaviour, and which for large enough values of $R$ can lead to substantial suppressions of the couplings. Another general pattern we find is that the bottom quark Yukawa coupling is small, implying a small tan~$\beta$ regime is perhaps more natural.  Both of these properties are favourable in terms of suppressing dimension-five proton decay. The suppression for the lighter generations is particularly important, allowing to generate the small numbers needed to qualitatively match dimension-five proton stability constraints (assuming that a similar suppression for up-type couplings holds).

\begin{figure}
\centering
\epsfxsize=11cm
\hspace*{0in}\vspace*{.2in}
\epsffile{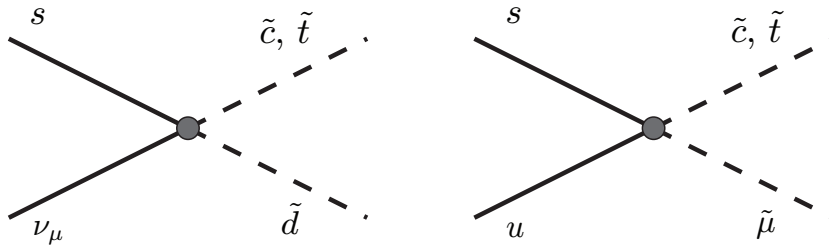}
\caption{Proton decay superpotential couplings involving only one lightest generation of quarks. The diagrams involving these operators are expected to dominate in the bulk of parameter space where the coupling to the heavier generations are only mildly suppressed.}
\label{fig:udsn}
\end{figure}

A quantitatively precise examination of these results would require the precise knowledge of the model dependent relation between the local scales $R_\perp$ and $R_\parallel$ and the global compactification scales parameterised by $M_{\rm Planck}$ and $\alpha_{\rm GUT}$. For nearly homogeneous compactifications, where small values of $R$ are most natural, such analysis reveals that within the bulk of the allowed local parameter space it is quite difficult to suppress the coupling to triplets sufficiently to weaken the proton stability constraints of minimal 4-dimensional GUTs that result from eqs.~(\ref{4dratios}).
The strongest constraints on the parameter space come from the relation to $\alpha_{\rm GUT}$ and from the requirement of a sufficiently large bottom Yukawa to be compatible with tan~$\beta>1$. Both of these are approximate and if we allow ourselves to go to the edge of parameter space, say with $R\sim25$, then further suppression is possible. This suppression of the couplings relative to the expected 4-dimensional values is most significant for the lighter generations which means that the strongest constraints would come from superpotential couplings involving as many heavier generations as possible, such as the presented in figure \ref{fig:udsn}.\footnote{This information is particularly non-trivial in the presence of additional selection rules which may, for example, forbid only the operators involving heavier generations but not lighter ones.}
%%%%%%%%%%%%%%%%%%%%%%%%%
%%%%%%%%%%%%%%%%%%%%%%%%%
We cannot rule out that allowing for order 1 factors and going to the edge of allowed parameter space, as well as assuming a similar suppression to the up-type Higgs triplet couplings, all the proton decay diagrams could be suppressed sufficiently to avoid experimental constraints in this class of models. However, while keeping this possibility in mind, within the framework we have used the most natural conclusion is that in F-theory models that are based on nearly homogeneous manifolds the coupling of the heavy triplets is of the same order as, or even more enhanced than, in minimal 4-dimensional field theory GUTs and further suppression of proton decay operators is required through additional mechanisms.

%%%%%%%%%%%%%%%%%%%%%%%%%%%%%%%%%%%%%%%%%%%%%%%%%%%%%%%%%%%%%%%%%%%%%%%%%%%%%
\subsection{Relation to Yukawa couplings}
\label{sec:phenoyuk}
%%%%%%%%%%%%%%%%%%%%%%%%%%%%%%%%%%%%%%%%%%%%%%%%%%%%%%%%%%%%%%%%%%%%%%%%%%%%%

We have computed the coupling of massless chiral modes to localised massive modes using wavefunction overlaps. So far our primary application of this calculation has been to proton decay through dimension-five operators. However the couplings we have calculated are important also for other physical quantities. In this section we present a brief discussion of their application to Yukawa couplings.

The nature of Yukawa couplings in F-theory GUTs depends on how the multiplicity of the generations arises. So far we have considered the case where the three matter generations arise from the multiplicity of Landau-level zero modes of a single curve. Yukawa couplings are obtained from zero-mode wavefunction overlaps and, as we have seen, for this setup they have a rank one structure due to the local U(1) symmetries \cite{Heckman:2008qa}. Yukawa couplings for the second and third generations should then arise from deformations of this structure. It was shown in \cite{Font:2009gq,Cecotti:2009zf,Conlon:2009qq} that such deformations cannot arise from worldvolume gauge fluxes, which are the only ingredients that we have included so far in our analysis. Instead, to generate a higher rank structure locally what is required is a non-commutative deformation of the local gauge theory. This can arise from two sources: either through Imaginary-Anti-Self-Dual (IASD) closed string fluxes of type $(2,1)$ \cite{Cecotti:2009zf}, or through non-perturbative effects such as instantons or gaugino condensation on some distant brane \cite{Marchesano:2009rz}. In \cite{Baumann:2009qx} it was argued that in fact the two possibilities are equivalent when it comes to their effect on the local theory since non-perturbative effects source IASD fluxes.

Calculations of Yukawa couplings within this framework were performed in \cite{Cecotti:2009zf,Aparicio:2011jx}. In particular it was shown in \cite{Aparicio:2011jx} that the deformed Yukawa couplings can be written as
\be
\label{noncomyuk}
Y_{\mathrm{total}} = Y^{(0)}_{\mathrm{tree}} + \epsilon \left( Y^{(1)}_{\mathrm{tree}} + Y^{(0)}_{\mathrm{n.p.}} \right) + ... \;.
\ee
Here the subscripts on the Yukawa coupling contributions denote the
type of operator in the gauge theory, with `tree' denoting the usual
Yukawa coupling arising from  $F \wedge \Phi$ as studied in this
paper, and `n.p.' denoting the higher dimension operator induced by
the non-perturbative effect (which takes the schematic form $\Phi F
\wedge F$). The parameter $\epsilon$ is related to the non-commutative
deformation or, equivalently, to the non-perturbative scale. The
superscripts on the operators denote to which order in the deformation
of the wavefunctions the operators are evaluated at. Thus, a
superscript 0 denotes the undeformed wavefunctions, while a
superscript 1 denotes that one of the three wavefunctions in the
triple overlaps is the first order deformed one. It was further shown
that the deformed wavefunction can be written as an explicit linear
combination of the massive wavefunctions of the undeformed
theory. These wavefunctions are exactly the wavefunctions we have been
calculating and the resulting coupling $Y^{(1)}_{\mathrm{tree}}$ is
directly given by the appropriate linear combination of the couplings
of two massless and one massive modes which we have
calculated. Therefore the calculations of massless and massive
wavefunction overlaps presented play an important role in calculating
also the Yukawa couplings. There are two direct applications of our
results. The first is to use within a Yukawa
coupling calculation analogous to \cite{Aparicio:2011jx} but for the more realistic SO(12) rather than U(3) setting (though in practice wavefunctions of both setups turn out to be rather similar). Secondly we see that there is a connection between Yukawa couplings and proton decay. In particular, within the region of parameter space where proton decay operators are suppressed also this contribution to the Yukawa coupling would be suppressed. Yukawa couplings would then primarily have to arise from the last term in (\ref{noncomyuk}).\footnote{Actually, as noticed in \cite{Aparicio:2011jx}, the coefficient of the last term in (\ref{noncomyuk}) vanishes for an SO(12) enhancement point and the non-trivial contribution to the Yukawa couplings is generated at higher order in the deformation. In that sense the suppression of the first term may be quite important.}

This model of flavour that we have been using is elegant and has
attractive features. Besides its simplicity, it provides a very
natural explanation for the observed hierarchy between the Yukawa
couplings of the third and the two lighter generations. It suffers,
however, from a number of difficulties in attempting a realistic
implementation. In particular,  no global compact stable backgrounds
are known, either in IIB or F-theory, which support the required
non-commutative deformation. Moreover, in order to obtain a realistic
flavour structure the Yukawa deformation should arise at order $\sim
0.2$, which seems too large to have a non-perturbative origin while
still maintaining a perturbative expansion. Our results showing that
within the local perturbative regime $R>1$ wavefunction overlaps are
generally suppressed are likely to make this problem even more
severe.\footnote{There is also a potential difficulty discussed in
  \cite{Hayashi:2009bt} which is that the initial rank one structure
  applies only locally to a single enhancement point of say
  SO(12). However, in concrete compactifications there are many such
  enhancement points, which are spatially separated, generically
  leading to a higher rank structure and destroying the hierarchical
  nature. The presence of a number of enhancement points however is
  not a no-go theorem as not all say SO(12) points must correspond to the down-type Yukawa but can amount to intersections of different curves.}

An alternative model of flavour was proposed in \cite{Dudas:2009hu} where each generation arises on a different matter curve and the flavour structure comes from embedding the SU(5) GUT into E$_8$ through the Froggatt-Nielsen mechanism. The calculations that we have performed in the previous sections are also relevant for these models, as Yukawa couplings in this case arise from higher dimension operators which, in turn, come from integrating out heavy modes. For example, starting from a bottom-type coupling to heavy modes and integrating out those, generates the $s$ quark Yukawa coupling. The relevant operators are
\be
W \supset \fb_{H_d}\fb^{\rm KK}_{b}\te_{t} + X \fb_{s} \f^{\rm KK}_{b} + M_{\rm KK}\fb^{\rm KK}_{b}\f^{\rm KK}_{b}\;,
\label{supfrog}
\ee
where we have denoted the heavy modes with the superscript `KK'. Integrating out the heavy modes gives
\be
\frac{X}{M_{\rm KK}} \fb_{H_d} \fb_{s} \te_{t} \;,
\ee
where $X$ is a GUT singlet that develops a vev. The first coupling of (\ref{supfrog}) is of the type we have been studying, with the small difference that the massive mode is taken for the matter curve rather than the Higgs curve. Since the flavour structure in the models of \cite{Dudas:2009hu} comes from the additional U(1) symmetries, it would require a large coupling to the massive modes and so in this case the, more natural, small $R$ region of parameter space is preferred.\footnote{Note that in the limit of constant fluxes, due to the local geometric U(1) selection rules, the leading such coupling would involve one of the massive towers which does not have a massless mode.} The second term in eq.~(\ref{supfrog}) is the type of coupling that we study in the next section. Finally note that in the context of proton decay the singlet vevs already account for the 4-dimensional coupling suppression as in (\ref{4dratios}), and so only a further mild suppression, coming for the difference between the doublet and triplet coupling as we have been studying, is required to avoid proton decay constraints. Given our results, which showed that couplings to heavy modes can be easily and generically much smaller than 1, such a possibility seems quite natural. However an explicit computation of the dependence on the generation structure would require going beyond the local SO(12) framework we have been using.

%%%%%%%%%%%%%%%%%%%%%%%%%%%%%%%%%%%%%%%%%%%%%%%%%%%%%%%%%%%%%%%%%%%%%%%%%%%%%
\subsection{Soft masses and FCNC}
\label{sec:phenoFCNC}
%%%%%%%%%%%%%%%%%%%%%%%%%%%%%%%%%%%%%%%%%%%%%%%%%%%%%%%%%%%%%%%%%%%%%%%%%%%%%

As another application of the type of computations we have performed we study a set of operators which may be responsible for soft masses after supersymmetry breaking. This is a particularly interesting set of operators to investigate because a well-known criticism of gravity mediated supersymmetry breaking is that FCNCs are expected to be generated after integrating out heavy modes. The coupling of the massless matter sector to heavy modes is therefore crucial in calculating this effect.\footnote{For non-perturbatively generated FCNC in string theory see \cite{Blumenhagen:2010dt}.}

More precisely, let $X = \ldots +\  \theta^2 F_X$ be a supersymmetry breaking superfield and $Q_i$ a chiral matter
superfield (baryonic or leptonic) belonging to the $i$-th generation. If the K\"ahler potential contains dimension-six operators of the type
\be
K \supset K_{ij X {\bar X}} \int d^4 \theta \ X^{\dagger} X  Q_j^{\dagger} Q_i \ , \label{fcnc1}
\ee
then soft masses are generated
\be
{\tilde m}_{ij}^2 \ = \ K_{ji X {\bar X}} \ |F_X|^2 \ . \label{fcnc2}
\ee
Denoting by $V_{q,ij} \simeq \delta_{ij} + \epsilon_{ij}$ the unitary
matrix which diagonalises the fermion mass matrix (with $\epsilon_{ij} << 1$ and $\epsilon_{ij}=0$ for $i=j$), the scalar soft masses in the basis where fermion masses and gaugino-fermion-scalar matrices are diagonal are
\be
m_{ij}^2 \simeq {\tilde m}_{ij}^2 + \epsilon_{ik} {\tilde m}_{kj}^2 - {\tilde m}_{il}^2  \epsilon_{lj} \ .  \label{fcnc3}
\ee
Experimental results on flavour violation in the Standard Model set severe bounds on the off-diagonal terms in the family space
$m_{ij}^2$ \cite{fcnc}.

\begin{figure}[ht!]
\begin{center}
{
\includegraphics[width=13cm]{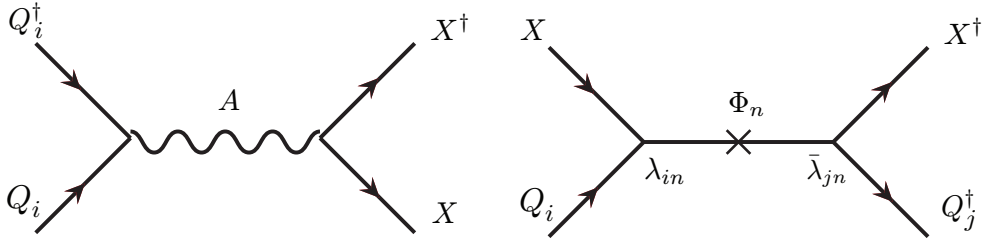}
 \caption{Feynman diagrams leading to dimension-six operators of the type (\ref{fcnc1}) by exchange of massive U(1) gauge bosons (left) or massive scalars (right).}
 \label{dim6}}
 \end{center}
\end{figure}

There are various possible sources for the dimension-six operators (\ref{fcnc1}) within the F-theory GUTs context. In figure \ref{dim6} we present the leading such contributions.
The first type of contribution, introduced in \cite{Heckman:2008qt}, is obtained by  integrating out a massive U(1) gauge boson.  This contribution has been argued to be flavour universal. However, besides this contribution there can also be non-universal contributions arising from the exchange of heavy Landau or KK states $\Phi_n$, which we study in what follows.

The relevant cubic coupling constants $\lambda_{j n}$ involved in the diagram of figure \ref{dim6} (right) are given by the overlap integral
\be
\lambda_{j n} \ = \ \int_S \Psi_X \Psi_j \Psi_n \ = \ \langle n | X
Q_j \rangle \ . \label{fcnc4}
\ee
In this expression we have used a convenient Dirac notation, where actually  $\langle x | n  \rangle = {\bar \Psi}_n $ denotes the complex conjugate of the wavefunction $\Psi_n$. Hence, the coefficient of the dimension-six operator eq.~(\ref{fcnc1}) which results from these contributing diagrams is
\be
K_{ji X {\bar X}} \ = \ \sum_n \frac{{\bar \lambda}_{i n} \lambda_{j n}}{M_n^2}  \ = \ \sum_n \frac{\langle X Q_i | n \rangle \langle n | X Q_j \rangle}{M_n^2}
 \ , \label{fcnc5}
\ee
with $M_n$ the mass of the state $\Phi_n$. Such a sum over mediating states has a lower bound obtained by keeping the lightest state $\Phi_1$ with mass $M_1$,
\be
 K_{ji X {\bar X}} \geq \frac{{\bar \lambda}_{i 1} \lambda_{j 1}}{M_1^2}  \ . \label{fcnc6}
\ee
Similarly, an upper bound can be obtained by using the inequality
\be
 K_{ji X {\bar X}} = \sum_n \frac{\langle X Q_i | n \rangle \langle n | X Q_j \rangle}{M_n^2}
 \ \leq \frac{\langle X Q_i | X Q_j \rangle}{M_1^2}
= \frac{1}{M_1^2} \int_S |\Psi_X|^2 {\bar \Psi}_i \Psi_j  \ , \label{fcnc7}
\ee
which results from making use of $M_n \geq M_1$ and the completeness of Landau levels.
It is important to recall here that eqs.~(\ref{fcnc6}) and (\ref{fcnc7}) put bounds on the particular contribution of the Landau states $\Phi_n$ to the operator (\ref{fcnc1}), and not on the operator itself, which in general may receive other contributions (for instance from string oscillators) not accounted for by these expressions.

In order to perform a more precise estimation of the quantities appearing in eqs.~(\ref{fcnc6}) and (\ref{fcnc7}) we have also to specify the origin of the SUSY breaking field $X$. If it is a modulus-like field, singlet under the
Abelian symmetries we are considering, local considerations do not determine its wavefunction and we cannot say much by using the methods of this paper. However, if $X$ is a charged field then it actually experiences the Higgs vev and/or the Abelian fluxes and it has some degree of localisation. This is the case
we consider in what follows.

Since the cubic coupling we are interested in involves two fields charged under the GUT group and one GUT singlet, the SO(12) model that we have been considering in previous sections is not accurate for this case. A more suitable toy model based on the gauge group U(3) and containing all the necessary ingredients was studied in \cite{Aparicio:2011jx}, that we consider in what follows.

In this model massive $\Phi_n$ states\footnote{We identify our fields
$X, Q_i$ and $\Phi_n$ with the fields in \cite{Aparicio:2011jx} according to
\begin{equation*}
X \ \rightarrow c^+ \quad , \quad Q_i \ \rightarrow b^{i,+}  \quad , \quad \Phi_n \ \rightarrow a_{m,n,l}^+ \ ,
\end{equation*}
and $M_x= - M_y = M < 0$.} are actually labeled by three quantum numbers $|\Phi_n\rangle\to |\Phi_{(m,n,l)}\rangle$. Their masses are given by
\be
M_{(m,n,l)}^2 \ = \ \frac{M_*^2}{R_\parallel^2}(m \lambda^+ + n |M| + l |\lambda^-|) \ , \label{fcnc8}
\ee
where $\lambda^\pm\equiv\frac12\left(M\pm\sqrt{M^2+4R^2m_\phi^4}\right)$, $M$ and $m_\phi$ are respectively flux and Higgs parameters and we have made explicit the dependence on the local scales $R_\parallel$ and $R_\perp$ introduced in section \ref{sec:effthe}.
We consider the minimal case in which the multiplicity of $X$ and
$\Phi_n$ is one, whereas the multiplicity of
$Q_i$ is taken to be three, $i=0,1,2$ with $i=0$ labeling the heaviest generation.

We denote by $\lambda^{i}_{(m,n,l)}$ the cubic coupling between $X$, $\Phi_{(m,n,l)}$ and $Q_i$. As it was the case in the SO(12) setup, the local geometric U(1) charge conservation (\ref{u1}) only allows for couplings which satisfy $i-m-n+l=0$. Thus, the lightest fields coupling to $X$ and to different generations of squarks and sleptons are
\begin{align*}
\textrm{1st generation:}\quad &\Phi_{(2,0,0)} \ , \ \Phi_{(0,2,0)} \ , \ \Phi_{(1,1,0)} \ , \ldots\\
\textrm{2nd generation:}\quad &\Phi_{(1,0,0)} \ , \ \Phi_{(0,1,0)} \ , \ldots\\
\textrm{3rd generation:}\quad &\Phi_{(0,0,0)} \ , \ldots
\end{align*}
In the toy model that we are considering $\Phi_{(0,0,0)}$ is massless. However, we
expect that in a more realistic setup there would not be such massless exotic state.
This mass lifting is not under control locally and therefore we have nothing to say about it here.

For simplicity we choose the fluxes of the model to be small enough $2 M^2 < R^2 m_\phi^4$, such that the
states with $m=l=0$ are lighter than the other states.
In that case cubic couplings are of the form
\begin{multline}
\lambda^{i}_{(0,n,0)} = \frac16 \int_S \left[ (\psi^{Q_i}_{\bar{1}} \psi^{X}_{\bar{2}} - \psi^X_{\bar{1}} \psi^{Q_i}_{\bar{2}}) \psi^{\Phi_n}_{\bar 3}
- (\psi^{Q_i}_{\bar 1} \psi^{\Phi_n}_{\bar{2}} - \psi^{\Phi_n}_{\bar{1}} \psi^{Q_i}_{\bar{2}}) \psi^{X}_{\bar 3}  \right. \\
 \quad\left. + (\psi^X_{\bar{1}} \psi^{\Phi_n}_{\bar{2}} -
   \psi^{\Phi_n}_{\bar 1} \psi^{X}_{\bar{2}})
\psi^{Q_i}_{\bar{3}} \right] \;
\label{fcnc09}
\end{multline}
and, in particular, the leading couplings to massive states for the first two generations are $\lambda^{2}_{(0,2,0)}$ and $\lambda^{1}_{(0,1,0)}$. Making use of the explicit expression of the wavefunctions in \cite{Aparicio:2011jx} we find
\bea
&& \lambda^{1}_{(0,1,0)} = - \frac{|M|^{1/2} \lambda^- (\lambda^- -
\sqrt{2}R m_\phi^2)}{N_{\Phi} N^1_{Q} N_XR^2m_{\phi}^4} \int_S \ z_1 {\bar z_2} \ e^{\lambda^- |z_1|^2 + \lambda^- |z_2|^2 - \frac{m_{\phi}^2}{\sqrt{2}} |z_1-z_2|^2 } \ , \label{fcnc9} \\
&& \lambda^{2}_{(0,2,0)} =  \frac{ |M|  \lambda^- (\lambda^- -
\sqrt{2} R m_\phi^2)}{\sqrt{2}N_{\Phi} N^2_{Q} N_XR^2m_{\phi}^4}
\int_S \ z_1^2 {\bar z_2}^2 \ e^{\lambda^- |z_1|^2 + \lambda^- |z_2|^2 - \frac{m_{\phi}^2}{\sqrt{2}} |z_1-z_2|^2 } \ , \nn
\eea
where $N_\Phi$, $N^j_Q$ and $N_X$ are wavefunction normalisation constants.
Using the integral (\ref{Ianswer}), together with eqs.~(\ref{fcnc6}) and (\ref{fcnc8}),
the corresponding terms in the K\"ahler potential are then readily found to be
\bea
&& K_{1 \bar 1 X {\bar X}} \simeq \frac{|\lambda^{1}_{(0,1,0)}|^2}{M_{(0,1,0)}^2}  =  \frac{ \pi^4 R_\parallel^{10}}{2M_*^2 (N_\Phi N^1_{Q} N_X)^2R^2m_{\phi}^4 \lambda^2 (\lambda - \sqrt{2} R m_\phi^2)^2} \ ,  \label{fcnc11}\\
&& K_{2 \bar 2 X {\bar X}} \simeq \frac{|\lambda^{2}_{(0,2,0)}|^2}{M_{(0,2,0)}^2}  =  \frac{\pi^4 R_\parallel^{10}|M|}{4M_*^2 (\lambda^-)^4 (N_\Phi N^2_{Q} N_X)^2(\lambda^- -
\sqrt{2} R m_\phi^2)^4} \ .  \nn
\eea
Hence, due to conservation of the geometric U(1) charge no off-diagonal terms are generated in the K\"ahler potential.
We know however from eq.~(\ref{fcnc3}) that this is not enough to suppress FCNCs, since unequal terms on the diagonal do generate FCNCs after diagonalising
the fermion mass matrices.

Indeed, from eqs.~(\ref{fcnc11}) we obtain
\begin{equation}
{\tilde m}_{22}^2 - {\tilde m}_{11}^2 \sim \frac{R^2 m_{\phi}^4 M^3R_\parallel^2}{8 N_X^2M_*^2 (\lambda^-)^6 (\lambda^- - \sqrt{2} R m_\phi^2)^4}
\left[- M^2 R^2m_\phi^4 + 4 (\lambda^-)^2 (\lambda^- - \sqrt{2}R m_\phi^2)^2  \right] |F_X|^2 \ ,\label{fcnc12}
\end{equation}
where we have used the normalisation of the wavefunctions
\be
N_{\Phi} \simeq N^j_{Q} \simeq R_\parallel^2\left(\frac{j!}{|M|^{j}}
 \frac{\pi^2  \lambda^-}{R^2m_{\phi}^4 M}\right)^{1/2} \ .
\label{norm}
\ee
For a suppression of the FCNCs the two mass terms ${\tilde m}_{11}^2$ and ${\tilde m}_{22}^2$ would have to be equal to a high accuracy.

A similar reasoning can be followed for the upper bound (\ref{fcnc7}) on the K\"ahler term. Starting from the cubic couplings (\ref{fcnc09})
and integrating out over the massive states, one has
\begin{multline}
 K_{ijX {\bar X}} \ \leq   K_{ijX {\bar X}}^{\rm max} \ = \ \frac{1}{M_{(0,i,0)}M_{(0,j,0)}} \int_S \left[ (\psi^{Q_i}_{\bar{1}} \psi^{X}_{\bar{2}} - \psi^X_{\bar{1}} \psi^{Q_i}_{\bar{2}})
 (\overline{\psi^{Q_j}_{\bar{1}} \psi^{X}_{\bar{2}} - \psi^X_{\bar{1}} \psi^{Q_j}_{\bar{2}}}) \right. \\
\left. +  (\psi^{Q_i}_{\bar{2}} \psi^{X}_{\bar{3}} - \psi^X_{\bar{2}} \psi^{Q_i}_{\bar{3}})
 (\overline{\psi^{Q_j}_{\bar{2}} \psi^{X}_{\bar{3}} - \psi^X_{\bar{2}} \psi^{Q_j}_{\bar{3}}}) +
 (\psi^{Q_i}_{\bar{3}} \psi^{X}_{\bar{1}} - \psi^X_{\bar{3}} \psi^{Q_i}_{\bar{1}})
 (\overline{\psi^{Q_j}_{\bar{3}} \psi^{X}_{\bar{1}} - \psi^X_{\bar{3}} \psi^{Q_j}_{\bar{1}}})    \right] \;,
\label{fcnc13}
\end{multline}
which, making use the explicit form of the wavefunctions, becomes
\begin{multline}
K_{ijX {\bar X}}^{\rm max}  =  \frac{1}{M_{(0,i,0)}M_{(0,j,0)} N_X^2 N_{Q}^iN_Q^j}
\left[
\frac{3 (\lambda^-)^2}{2 R^2m_{\phi}^4} - \frac{\sqrt{2} \lambda^-}{R m_{\phi}^2} + 1 \right]\\
\times \int_S z_1^i \bar z_1^j \ e^{M |z_1|^2 - \sqrt{M^2+ 4 R^2m_{\phi}^4} |z_2|^2 - \sqrt{2} R m_{\phi}^2 |z_1-z_2|^2}  \label{fcnc14}
\end{multline}
This integral can be computed as before. We do not write here its explicit analytic expression. Notice however that, similarly to the lower bound that we have just discussed, the resulting K\"ahler terms are again diagonal in the generation space due
to geometric U(1) charge conservation, but, importantly, not degenerate.

Since this is a toy model we do not perform an analysis of eqs.(\ref{fcnc11}) and (\ref{fcnc14}) on their parameter space. Our main emphasis is to show how contributions of the form (\ref{fcnc1}) can be calculated and that indeed for generic parameters they are not flavour-universal and lead to large FCNCs. In particular,  there is a priori no natural reason why the flavour-universal contributions from U(1) gauge boson exchange should dominate the contributions studied here, since the masses of the corresponding gauge bosons and of the scalar fields are expected to be at the same scale. However, having shown how such contributions can be explicitly calculated, it would be very interesting to explore the full parameter space, particularly within a more realistic model, and see if there are regions where the FNCN operators can be sufficiently suppressed.

Finally, let us raise a further point regarding the U(1) boson exchange contribution itself. The flavour universality in this case arises because the gauge boson is taken to couple in the K\"ahler potential in such way that diagonalising the kinetic terms for the matter fields also diagonalises the coupling to the gauge boson. However, in general this is only possible if the wavefunction profile of the gauge boson is constant. We therefore expect flavour non-universal couplings to higher KK states of the U(1) gauge boson, as the wavefunctions of these states are not constant. Once more studying the nature of these FCNCs reduces to studying massive mode wavefunction integrals, but this time for the U(1) gauge bosons. Such studies are however beyond the scope of this work.

%%%%%%%%%%%%%%%%%%%%%%%%%%%%%%%%%%%%%%%%%%%%%%%%%%%%%%%%%%%%%%%%%%%%%%%%%%%%%
\section{Summary}
\label{sec:summary}
%%%%%%%%%%%%%%%%%%%%%%%%%%%%%%%%%%%%%%%%%%%%%%%%%%%%%%%%%%%%%%%%%%%%%%%%%%%%%

In this work we have studied the coupling of matter fields to heavy modes through cubic superpotential interactions in the context of SU(5) F-theory GUTs. We have considered, in particular, fields that are localised within a patch around a point of SO(12) enhancement and computed the local form of their wavefunctions, for both massless and massive modes. Down-type Yukawa couplings are obtained by integrating overlaps of three massless wavefunctions, whereas couplings of two matter fields to the heavy coloured triplets are given by integrating overlaps of one massive and two massless wavefunctions.

The coupling of MSSM fields to heavy triplets is an important ingredient in the study of proton stability through dimension-five effective operators that result from integrating out the heavy triplets. In this context our calculations are a prerequisite to constraining F-theory GUTs through dimension-five proton stability. We find that, analogously to minimal 4-dimensional GUTs, the coupling of the heaviest generation to the massive triplets is of the same order as the Yukawa coupling, and that the coupling of the lighter generations to the triplets are suppressed with respect to the heavier ones. The quantitative analysis depends on few local parameters associated to the flux and geometry, which are in turn constrained by the measured values of $\alpha_{\rm GUT}$ and $M_{\rm Planck}$ and also by keeping the bottom quark Yukawa not too small so that we can remain within the tan~$\beta>1$ regime favoured by unification. The detailed relation between the local scales and the global ones is however very complicate and model dependent, and given a particular compactification it is in general out of the scope of the present techniques to obtain such relation in a precise way. Once the relation between local and global scales is understood in a given model, the constraints on the local parameter space coming from the observed values of $M_{\rm Planck}$ and $\alpha_{\rm GUT}$ would be made precise within that model.

For the particular case of nearly homogeneous compactifications we find that the relation between local and global scales is such that for the most natural values of the local parameters the couplings to the heavy triplets is typically the same or, in the case of the heaviest generations, stronger than in minimal 4-dimensional GUTs. Thus, constraints on the coefficient of dimension-five proton decay operators involving heavy generations in this class of models are stronger than in minimal 4-dimensional GUTs. We can also identify regions on the edge of local parameter space where the couplings to the triplets are more suppressed than in 4-dimensional GUTs for the lighter generations, thus weakening constraints from proton decay on those operators. However, within these region tan~$\beta$ is expected to be very small.

The computation that we have performed is only a part of the complete computation for dimension-five proton stability since cubic couplings to up-type triplets localised at points of E$_6$ enhancement also participate in the above operators. We expect a similar computation of those couplings to be technically more involved than the one that we have carried out here, mainly because of the presence of local monodromies. It would certainly be very interesting to perform such a computation. In particular, note that the up-type ratios of couplings analogous to eqs.~(\ref{4dratios}) are actually significantly smaller than the down-type ones, and therefore even stronger suppression in the F-theory couplings near an E$_6$ point would be required to match or to overcome the suppression in minimal 4-dimensional field theory GUTs.

The coupling of massive modes to massless ones has a number of phenomenologically interesting applications, some of which we have already discussed. Within flavour physics the computation features in that of Yukawa couplings as studied in \cite{Aparicio:2011jx}. It is also a key computation in realising the Froggatt-Nielsen model of flavour proposed in \cite{Dudas:2009hu}, where the higher dimension operators come from coupling of two massless modes to one heavy mode. Interestingly, since in the above model of flavour the suppression of the coupling to the heavy Higgs triplets is due to the U(1) symmetries at least as strong as in minimal 4-dimensional GUTs, only a slight suppression of the coupling to the triplets relative to the doublets is required to evade constraints from proton decay. Given that the coupling to heavy triplets can be naturally suppressed it would be interesting to study if the required small suppression could be realised within this framework.

Another application  of this type of couplings that we have discussed is in understanding FCNCs induced by supersymmetry breaking. Within a toy model we were able to give expressions for the magnitude of FCNCs induced by integrating out heavy modes. We find that FCNCs are induced by tree-level diagrams and are suppressed by the mass scale of the heavy modes. In particular we find no natural suppression relative to the flavour-universal contribution coming from heavy U(1) gauge bosons studied in \cite{Heckman:2008qt}.

There are further phenomenological applications that one might think of, such as the generation of neutrino masses through the seesaw mechanism. Since the right-handed neutrinos are GUT singlets the neutrino Dirac masses would be of the same form as the coupling studied in section \ref{sec:phenoFCNC}.\footnote{For the type I seesaw mechanism the Majorana nature of right-handed neutrinos would however require different techniques to the ones that we have been using in this work.} More generally, we believe that the local approach opens up a calculational framework for studying the interactions with massive modes within a phenomenologically realistic and very general setting. This is important because it is precisely in understanding the interaction between infrared and ultraviolet physics that string phenomenology as a subject has its strongest claim as a necessary tool in physics.

\vspace{0.3cm}

%%%%%%%%%%%%%%%%%%%%%%%%%%%%%%%%%%%%%%%%%%%%%%%%%%%%%%%%%%%%%%%%%%%%%%%%%%%%%%%%%%%%%%%%%%%%%%%%%%%%%%%%%%%%%%%%%%%%%%%%%
\begin{center}\subsection*{Acknowledgments}\end{center}
%%%%%%%%%%%%%%%%%%%%%%%%%%%%%%%%%%%%%%%%%%%%%%%%%%%%%%%%%%%%%%%%%%%%%%%%%%%%%%%%%%%%%%%%%%%%%%%%%%%%%%%%%%%%%%%%%%%%%%%%%

We thank J.~Conlon, A.~Font, L.~Ib\'a\~nez and F.~Marchesano for
useful discussions. The work presented was
supported in part by the European ERC Advanced Grant 226371 MassTeV,
by the CNRS PICS no. 3059 and 4172, by the
grants ANR-05-BLAN-0079-02, ANR TAPDMS ANR-09-JCJC-0146, the PITN
contract PITN-GA-2009-237920 and the project AGAUR
2009-SGR-168. The research of EP is supported by a Marie Curie  Intra European
Fellowship within the 7th European Community
Framework Programme.

\newpage

\appendix

\section{Wavefunctions for oblique hypercharge flux}
\label{app:oblique}

In this appendix we consider a slightly more general set of U(1) fluxes, which in addition to eq.~(\ref{fluxansatz}) includes also an oblique component for the hypercharge flux, given by
\begin{equation}
F^{Y}_{\rm obliq.}=\frac{\alpha M_{*}^2}{R_\parallel^2}(dz_1\wedge d\bar z_2+d\bar z_1\wedge dz_2)Q_Y \;.
\end{equation}
We choose to integrate this flux as,
\begin{equation}
A^Y_{\rm obliq.}=\frac{\alpha M_{*}}{2R_\parallel}(z_1\wedge d\bar z_2+\bar z_1\wedge dz_2-\bar z_2\wedge dz_1- z_2\wedge d\bar z_1)Q_Y \;.
\end{equation}
Wavefunctions can be solved by following the general procedure
described in
subsection \ref{sublocal}. We find that oblique fluxes do not affect the physics of the wavefunction in a qualitatively important way.

\subsection{Wavefunctions for the $\fb_M$ matter curve}

In this case the gauge covariant derivatives appearing in eq.~(\ref{massiveeq}) get modified to
\begin{align}
D_1&=\frac{M_{*}}{R_\parallel}\left(\partial_1 + \frac12\textrm{Re}(\a_1)\bar z_1 - \frac{i}{2} \textrm{Im}(\a_1)\bar z_2\right)\ , \\
D_2&=\frac{M_{*}}{R_\parallel}\left(\partial_2 - \frac12\textrm{Re}(\a_1)\bar z_2 + \frac{i}{2} \textrm{Im}(\a_1)\bar z_1 \right)\ , \nn\\
D_3&=-\frac{M_*R_\perp}{v_1} \bar z_1 \nn
\end{align}
where we have defined $\alpha_1 \equiv 2\tilde{M}_1 - iq^Y_1 \alpha$. The corresponding matrix $\mathbb{B}$ reads
\begin{equation}
\mathbb{B}=\frac{M_*^2}{R_\parallel^2}\begin{pmatrix}0&0&0&0\\
0& -\textrm{Re}(\a_1) & -i\textrm{Im}(\a_1) & \frac{R}{v_1} \\
0& i\textrm{Im}(\a_1) & \textrm{Re}(\a_1) & 0 \\
0& \frac{R}{v_1} & 0 & 0 \end{pmatrix}
\end{equation}
with dimensionless eigenvalues given by the roots of the following depressed cubic equation,
\begin{equation}
\left(\lambda_p^{\fb_M}\right)^3-\lambda_p^{\fb_M}\left(\frac{R^2}{v_1^2}+|\alpha_1|^2\right)+\frac{R^2}{v_1^2}\textrm{Re}(\a_1)=0\ , \qquad p=1,2,3
\end{equation}
and by $\lambda_0^{\fb_M}=0$. The eigenvectors are,
\begin{equation}
\xi_0^{\fb_M}=\begin{pmatrix}1\\ 0\\ 0 \\ 0\end{pmatrix}\ ,\qquad \xi_p^{\fb_M}=\begin{pmatrix}0\\ k^{\fb_M}_{(p,1)} \\  k^{\fb_M}_{(p,2)}  \\ \frac{2iR}{v_1} \textrm{Im}(\a_1) \end{pmatrix}\ ,\quad p=1,2,3
\end{equation}
where
\bea
k^{\fb_M}_{(p,1)} &=& 2i\lambda^{\fb_M}_p \textrm{Im}(\a_1) \;, \nn \\
k^{\fb_M}_{(p,2)} &=& -2\left(\lambda^{\fb_M}_p\right)^2 - 2\lambda^{\fb_M}_p \textrm{Re}(\a_1) + \frac{2R^2}{v_1^2}  \;.
\eea
In terms of these quantities the solution to eqs.~(\ref{zero00}) for $\tilde{M}_1 > 0$ leads to the wavefunctions for the ground state of each of the four towers of fields localised in the $\fb_M$ matter curve in the presence of oblique hypercharge flux. The expressions are given by eq.~(\ref{massless1}) with,
\begin{align}
p_1^{\fb_M} &= -\lambda_1^{\fb_M} - \frac12\textrm{Re}(\alpha_1) \;, &
p_2^{\fb_M} &= \frac12\textrm{Re}(\alpha_1) \;, \\
p_3^{\fb_M} &= \frac{i}{2}\textrm{Im}(\alpha_1) \;, &
p_4^{\fb_M} &= \frac{i}{2} \frac{\lambda_1^{\fb_M} + \textrm{Re}(\alpha_1)}{\lambda_1^{\fb_M} - \textrm{Re}(\alpha_1)} \textrm{Im}(\alpha_1)\;\nn
\end{align}
and $k^{\fb_M}_{i}=k^{\fb_M}_{(1,i)}$, $i=1,2$.

\subsection{Wavefunctions for the $\te_M$ matter curve}

Wavefunctions for the $\te_M$ matter curve can be obtained from the $\fb_M$ wavefunctions by applying the transformations (\ref{trans}). We obtain that the ground state wavefunctions for $\tilde{M}_2>0$ are given by the general expression  eq.~(\ref{massless1}) with
\begin{align}
p_1^{\te_M} &= \frac12\textrm{Re}(\alpha_2) \;, &
p_2^{\te_M} &= -\lambda_1^{\te_M} - \frac12\textrm{Re}(\alpha_2) \;, \\
p_3^{\te_M} &= \frac{i}{2} \frac{\lambda_1^{\te_M} + \textrm{Re}(\alpha_2)}{\lambda_1^{\te_M} - \textrm{Re}(\alpha_2)} \textrm{Im}(\alpha_2)\;, &
p_4^{\te_M} &= \frac{i}{2}\textrm{Im}(\alpha_2) \;, &
\end{align}
where we have defined $\alpha_2\equiv 2\tilde{M}_2+iq_Y\alpha$. Eigenvalues are given by the roots of the cubic equation,
\begin{equation}
\left(\lambda_p^{\te_M}\right)^3-\lambda_p^{\te_M}\left(\frac{R^2}{v_2^2}+|\alpha_2|^2\right)+\frac{R^2}{v_2^2}\textrm{Re}(\a_2)=0\ , \qquad p=1,2,3
\end{equation}
and by $\lambda_0^{\te_M}=0$. The corresponding eigenvectors are
\begin{equation}
\xi_0^{\te_M}=\begin{pmatrix}1\\ 0\\ 0 \\ 0\end{pmatrix}\ ,\qquad \xi_p^{\fb_M}=\begin{pmatrix}0\\   k^{\te_M}_{(p,2)}  \\ k^{\te_M}_{(p,1)} \\ \frac{2iR}{v_2} \textrm{Im}(\a_2) \end{pmatrix}\ ,\quad p=1,2,3
\end{equation}
where we have defined
\bea
k^{\te_M}_{(p,1)} &=& 2i\lambda^{\te_M}_p \textrm{Im}(\a_2) \;, \nn \\
k^{\te_M}_{(p,2)} &=& -2\left(\lambda^{\te_M}_p\right)^2 - 2\lambda^{\te_M}_p \textrm{Re}(\a_2) + \frac{2R^2}{v_2^2}  \;
\eea
and $k^{\te_M}_{1}=k^{\fb_M}_{(1,2)}$, $k^{\te_M}_{2}=k^{\fb_M}_{(1,1)}$.

\subsection{Wavefunctions for the $\fb_H$ Higgs curve}

In the presence of oblique hypercharge flux the gauge covariant derivatives in the $(u,v)$-basis are deformed to,
\begin{equation}
D_u = \frac{M_{*}}{R_\parallel}\left(\partial_u - \frac12 \alpha_h \bar{w} \right) \;, \quad
D_w = \frac{M_{*}}{R_\parallel}\left(\partial_w - \frac12 \bar\alpha_h \bar{u} \right) \;, \quad
D_3 = M_{*}R_\perp\left( \frac{\bar{w}}{v_+} + \frac{\bar{u}}{v_-}\right) \;,
\end{equation}
where $\alpha_h = 2\tilde{M}_{12} - iq_Y \alpha$. The matrix $\mathbb{B}$ now reads
\begin{equation}
\mathbb{B}=\frac{M_*^2}{R_\parallel^2}\begin{pmatrix}0&0&0&0\\
0&0&\alpha_h^*& -\frac{R}{v_-} \\
0&\alpha_h&0&-\frac{R}{v_+}\\
0&-\frac{R}{v_-}&-\frac{R}{v_+}&0\end{pmatrix}\label{bpr}
\end{equation}
The eigenvalues are given by the roots of
\begin{equation}
\left(\lambda^{\fb_H}_p\right)^3 - \lambda^{\fb_H}_pR^2\left(\frac{1}{v_+^2} + \frac{1}{v_-^2} + |\alpha_h|^2 \right) - \frac{2R^2}{v_+v_-} \textrm{Re}(\alpha_h) =0 \;, \qquad p=1,2,3
\end{equation}
and by $\lambda^{\fb_H}_0=0$. The corresponding eigenvectors are,
\begin{equation}
\xi^{\fb_H}_0=\begin{pmatrix}1\\ 0\\ 0\\ 0\end{pmatrix}\ , \qquad \xi^{\fb_H}_p=\begin{pmatrix} 0\\ k^{\fb_H}_{(p,u)} \\ k^{\fb_H}_{(p,w)} \\ -|\alpha_h|^2+\left(\lambda_p^{\fb_H}\right)^2  \end{pmatrix} \;,\quad p=1,2,3
\end{equation}
where now,
\begin{equation}
k^{\fb_H}_{(p,u)} = -R\left(\frac{\alpha_h^*}{v_+}+\frac{\lambda_p^{\fb_H}}{v_-}\right) \;, \qquad
k^{\fb_H}_{(p,w)} =  -R\left(\frac{\alpha_h}{v_-}+\frac{\lambda_p^{\fb_H}}{v_+}\right)\;,
\end{equation}
With this information we can solve eqs.~(\ref{zero00}) and look for the wavefunctions of the ground state fields localised in the $\fb_H$ curve. For simplicity here we only solve for the case $\tilde{M}_{12}>0$. However, note from eq.~(\ref{bpr}) that
\begin{equation}
\lambda_1^{\fb_H}\lambda_2^{\fb_H}\lambda_3^{\fb_H}=\frac{4\tilde{M}_{12}R^2}{v_+v_-}
\end{equation}
is independent of the oblique components of the hypercharge flux, and therefore the latter do not affect the chirality of the zero mode and therefore to double-triplet splitting. We find that in the presence of oblique hypercharge flux the ground state wavefunctions for each of the four towers of fields in the $\fb_H$ are given by eq.~(\ref{higgsgen}) with $k^{\fb_H}_{u}=k^{\fb_H}_{(1,u)}$, $k^{\fb_H}_{w}=k^{\fb_H}_{(1,w)}$ and
\begin{align}
p^{\fb_H}_1 &= -\frac{(\a_h v_+v_- +  \lambda^{\fb_H}_1v_-^2)\lambda^{\fb_H}_1}{2\textrm{Re}(\a_h)  + (v_+^2 + v_-^2)\lambda^{\fb_H}_1} \;, &
p^{\fb_H}_2 &= -\frac{(\bar\a_h v_+v_- +  \lambda^{\fb_H}_1v_+^2)\lambda^{\fb_H}_1}{2\textrm{Re}(\a_h)  + (v_+^2 + v_-^2)\lambda^{\fb_H}_1} \;,  \\
p^{\fb_H}_3 &= \frac{(\a_h v_+^2 +  \lambda^{\fb_H}_1v_+v_-)\lambda^{\fb_H}_1}{2\textrm{Re}(\a_h)  + (v_+^2 + v_-^2)\lambda^{\fb_H}_1} - \frac{\a_h}{2}\;, &
p^{\fb_H}_4 &= \frac{(\bar \a_h v_-^2 +  \lambda^{\fb_H}_1v_+v_-)\lambda^{\fb_H}_1}{2\textrm{Re}(\a_h)  + (v_+^2 + v_-^2)\lambda^{\fb_H}_1} - \frac{\bar \a_h}{2} \;.\nn
\end{align}

\subsection{Wavefunction overlaps}

The wavefunction overlaps are calculated as in the main text and we simply display the results here.
\begin{multline}
N_1^{\rm cubic} = \frac{i\sqrt{2}k^{\te_M}_{(1,1)}R}{v_1}\left(k^{\fb_H}_{(1,u)} + k^{\fb_H}_{(1,w)}\right) \textrm{Im}(\alpha_1) -
    \frac{i\sqrt{2}k^{\fb_M}_{(1,1)}R}{v_2}\left(k^{\fb_H}_{(1,u)} - k^{\fb_H}_{(1,w)}\right) \textrm{Im}(\alpha_2)   \\
  +R^2\left(\frac{\alpha_h^*}{v_+^2} - \frac{\alpha_h}{v_-^2} \right) k^{\fb_M}_{(1,1)} k^{\te_M}_{(1,1)} -
    \frac{i\sqrt{2}k^{\fb_M}_{(1,2)}R}{v_2}\left(k^{\fb_H}_{(1,u)} + k^{\fb_H}_{(1,w)}\right) \textrm{Im}(\alpha_2) \\
  +\frac{i\sqrt{2}k^{\te_M}_{(1,2)}R}{v_1}\left(k^{\fb_H}_{(1,u)} - k^{\fb_H}_{(1,w)}\right) \textrm{Im}(\alpha_1) -
    R^2\left(\frac{\alpha_h^*}{v_+^2} - \frac{\alpha_h}{v_-^2} \right) k^{\te_M}_{(1,2)} k^{\fb_M}_{(1,2)} \;.
\end{multline}
To calculate the leading contribution triple couplings we rewrite the polynomial prefactors in the wavefunctions in the coordinates $u$ and $w$ and then expand them collecting factors of $w\bar{w}$, $u\bar{u}$, $u\bar{w}$ and $w\bar{u}$. Then making use of the integral (\ref{int1}) we can write this as
\begin{multline}
Y^{(p,q)}_{1,(0,m,l)} = \frac{N^{\rm cubic}_r}{N^{\fb_M,p}_1N^{\te_M,q}_1N^{\fb_H,0}_1\sqrt{2^ {p+q}m!l!}\left(\lambda^{\fb_H}_2\right)^{m/2}\left(\lambda^{\fb_H}_3\right)^{l/2}}\times \\
 \sum^p_{k_a=0} \sum^q_{k_b=0} \sum^m_{k_c=0} \sum^n_{k_d=0}
(-1)^{k_a}\begin{pmatrix}p\\ k_a\end{pmatrix}\begin{pmatrix}q\\ k_b\end{pmatrix}\begin{pmatrix}m\\ k_c\end{pmatrix}\begin{pmatrix}n\\ k_d\end{pmatrix} \left(\tilde{d}^2_w\right)^{m-k_c} \left(\tilde{d}^2_u\right)^{k_c}
    \left(\tilde{d}^3_w\right)^{n-k_d} \left(\tilde{d}^3_u\right)^{k_d}\\
 \left(k^{\fb_M}_{(1,1)} - k^{\fb_M}_{(1,2)}\right)^{p-k_a} \left(k^{\fb_M}_{(1,1)} + k^{\fb_M}_{(1,2)}\right)^{k_a}
\left(k^{\te_M}_{(1,1)} - k^{\te_M}_{(1,2)}\right)^{q-k_b} \left(k^{\te_M}_{(1,1)} + k^{\te_M}_{(1,2)}\right)^{k_b} \\
I\left(p+q-\mathrm{max}\left[k_a+k_b,k_c+k_d\right],\mathrm{min}\left[k_a+k_b,k_c+k_d\right], \delta_k\Theta\left(\delta_k\right),
  -\delta_k\Theta\left(-\delta_k\right)\, ;\, \mathfrak{p}_1,\mathfrak{p}_2,\mathfrak{p}_3,\mathfrak{p}_4 \right)
\end{multline}
Where $\Theta$ is the Heaviside theta functions, $\delta_k\equiv k_a+k_b-k_c-k_d$ and we have defined the quantities
\bea
\mathfrak{p}_1 &=& p_1^{\fb_H} + \frac12 \left[\left(p_1^{\fb_M}+p_2^{\fb_M}-p_3^{\fb_M}-p_4^{\fb_M}\right) + \left(p_1^{\te_M}+p_2^{\te_M}-p_3^{\te_M}-p_4^{\te_M}\right) \right] \;, \nn \\
\mathfrak{p}_2 &=& p_2^{\fb_H} + \frac12 \left[\left(p_1^{\fb_M}+p_2^{\fb_M}+p_3^{\fb_M}+p_4^{\fb_M}\right) + \left(p_1^{\te_M}+p_2^{\te_M}+p_3^{\te_M}+p_4^{\te_M}\right) \right] \;, \nn \\
\mathfrak{p}_3 &=& p_3^{\fb_H} + \frac12 \left[\left(-p_1^{\fb_M}+p_2^{\fb_M}-p_3^{\fb_M}+p_4^{\fb_M}\right) + \left(-p_1^{\te_M}+p_2^{\te_M}-p_3^{\te_M}+p_4^{\te_M}\right) \right] \;, \nn \\
\mathfrak{p}_4 &=& p_4^{\fb_H} + \frac12 \left[\left(-p_1^{\fb_M}+p_2^{\fb_M}+p_3^{\fb_M}-p_4^{\fb_M}\right) + \left(-p_1^{\te_M}+p_2^{\te_M}+p_3^{\te_M}-p_4^{\te_M}\right) \right] \;, \nn
\eea
and
\bea
\tilde{d}^i_w &=& \frac{1}{||\xi_i^{\fb_H}||}\left[k^{\fb_H}_{(i,u)} \left(p_3^{\fb_H}-\frac12 \alpha_h \right) - p_1^{\fb_H} k^{\fb_H}_{(i,w)} + \frac{R^3}{v_-}\left( \frac{\alpha_h}{v_-^2} - \frac{\alpha_h^*}{v_+^2} \right)\right] \;, \nn \\
\tilde{d}^i_u &=& \frac{1}{||\xi_i^{\fb_H}||}\left[k^{\fb_H}_{(i,w)} \left(p_4^{\fb_H}-\frac12 \alpha^*_h \right) - p_2^{\fb_H} k^{\fb_H}_{(i,u)} +\frac{R^3}{v_+} \left( \frac{\alpha_h}{v_-^2} - \frac{\alpha_h^*}{v_+^2} \right)\right] \;.
\eea

%%%%%%%%%%%%%%%%%%%%%%%%%%%%%%%%%%%%%%%%%%%%%%%%%%%%%%%%%%%%%%%%%
\section{Delocalisation and normalisation of wavefunctions}
\label{app:norm}
%%%%%%%%%%%%%%%%%%%%%%%%%%%%%%%%%%%%%%%%%%%%%%%%%%%%%%%%%%%%%%%%%

The local form of the wavefunctions (\ref{fz}) exhibits exponential localisation along all four real directions in $S$. Localisation onto a complex curve is induced by the Higgs vev, while localisation within the curve is induced by the flux. However the general form of the wavefunction (\ref{vphi}) need not be fully localised in this way. Indeed the arbitrary holomorphic prefactor can include an exponential so as to cancel the flux induced exponential localisation along one real direction of the matter curve. This situation is rather general for instance in toroidal compactifications \cite{Cremades:2004wa}. It is clear that such delocalisation cannot occur over the full complex curve since a holomorphic function cannot cancel the non-holomorphic flux induced localisation. The local form of the wavefunctions (\ref{fz}) are defined as the appropriate linear combinations of the, possibly delocalised, wavefunctions such that their expansion around the origin begins with increasing powers of $z$. This is a valid form to take for calculating the triple overlap since then the localisation of the other wavefunctions ensures that only this local form of the wavefunction contributes to the integral. However, for calculating the normalisation of the wavefunction the possible delocalisation can affect the results. Indeed the normalisation integral can diverge as an integral over ${\mathbb C}^2$ and must be cutoff at the KK scale of $S$ which for a homogenous manifold matches the local scale $R_{\parallel}$.

\begin{figure}[ht!]
{\begin{center}
                      \includegraphics[width=.45\textwidth]{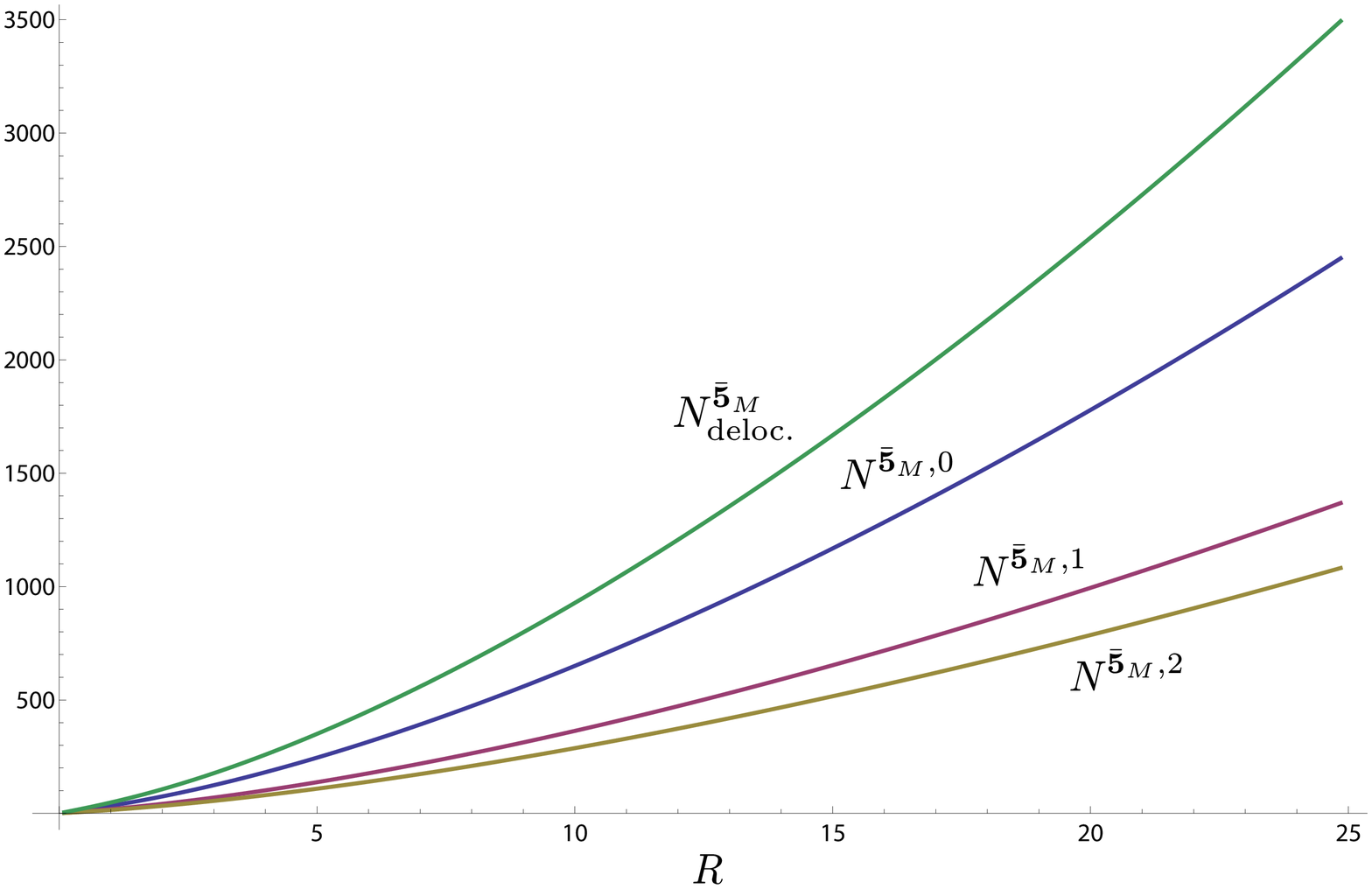}
\hspace{.3in}
 \includegraphics[width=.45\textwidth]{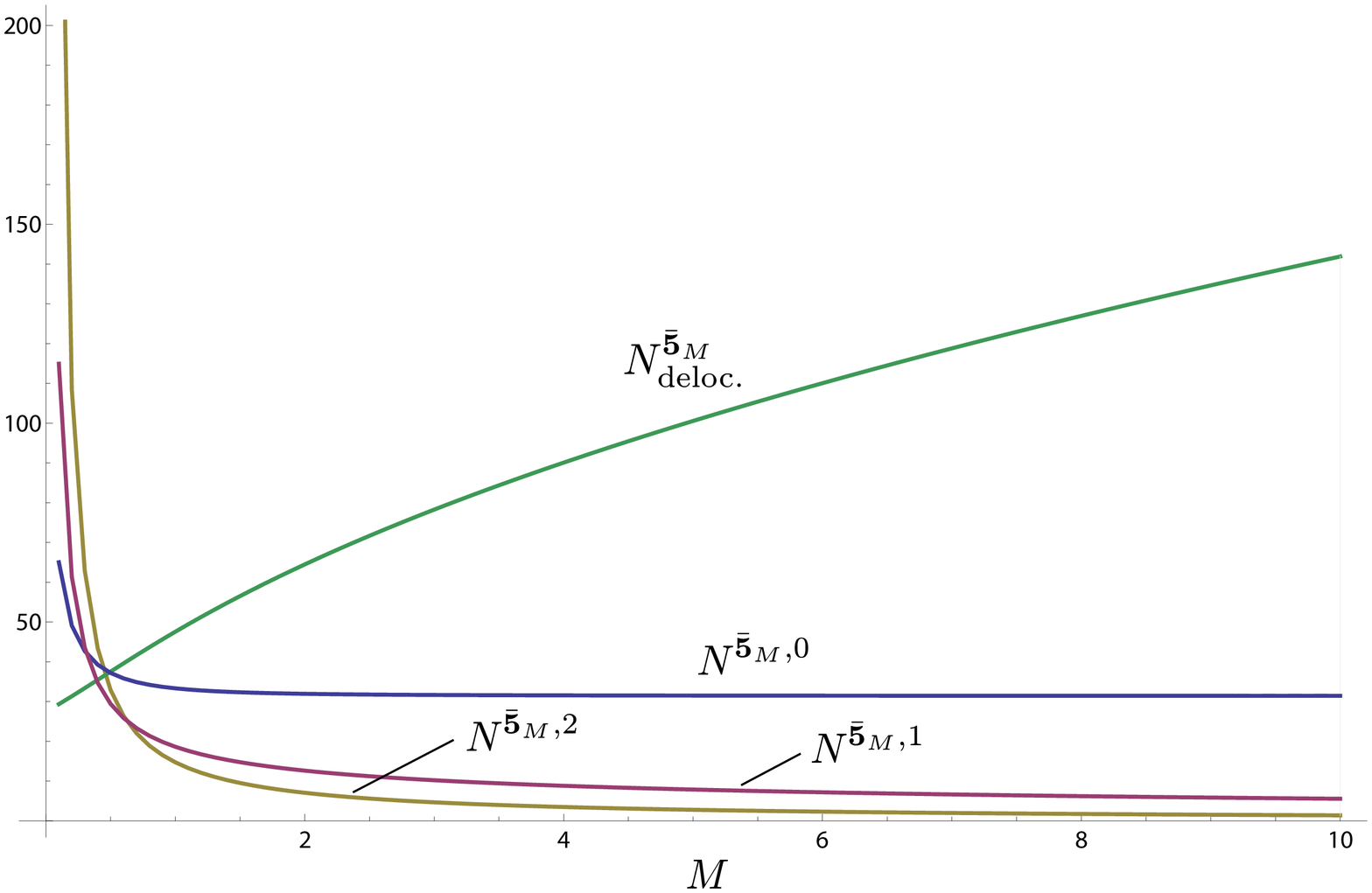}
 \caption{Normalisation factors for localised and delocalised wavefunctions for flux values $v_1=5/6$, $v_2=5/4$, $M_1=1.6 M$, $M_2=2 M$, $\gamma=0$, $\varepsilon=1/10$. The first plot fixes $M=1$ and varies $R$ while the second plot fixes $R=1$ and varies $M$.}
 \label{normrm}
 \end{center}}
\end{figure}

\begin{figure}[ht!]
{\begin{center}
 \includegraphics[width=12cm]{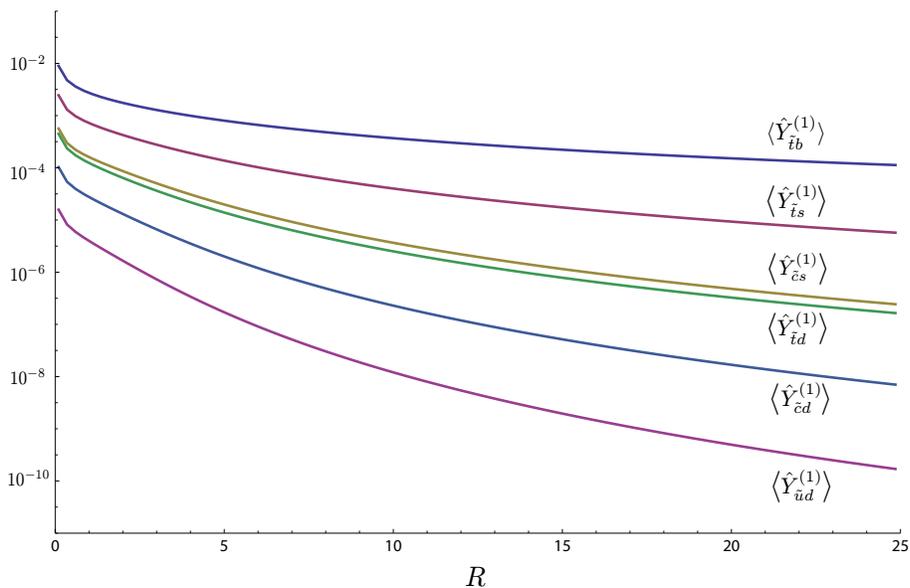}
 \caption{Plot showing the coupling as in figure \ref{figtriplewithoutflux} but with the normalisation of the matter curve wavefunctions calculated using the geometric average of the localised and delocalised normalisations.}
 \label{geoavnormplot}
 \end{center}}
\end{figure}

Determining whether a wavefunction delocalises along one real direction in this way is of course a global question and not answerable in general from a local perspective. Therefore the normalisation integral of the wavefunctions carries an ambiguity. In this appendix we study this ambiguity quantitatively. Our aim is to quantify how much the normalisation integral can change when a wavefunction delocalises in this way. We do this by plotting the normalisation integral as evaluated in (\ref{norma1}) for the different generations for different values of the flux and of $R$. We also plot for the same values of flux and $R$ the normalisation integral but with the exponential localisation factor along the curve dropped, thereby modeling the delocalisation effect. Of course dropping the exponential factor causes the wavefunction to delocalise along the full curve while we expect only a possible delocalisation along one real direction. But then this half-delocalistion can be expected to be somewhere close to the geometric average of the fully localised wavefunction and the wavefunctions with the exponential localisation along the curve dropped. Dropping the exponential localisation along the curve means that we perform the normalisation integral by explicit integration perpendicular to the curve while for the integral along the curve we simply take the homogenous answer of $R_{\parallel}^2$, similarly to the Higgs case in (\ref{wavenofluxnorm}). We plot the resulting normalisation factors in figure \ref{normrm}.
%%%%%%%%%%%%%%%%%%%%%%%%%%%%%%%%%%%%%%%%%%%%%%%%%%%%%%%%%%%%%%%%%%%%%%%%%%%%%%%%%%%
%%%%%%%%%%%%%%%%%%%%%%%%%%%%%%%%%%%%%%%%%%%%%%%%%%%%%%%%%%%%%%%%%%%%%%%%%%%%%%%%%%%%
As can be seen the ratio of the delocalised wavefunction to the localised ones does not depend strongly on $R$. The dependence on $M$ however is stronger which is expected given that the amplitude of the flux measures the degree of localisation along the curve. For large flux values the delocalised normalisation and localised ones, particularly for the lighter generations, differ significantly. However in the actual models we study the fluxes are fixed to be of order one, which means that the wavefunctions are already spread out to a significant fraction of $R_{\parallel}$ and hence, as can be seen in the plot, the normalisations do not differ significantly. This means that we can be confident that the normalisation of the wavefunctions that we use are correct up to order one factors even if the wavefunctions delocalise. To illustrate this we plot in figure \ref{geoavnormplot} the same couplings as in figure \ref{figtriplewithoutflux} except now we take the geometric average of the localised normalisation and the completely delocalised one, this being a good measure of delocalisation along one real direction. We see that the results do not differ significantly from figure \ref{figtriplewithoutflux}.
%%%%%%%%%%%%%%%%%%%%%%%%%%%%%%%%%%%%%%%%%%%%%%%%%%%%%%%%%%%%%%%%%%%%%%%%%%%%%%%%%%%%
%%%%%%%%%%%%%%%%%%%%%%%%%%%%%%%%%%%%%%%%%%%%%%%%%%%%%%%%%%%%%%%%%%%%%%%%%%%%%%%%%%%%
Note however that if taking the delocalised integral to be of order $R_{\parallel}$ is not accurate up to order one factors then these results may be modified. This can occur if the manifold is strongly inhomogeneous. In that case a more global calculation would be required to obtain more accurate results.

%******************************************************* BIBLIOGRAPHY **************************************************************************

\end{document}